\documentclass[12pt]{article}

\usepackage{latexsym,amsmath,amsfonts,amssymb}
\usepackage[latin1]{inputenc}
\usepackage[american]{babel}
\usepackage[dvips]{graphicx}
\usepackage{bbm}
\usepackage{bm}
\usepackage{subfigure}
\usepackage{url}
\usepackage{multirow}

\renewenvironment{thebibliography}[1]{%
\begin{oldthebibliography}{#1}%
\setlength\itemsep{-0.5mm}%
}%
{%
\end{oldthebibliography}%
}

\newcommand{\rep}[1]{\mathbf{#1}}

\numberwithin{equation}{section}

\newcommand{\nn}{\nonumber}
\newcommand{\mat}[1]{\begin{pmatrix} #1 \end{pmatrix}}
\newcommand{\be}{\begin{equation}} \newcommand{\ee}{\end{equation}}
\newcommand{\bea}{\begin{equation} \begin{aligned}} \newcommand{\eea}{\end{aligned} \end{equation}}
\newcommand{\tabs}{\rule[-1ex]{0pt}{3.5ex}}

\newcommand{\cA}{\mathcal{A}}

\newcommand{\cC}{\mathcal{C}}

\newcommand{\cE}{\mathcal{E}}

\newcommand{\cG}{\mathcal{G}}

\newcommand{\cI}{\mathcal{I}}

\newcommand{\cM}{\mathcal{M}}
\newcommand{\cN}{\mathcal{N}}
\newcommand{\cO}{\mathcal{O}}

\newcommand{\cQ}{\mathcal{Q}}

\newcommand{\cX}{\mathcal{X}}

\newcommand{\cZ}{\mathcal{Z}}

\newcommand{\bC}{\mathbb{C}}
\newcommand{\bF}{\mathbb{F}}

\newcommand{\bP}{\mathbb{P}}

\newcommand{\bR}{\mathbb{R}}
\newcommand{\bZ}{\mathbb{Z}}

\newcommand{\CP}{\bC\bP}

\DeclareMathOperator{\Tr}{Tr}

\begin{document}

\makeatletter \@addtoreset{equation}{section} \makeatother
\renewcommand{\theequation}{\thesection.\arabic{equation}}
\pagestyle{empty}
\rightline{WIS/02/12-JAN-DPPA}

\vspace{0.8cm}
\begin{center}
{\LARGE{\bf Seiberg duality for Chern-Simons quivers\\  and D-brane mutations
 \\[10mm]}} {\large{ Cyril Closset \\[5mm]}}
{\small{
{} Department of Particle Physics and Astrophysics \\
Weizmann Institute of Science, Rehovot 76100, Israel \\
}}

\medskip

\medskip

\medskip

\medskip

\vskip5cm

{\bf Abstract}
\vskip 20pt
\begin{minipage}[h]{16.0cm}
Chern-Simons quivers for M2-branes at Calabi-Yau singularities are best understood as the low energy theory of D2-branes on a dual type IIA background. We show how the D2-brane point of view naturally leads to three dimensional Seiberg dualities for Chern-Simons quivers with chiral matter content: They arise from a change of brane basis (or mutation), in complete analogy with the better known Seiberg dualities for D3-brane quivers. This perspective reproduces the known rules for Seiberg dualities in Chern-Simons-Yang-Mills theories with unitary gauge groups. We provide  explicit examples of dual theories for the quiver dual to the $Y^{p,q}(\CP^2)$ geometries. We also comment on the string theory derivation of CS quivers dual to massive type IIA geometries.

\end{minipage}
\end{center}

\newpage

\setcounter{page}{1} \pagestyle{plain}
\renewcommand{\thefootnote}{\arabic{footnote}} \setcounter{footnote}{0}

{
\tableofcontents}

\vspace*{1cm}


\section{Introduction}
Seiberg duality \cite{Seiberg:1994pq} is a powerful tool to study supersymmetric theories with four supercharges. When engineering a supersymmetric theory on D-branes in string theory, Seiberg duality typically arises as a change of ``brane basis'' of some sort.

In three dimensional theories with Chern-Simons (CS) interactions, some Seiberg dualities have been proposed by using branes suspended between branes in type IIB string theory \cite{Aharony:2008gk, Giveon:2008zn, Amariti:2009rb, Niarchos:2009aa}. In all these examples, the field theory has matter in real representations of the gauge group, and the duality is suggested by the brane creation effect when 5-branes cross each other \cite{Hanany:1996ie}.\footnote{One notable exception is \cite{Cremonesi:2010ae} which studies some theories with fundamental matter in non-real representations from the IIB setup as well.}

In this paper we study Seiberg duality for CS quiver theories constructed from type IIA string theory, following \cite{Aganagic:2009zk, Benini:2009qs, Jafferis:2009th, Benini:2011cma,Closset:2012ep}. There are at least two advantages to the type IIA approach. Firstly, it allows to engineer quivers with chiral matter content (by ``chiral'' we mean matter fields in non-real representations). Secondly, the connection between quivers and M2-branes on a Calabi-Yau (CY) fourfold singularity, which is the motivation to study such quivers \cite{Aharony:2008ug}, is rather transparent: The IIA setup is the KK reduction of the M-theory setup along a wisely chosen M-theory circle, and the quiver is the one describing D-branes at a CY threefold singularity in type IIA.

Recently, precise rules for Seiberg duality in three dimensional chiral quiver gauge theories were found in \cite{Benini:2011mf} from a careful field theory analysis. They generalize the Giveon-Kutasov rules \cite{Giveon:2008zn} that govern the non-chiral case. In this work we will provide a string theory explanation for such rules. The idea is very simple, once we review in some detail the beautiful relation between supersymmetric quiver theories and D-branes at Calabi-Yau threefold singularities: 3d Seiberg duality is a \emph{double} brane mutation. This point of view builds on a similar understanding for 4d Seiberg duality \cite{Aspinwall:2004vm, Herzog:2004qw}.

Consider a conical CY$_3$ $Y$ and a partial resolution $\pi : \tilde{Y}\rightarrow Y$. On any $\tilde{Y}$ we can consider probe branes (D2-branes in particular), and there exists a well-known relation between the branes on $\tilde{Y}$ and a quiver $\cQ$, defined abstractly. Any D-brane corresponds to some ``representation'' of the quiver. In that context Seiberg duality for D-branes on $\tilde{Y}$ is well understood as a change of brane basis \cite{Brandhuber:1997ta, Elitzur:1997hc, Cachazo:2001sg, Feng:2002kk}; in the case of interest to us the change of basis will be related to sheaf mutations \cite{Aspinwall:2004vm, Herzog:2004qw}.

The CS quivers we consider are engineered by considering D2-branes on a CY$_3$ $Y$ fibered on a line $\{r_0\} = \bR$. There are rather generic RR fluxes, including the $F_2$ flux corresponding to a non-trivial M-theory fibration, and the whole setup corresponds to M2-branes at the tip of a CY$_4$ cone. The CY$_3$ $Y$ is singular at the origin $r_0 =0$, but it is partially resolved for any $r_0 \neq 0$. Considering $Y$ as an algebraic variety, the partial resolution $\tilde{Y}$ might not be the same at $r_0>0$ or $r_0<0$, and we have thus two distinct CY$_3$  varieties $\tilde{Y}_{\pm}$. The CS quiver gauge theories for D-branes on such a setup can be found by taking an appropriate ``average'' of the theories we would obtain from $\tilde{Y}_-$ and $\tilde{Y}_+$ separately. The Chern-Simons levels are induced on the branes by the RR background fluxes.

In that setup the 3d Seiberg duality is a \emph{doubling} of the mutation procedure, performed both on $\tilde{Y}_-$ and $\tilde{Y}_+$. We will explain this in detail after introducing the necessary formalism.

In section  \ref{section: Dbranes and SD, general} we review the deep relationship that exists between D-branes at $Y$ and a quiver $\cQ$.
Considered abstractly, the quiver is just an oriented graph. To that abstract quiver we can associate a choice of gauge group $\prod_i U(N_i)$, which is part of a choice of \emph{quiver representation}: different representations correspond to different D-brane configurations on $Y$. When dealing with D3-brane quivers this point of view is a bit less useful, because the choice of ranks $N_i$ are essentially fixed by anomaly cancelation. Our 3d theories have completely generic ranks $N_i$, which makes the abstract point of view more necessary. More background material is provided in several Appendices.

In section \ref{sec: SD for 3d CS quivers} we start by reviewing some relevant facts about the moduli space of Chern-Simons quivers, following the recent work \cite{Closset:2012ep}. We then present the formalism of \cite{Benini:2011cma, Closset:2012ep} providing the map  from (some) M-theory/IIA backgrounds to CS quivers. We next proceed to state our main result, relating 3d Seiberg duality for chiral quivers to brane mutations.

In section \ref{examples: dP0 Seiberg duals} we apply our formalism to the CS theories dual to $Y^{p,q}(\CP^2)$, we derive Seiberg dual quivers for the whole family of theories of \cite{Benini:2011cma}, and we provide some consistency checks of the result. Our mutation perpective on 3d Seiberg duality has also been applied to the study of many toric examples in \cite{Closset:2012ep}.

Finally, in section  \ref{section: brane charges}, which is independent from the rest of the paper, we briefly comment on the correct brane charge to use in the formalism of \cite{Benini:2011cma, Closset:2012ep}, allowing to generalize those works to cases when the IIA background contains $F_0$ flux and has no eleven dimensional M-theory dual \cite{Aharony:2010af}.

\section{D-branes and Seiberg duality}\label{section: Dbranes and SD, general}
In this section we review supersymmetric quiver theories from the perspective of D-branes and the precise sense in which Seiberg duality for quivers arises as a change of ``brane basis''.

\subsection{$CY_3$ Quivers and Seiberg duality}
A quiver $\cQ$ is an oriented graphs which can have multiples arrows between the nodes. It also allows for closed loops. Let $Q_0= \{1, 2, \cdots, G \}$ be the set of nodes (indexed by the letters $i,j$) and $Q_1= \{a_1,a_2, \cdots, a_E \}$ the set of arrows; the notation $a_{ij}$ denotes an arrow from $i$ to $j$. We are interested in quivers with superpotential $(\cQ, W)$. The superpotential is a formal sum of words (up to cyclic permutation),
\be
W = \sum_{L} a_{i_1 i_2} a_{i_2 i_3}\cdots a_{i_{|L|}i_1}\, ,
\ee
where the summands correspond to closed loops $L=\{i_1 i_2\cdots i_{|L|}\}$ in the quiver. We denote by $Q_2$ the set of all loops appearing in $W$. The quiver with superpotential is a quiver with relations $\{F_a=0\}$, where the relations between the path are given by formal derivation
\be\label{F-term rels}
F_a = \frac{\partial}{\partial a}W \, ,\qquad \qquad \forall \, a\in Q_1\, .
\ee
In fact the quivers we consider are even more restricted and correspond to $CY_3$ quivers. We define (rather tautologically)  a $CY_3$ quiver as a quiver that arises from D-branes at a CY$_3$ cone; see \cite{2006math.....12139G} and references therein for a mathematical definition of the concept%
\footnote{From now on when we write $\cQ$ we mean a $CY_3$ quiver $(\cQ, W)$.}.

Our motivation to study $\cQ$ is that it summarizes the classical Lagrangian of a supersymmetric field theory, which is itself the low energy theory on D-branes on a Calabi-Yau threefold. The more abstract point of view that is taken here is however very useful even for physicists --- see \emph{e.g.} \cite{Douglas:2000qw, Berenstein:2002fi}.

The basic notion that we will need is that of a \emph{quiver representation}. Some general facts on quivers are reviewed in Appendix \ref{appendix: on quiver and such}. A representation $X$ of $\cQ$ is a choice of vector space $V_i$ for each node and a choice of linear map%
\footnote{The map $X_a: i\rightarrow j$ goes in the same direction as the corresponding arrow in the quiver; if written $X_{ij}$ it goes from left to right in the subscript. This gives us the unusual convention that the composition of maps is written $X_a X_b : i \rightarrow k$ for $X_a: i\rightarrow j$ and $X_b: X_j\rightarrow k$. }
 $X_a: V_i \rightarrow V_j$ for each arrow $a=a_{ij}$, such that the $X_a$'s satisfy the relations (\ref{F-term rels}). More concretely, we consider only complex vector spaces $V_i = \bC^{N_i}$ and thus the linear maps $X_a$ are complex-valued $N_j \times N_i$ matrices.

A \emph{morphism} $\phi$ between two quiver representations $X$ and $X'$ is a set of $G$ linear maps $\phi_i : V_i \rightarrow V_i'$ such that
\be
\phi_i X_a' = X_a \phi_j \, , \qquad \forall\, a=a_{ij} \in Q_1\, .
\ee
If $\phi$ is injective, we say that $X$ is a \emph{subrepresentation} of $X'$. On the other end two representations $X$, $X'$ are called isomorphic (or gauge equivalent) if there exist a bijective morphism $\phi: X\rightarrow X'$. We should really only consider equivalence classes under such relations. The dimension vector of any representation $X$ is defined as
\be
\text{dim}\, X\equiv (\text{dim}\, V_1,\cdots, \text{dim}\, V_G ) = (N_1, \cdots, N_G) \equiv \bm{N}.
\ee

A choice of dimension vector determines a $\cN=1$ supersymmetric quiver theory $(\cQ, \bm{N})$ in four dimensions, and more generally a theory with four supercharges in dimension $\leq 4$. The gauge group is
\be\label{generic gauge group of Q}
\cG = U(N_i)\times \cdots \times U(N_G)
\ee
and the arrows $a_{ij}\in Q_1$ correspond to chiral superfields $X_{ij}$ in  bifundamental representations. In the following we will assume there are no arrow $a: i\rightarrow i$ (no chiral multiplets in adjoint representations), although this case can be discussed as well \cite{Berenstein:2002fi}. So far everything in the discussion is holomorphic, and it is well known that as far as the F-terms are concerned the actual gauge group of the theory $(\cQ, \bm{N})$ is complexified, $\cG_{\bC}= \prod_{i=1}^G\, GL(N_i, \bC)$. The set of all quiver representations (more precisely $\cG_{\bC}$ equivalence classes of representations) with fixed dimension vector  correspond to all possible VEVs for the fields $X_{ij}$ which satisfy the F-term relations (\ref{F-term rels}), quotiented by $\cG_{\bC}$. In other words, the moduli space of quiver representations of fixed dimension vector $\bm{N}$ is the same as the vacuum moduli space of physicists. For the moment we leave aside the issue of how exactly one solves the D-term equations of the quiver theory, content with the fact that such a solution always exists along the complexified gauge orbits \cite{Luty:1995sd} (in the absence of FI parameters).

To discuss Seiberg duality for quivers in full generality, we would need to introduce a few more abstract facts about quivers. In the following we will discuss some simpler notions of Seiberg duality, which are less general, but for completeness we provide a summary of the general case in Appendix \ref{appendix: on quiver and such}.

\subsection{Purely chiral quivers and quiver mutations}\label{subsec: quiver mutations}
The approach to Seiberg duality of \cite{Berenstein:2002fi}, reviewed in Appendix, is very beautiful and general, but explicit computations are a bit hard for anyone not familiar with higher mathematics. In the following we present a more concrete definition of Seiberg duality based on the mathematical work of \cite{2007arXiv0704.0649D}. It only applies to a subset of CY$_3$ quivers, but it will turn out to be the most interesting ones for our later purposes.%
\footnote{It has been shown  that the approach of \cite{2007arXiv0704.0649D} leads to a derived equivalence \cite{2009arXiv0906.0761K} , and according to Rickard's theorem it is thus a tilting equivalence, in agreement with the Douglas-Berenstein picture of Seiberg duality.}

Let us denote by $A$ the antisymmetric adjacency matrix of the quiver $\cQ$, defined by
\be
A_{ij} = n_{ij}-n_{ji}\, ,
\ee
where $n_{ij}$ is the number of arrows $a_{ij}$ from $i$ to $j$. We call a quiver ``purely chiral'' if $A$ is enough to reconstruct $\cQ=(Q_0, Q_1)$, namely if there are no loops of two arrows ($ab$ such that $a: i\rightarrow j$ and $b: j\rightarrow i$) and no arrow $a: i\rightarrow i$ from a node to itself.

A quiver mutation $\mu_k(\cQ)$ on a node $k\in Q_0$ is defined as expected from physics: We reverse all the arrows of the form $a_{ik}$ and $a_{kj}$, introduce new arrows $m_{ij}=a_{ik}a_{kj}$ and a new superpotential $\mu_k(W)=W+ \tilde{a}_{ki}\, m_{ij}\, \tilde{a}_{jk}$ written in term of the new arrows, and finally remove any arrows that appear in a term of order $2$ in the new $W$ while imposing the corresponding relations $F_a=0$. Of course this is nothing but the usual prescription for Seiberg duality \cite{Seiberg:1994pq}, including the integrating out of massive mesons. The interest for us is that is has been abstracted to a quiver $\cQ$ instead of any particular quiver representation $(\cQ, \bm{N})$.
The effect on the adjacency matrix is a so-called matrix mutation at position $k$ \cite{2007arXiv0704.0649D}
\be\label{quiver mutation on A}
\mu_k \left(A_{ij}\right) \, =\, \begin{cases}  -A_{ij}  \qquad & \quad \text{if} \; i=k \; \text{or}\; j=k\\
                                                A_{ij}+ \frac{1}{2}\left(|A_{ik}|A_{kj} +A_{ik}|A_{kj}|\right) &\quad \text{otherwise}\end{cases}
\ee
For a given supersymmetric theory $(\cQ, \bm{N})$, the mutation $\mu_k$ can give rise to different actions on the dimension vector $\bm{N}$ depending on the precise vacuum of $(\cQ, \bm{N})$ \cite{2009arXiv0906.0761K}. Two cases are of particular physical interest:
\bea\label{L and R mutation on Q reps}
\mu_k^{L}\, :&\,\quad  N_i' = N_i \, , \qquad \text{if}\; i\neq k\, , \qquad \quad N_k' = -N_k + \sum_i\, [A_{ik}]_+ N_i\, ,\\
\mu_k^{R}\, :&\,\quad N_i' = N_i \, , \qquad \text{if}\; i\neq k\, , \qquad \quad N_k' = -N_k + \sum_j\, [A_{kj}]_+ N_j\,
\eea
 (with the definition $[x]_+ = \max(x,0)$), which we call left- and right-mutation of quiver representation, respectively. This reproduces the Seiberg duality of four dimensional quivers (at the classical level), but it is also more general. In four dimensions, anomaly cancelation restricts the allowed theories $(\cQ, \bm{N})$ according to the condition
\be
A\bm{N}=0\, .
\ee
In three dimensions there is no such restriction.

\subsection{From D-branes to quivers}\label{subsec: from Dbrane to quivers}
Consider a background space-time $\bR^{1,3}\times Y$ in type II string theory, where $Y$ is a non-compact conical Calabi-Yau threefold, together with some D-branes transverse to $Y$ ---for our purposes these will be D3-branes in type IIB or D2-branes in type IIA. At the conical singularity of $Y$, these D-branes usually decay marginally into a bound state of so-called fractional branes, which we denote $\{E^{\vee}_i\}_{i=1, \cdots, G}$. The fractional branes can be loosely thought of as D-branes wrapped on vanishing cycles \cite{Diaconescu:1997br}. The local dynamics of the fractional branes in well encoded in a supersymmetric quiver gauge theory, which describes the massless open string sector of the theory.

In general we cannot compute the open string spectrum at the CY singularity $Y$ directly since we lack a perturbative definition of string theory on such spaces%
\footnote{The exception being $Y= \bC^3/\Gamma$ an orbifold of flat space, in which context D-brane quivers were first discovered \cite{Douglas:1996sw}.}. Nevertheless the question of finding the quiver $\cQ_Y$ corresponding to any given $Y$ has been much studied in the past decade \cite{Cachazo:2001sg,Wijnholt:2002qz, Herzog:2003zc, Herzog:2005sy}, and it has been solved in the toric case \cite{Franco:2005sm, Hanany:2005ss, Kennaway:2007tq, Gulotta:2008ef}.

Consider a crepant resolution $\tilde{Y}$ of the cone $Y$,
\be
\pi : \tilde{Y}\rightarrow Y \, , \qquad \qquad \text{with}\qquad \cO_{\tilde{Y}}(K) \cong \pi^*\cO_{Y}(K)\, ,
\ee
where $\cO_M(K)$ denotes the canonical line bundle (or canonical sheaf) of a variety $M$. For simplicity we will restrict our study to the case when $\tilde{Y}$ has one and only one exceptional divisor $B_4$. The surface $B_4$ is then a 2-complex dimensional Fano variety (not necessarily smooth), and we have
\be\label{Y as O(K)}
\tilde{Y}\,=\, \cO_{B_4}(K)\, .
\ee
The main reason to write $\tilde{Y}$ as (\ref{Y as O(K)}) is that fractional branes must be wrapped on compact cycles, and therefore correspond to branes on $B_4$. As reviewed in Appendix \ref{sec:App:sheaves and quivers}, our branes are generally chain complexes of sheaves (B-branes),
\be
\label{generic element of D(Y) 00}
\mathsf{E}\;=\; \cdots \rightarrow E_{(-2)} \rightarrow E_{(-1)} \rightarrow E_{(0)} \rightarrow E_{(1)} \rightarrow \cdots \;.
\ee
In this work we will only be concerned with the \emph{charges} of the branes, so that we can skip most of the B-brane category mumbo-jumbo. The charge of a B-brane (\ref{generic element of D(Y) 00}) is defined as its Chern character
\be\label{Chern character of B brane}
ch(\mathsf{E})=\sum_n (-1)^n \,ch(E_{(n)}) \;,
\ee
where $ch(E_{(n)})$ are the Chern characters of the individual sheaves. We refer to section \ref{section: brane charges} for a more careful discussion of brane charges. To discuss the most general brane charges we consider a full resolution $\pi : \tilde{B}_4\rightarrow B_4$, which is what the B-model probes. Let us denote
\be\label{def 2 cycles and cI}
\cC_{\alpha} \in H^2(\tilde{B}_4, \bZ)\, , \qquad \alpha= 1, \cdots, m\, , \qquad \quad\quad \cC_{\alpha}\cC_{\beta}= \cI_{\alpha\beta}\, .
\ee
a primitive basis of 2-cycles, with $\cI$ the intersection matrix. There are $G\equiv m+2$ charges for the compactly supported branes (branes wrapping $B_4$, $\cC_{\alpha}$ or a point), and we denote the Chern character (\ref{Chern character of B brane}) of a generic brane $\mathsf{E}$ by the covector
\be\label{ch as vector}
\bm{Q}_{\text{branes}}(\mathsf{E}) \,=\, ch(\mathsf{E})\,=\, (rk(\mathsf{E}), c_1(\mathsf{E}), ch_2(\mathsf{E}))\, .
\ee
The Euler character of a pair of branes is defined by
\be
\chi(\mathsf{E}_i, \mathsf{E}_j)\,\equiv\, \sum_q (-1)^q \, \mathrm{dim}\, \mathrm{Ext}^q(\mathsf{E}_i, \mathsf{E}_j) \;,
\ee
and can be computed by the Riemann-Roch theorem:
\bea\label{pairing of E, F}
\chi(\mathsf{E}_i, \mathsf{E}_j) \,=\, \int_{\tilde{B}_4} \,ch(\mathsf{E}_i^*)\, ch(\mathsf{E}_j)\, Td(\tilde{B}_4) \;.
\eea
We can conveniently rewrite this in matrix notation,
\be
\label{definition of X}
\chi(\mathsf{E}_i,\mathsf{E}_j) = ch(\mathsf{E}_i) \, \mathbf{X}_{\tilde{B}_4} \, ch(\mathsf{E}_j)^T \;, \qquad \quad \text{with}\qquad
\mathbf{X}_{\tilde{B}_4} = \mat{ 1 & \frac12 c_1 & 1 \\ - \frac12 c_1 & - \cI & 0 \\ 1 & 0 & 0 }\, ,
\ee
where $\cI$ is defined in (\ref{def 2 cycles and cI}) and $c_1= c_1(\tilde{B}_4)$; thus the matrix $\mathbf{X}_{\tilde{B}_4}$ is intrinsic to the geometry we consider.

Suppose that we are given a collection of $G$ sheaves
\be\label{tilting coll notation}
\cE=\{E_1,E_2,  \cdots, E_G\}
\ee
which generates all the B-branes on $\tilde{B}_4$. Such a collection is called a \emph{tilting collection} if moreover
\be\label{condition tilting coll 000}
\text{Ext}^q(E_i, E_j)= 0\, \qquad  \qquad\forall q>0\, .
\ee
We refer to \cite{Aspinwall:2004jr} for a good introduction to $\text{Ext}$ groups.
Given the tilting collection (\ref{tilting coll notation}), we expect that the sheaves
\be
\mathsf{P}_i \equiv \pi^{\ast}E_i
\ee
form a tilting collection on the resolved cone $\pi:\tilde{Y}\rightarrow Y$; see for instance \cite{Herzog:2006bu} for similar relations. We will leave this as a conjecture.%
\footnote{The more precise statement of the conjecture is that there is a one-to-one equivalence between $\{\mathsf{E}_i\}$ being a tilting collection with respect to the B-brane category on $B_4$ and $\{\mathsf{P}_i\}$ being a tilting collection with respect to the category of \emph{compactly supported} B-branes on $\tilde{Y}$.}
For $\{\mathsf{P}_i\}$ a tilting collection is follows very generally that the algebra
\be\label{Q algebra from Es}
\cA= \mathrm{End}(\, \oplus_{i=1}^G\, \mathsf{P}_i)^{\, \text{op}}
\ee
is the path algebra of a fractional brane quiver.%
\footnote{See the Appendix for more background on this.  Under the correspondence $D(\text{Coh}Y)\rightarrow D(\cA_{\cQ}\text{-mod})$ the sheaves $\mathsf{P}_i$ map to the projective modules $P_i$ of the quiver. However we are more interested in the B-branes corresponding to the simple $\cA$-modules $e_i$ ---the fractional branes---, which we denote $\mathsf{E}_i^{\vee}$ in the following. The main reason we had to restrict our analysis to CY threefolds of the type (\ref{Y as O(K)}) is because only in that case do we known how to reconstruct $\{\mathsf{E}_i^{\vee}\}$ from the $\mathsf{P}_i$'s.} We are more interested in the collection of \emph{fractional brane}, denoted
\be\label{frac brane collection Evee}
\cE^{\vee}= \{\mathsf{E}^{\vee}_1, \mathsf{E}^{\vee}_2, \cdots, \mathsf{E}^{\vee}_G\}\, .
\ee
In some simple cases they can be written explicitly through sheave mutations \cite{Herzog:2003zc, Aspinwall:2004vm, Herzog:2004qw, Herzog:2005sy}. To compute their charge we only need to know that they are dual to the $E_i$'s in the sense of the Euler character,
\be\label{E Ev Chi-dual}
\chi(E_i, \mathsf{E}_j^{\vee})= \delta_{ij}\, ,
\ee
and we take (\ref{E Ev Chi-dual}) as our practical definition of (\ref{frac brane collection Evee}).
Let us also define the matrix
\be\label{definition S}
S_{ij} \equiv \chi(E_i, E_j)= \mathrm{dim}\, \text{Hom}(E_i, E_j) \;,
\ee
where the second equality holds because of (\ref{condition tilting coll 000}).
We introduce the two $G\times G$ matrices
\be\label{definition Q Qv}
Q = \mat{ ch(E_1) \\ \vdots \\ ch(E_G) } \;,\qquad\qquad
    Q^\vee = \mat{ ch(\mathsf{E}_1^\vee) \\ \vdots \\ ch(\mathsf{E}_G^\vee) } \;.
\ee
In term of these charge matrices we can rewrite (\ref{E Ev Chi-dual}) and (\ref{definition S}) in a compact way:
\be
Q^{\vee T }= (\mathbf{X}_{\tilde{B}_4})^{-1}\, Q^{-1} \, , \qquad \qquad S= Q\, \mathbf{X}_{\tilde{B}_4}\,  Q^T\, .
\ee
The antisymmetric adjacency matrix $A$ of the quiver $\cQ$ associated to (\ref{Q algebra from Es}) can be found from $S$, according to
\be\label{adj matrix from S}
A\equiv S^{-1 T}- S^{-1}\, .
\ee
Assuming the correct quiver is purely chiral, this completely determines it as a graph%
\footnote{To work out the more general case, one should compute separately the various $Ext$ groups instead of just $\chi(\mathsf{E}_i,\mathsf{E}_j)$.}. This simple technique does not determine the quiver superpotential, however, which can in principle be extracted by more careful computations in the B-brane category \cite{Aspinwall:2005ur}.

The matrix $Q^{\vee}$ is the \emph{dictionnary} that allows to translate between the brane charge basis (\ref{ch as vector}) and the fractional brane basis (ranks of the gauge groups in the quiver), according to
\be\label{rel N and Qbrane}
\bm{Q}_{\text{branes}} = \bm{N} \, Q^{\vee}\, .
\ee
In particular, for a point-like D-brane $\cO_p$, we have $\bm{Q}_{\text{branes}}= (0,\cdots, 0, 1)$, and
\be\label{definition of ri}
\bm{N} \,=\,  (0,\cdots, 0, 1)(Q^{\vee})^{-1}\, = (rk(E_1), \cdots, rk(E_G)) \equiv (r_1, \cdots, r_G) = \bm{r}\, ,
\ee
namely the ranks of the supersymmetric quiver are given by the ranks of the sheaves in the tilting collection (\ref{tilting coll notation}), which we will denote $(r_i)$. Remark that the theory on  $\cO_p$ will have an Abelian gauge group if and only if the corresponding tilting collection is a collection of line bundles ---for instance this is what happens in so-called toric quivers \cite{Hanany:2006nm, 2009arXiv0909.2013B}.

\subsection{Mutations, branes and quivers}\label{sec: mutation, branes and Qs}
Given a particular tilting collection on $B_4$ called a (complete) strongly exceptional collection%
\footnote{This is an ordered tilting collection such that the matrix $S$ in (\ref{definition S}) is upper-triangular (and with $1$'s on the diagonal). In many cases we can order the nodes such that $S$ has the correct form, and the results of \cite{Herzog:2004qw} follow, but we will need a more general conjecture.},
it is known that one can generate another strongly exceptional collection through so-called sheaf mutations \cite{Herzog:2004qw}.

We conjecture that, given a tilting collection (\ref{tilting coll notation}), it is possible to obtain another collection of B-branes which is again tilting by a similar mutation of sheaves, which exactly parallels the notion of quiver mutation. Let $\cQ$ be the quiver associated to the tilting collection $\cE$ and let us choose an element $E_k \in \cE$. Let us denote by $I_L\subset Q_0$ (resp. $I_R\subset Q_0$) the set of nodes which are connected to $k$ by incoming (resp. outgoing) arrows in $\cQ$. At the level of  branes charges, a ``left (resp. right) mutation'' on $E_k$ is defined by
\bea\label{L and R mutation on sheaves}
\mu_k^{L}\, :&\,\quad  ch(E_i') = ch(E_i) \, , \quad \text{if}\; i\neq k\, , \qquad \quad ch(E_k') = -ch(E_k) + \sum_{i \in I_L}\, A_{ik}\, ch(E_i)\, ,\\
\mu_k^{R}\, :&\,\quad ch( E_i') = ch(E_i) \, , \quad \text{if}\; i\neq k\, , \qquad \quad ch(E_k') = -ch(E_k) + \sum_{j \in I_R}\, A_{kj}\, ch(E_j)\, ,
\eea
where $A$ is defined in (\ref{adj matrix from S}). One can consider (\ref{L and R mutation on sheaves}) as our  working definition of brane mutation, at the level of charges (which is all we need in this work).

We introduce a matrix $M_k$ (also written $M_{(k;L)}$ or $M_{(k;R)}$ if we need to distinguish between left or right mutation) such that (\ref{L and R mutation on sheaves}) can be written
\be
\mu_k^{L,R}\, : \; Q \;\mapsto\; Q' \,=\, M_{(k; L,R)}\,  Q\, ,
\ee
in term of the charge matrix  $Q$ in (\ref{definition Q Qv}). Remark that $M_k^2=1$, and thus applying the mutation twice at the same node gives back the original collection; however if $M_k$ is a left mutation of the collection $\cE$ it will be a right mutation of the collection $\cE'=\mu_k(\cE)$, and vice versa. We can easily see that
\be
S' \,=\, M_k\, S\, M_k^T\, , \,\qquad \qquad A' = M_k^T\, A\, M_k\, ,
\ee
and
\be\label{mutation of dictionary Qv}
Q^{\vee}\phantom{}'= M^T_k\, Q^{\vee}\, .
\ee
Comparing to (\ref{quiver mutation on A}) one can see that brane mutation and quiver mutation of a purely chiral quiver are one and the same thing.

From (\ref{mutation of dictionary Qv}) and (\ref{rel N and Qbrane}) it also follows that the dimension vector of a quiver representation transforms as
\be
\bm{N}' = M_k \bm{N}\, ,
\ee
in agreement with the definitions (\ref{L and R mutation on Q reps}).

\subsection{K\"ahler moduli space and quiver locus}\label{subsection: KMS and quiver locus}
In the previous subsections we introduced some B-branes states $\cE= \{\mathsf{E}_i^{\vee}\}$ on a resolved cone $\tilde{Y}$ and we reviewed how such a set of branes is related to a supersymmetric quiver theory. To relate this to the intuitive notion of fractional branes at a singularity, we should show that these $\mathsf{E}^{\vee}$'s are indeed individually BPS and mutually BPS, when placed at the tip of $Y$. While in some rare cases \cite{Diaconescu:1999dt, Aspinwall:2004vm} this can be proven directly using mirror symmetry, in general we will simply assume that this is true and see what it entails.

The K\"ahler moduli space $\cM_K$ of $Y$ has real dimension $2m$.
Let $Z(\mathsf{E})$ be the central charge of a B-brane $\mathsf{E}$ (see Appendix \ref{App: rel FI and Kparam} for more details).
A B-brane is BPS at a given point in $\cM_K$ if it is $\Pi$-stable \cite{Douglas:2000ah}; we do not review this notion here, but we will review the quiver analog of  $\Pi$-stability in the following. The point-like D-brane $\cO_p$ is stable for any value of the K\"ahler moduli, and by convention we set
\be
Z(\cO_p)=1\, .
\ee
Assuming they are both $\Pi$-stable, two different D-branes $\mathsf{E}^{\vee}_i$ and $\mathsf{E}^{\vee}_j$ are mutually BPS if and only if their central charges have the same phase,
\be
\arg(Z(\mathsf{E}^{\vee}_1)) =\arg(Z(\mathsf{E}^{\vee}_2))\, .
\ee
In particular, a D-brane state $\mathsf{E}^{\vee}$ is mutually BPS with $\cO_p$ if and only if $Z(\mathsf{E}^{\vee})$ is real and positive. At small volumes (and in particular in the conical limit $\tilde{Y}\rightarrow Y$), $\alpha'$ corrections become large and the notion of D-branes as sheaves ceases to be valid, while the quiver description is the correct one. The \emph{quiver locus} $\cM_{Q}$ in K\"ahler moduli space is the subspace of $\cM_K$ where the $G= m+2$ fractional branes (\ref{frac brane collection Evee}) are mutually BPS. Requiring the phases of  all $Z(\mathsf{E}_i^{\vee})$ ($i=1, \cdots, m+2$) to align cuts out a subspace of codimension $m+1$ in $\cM_K$, $\cM_{Q} \subset \cM_K$ \cite{Aspinwall:2004vm}.%
\footnote{
As a simple example, consider the case $B= \CP^2$ studied in detail in \cite{Diaconescu:1999dt, Douglas:2000qw}. In that case $\cM$ has real dimension two, parameterized by $t= \int_H (B+iJ)$, and the quiver locus $\cM_Q$ is just a point (usually called the orbifold point).}
In general, the quiver locus will be located at
\be
\chi_{\alpha} = 0\, ,
\ee
(that is, $Y$ is conical), together with one particular constraint on the $m$ B-field periods $b_{\alpha}$, which we denote $\chi_0=0$.
D-branes on the quiver locus are described by the quiver representations. The closed string modes $\chi_{\alpha}$, $\chi_0$ couple to the D-branes as Fayet-Iliopoulos (FI) terms \cite{Douglas:1996sw}, which allow to probe $\cM_K$ transversally to $\cM_Q$%
\footnote{On the other hand moving along $\cM_Q\subset \cM_K$ corresponds to some motion on the conformal manifold of the low energy CFT on the D-branes, when it exists.}.
Let  $\bm{\xi}= (\xi_i)$ be the FI terms associated to the fractional branes $\mathsf{E}_i^{\vee}$, and let us define
$\bm{\chi} = (\chi_0, \chi_{\alpha}, 0)$, corresponding to the $m+1$ K\"ahler parameters transverse to $\cM_Q$. We show in the Appendix that
\be\label{dico xi to chi 0}
\bm{\xi} \, = \, Q^{\vee}\, \bm{\chi}\, ,
\ee
where $Q^{\vee}$ is the brane charge dictionary introduced in section \ref{subsec: from Dbrane to quivers}. One can easily see that $\sum_i r_i\, \xi_i=0$ by construction, with the $r_i$ defined in (\ref{definition of ri}).

\subsection{$\theta$-stability, moduli space and K\"ahler chambers}\label{sec: stability}
Near the quiver locus, the best handle we have on the D-branes is the quiver itself. A brane is stable is it corresponds to a quiver representation which is $\theta$-stable in the sense of King \cite{King1994}%
\footnote{See in particular \cite{Aspinwall:2004mb}. We define $\theta$-stability with $<0$ rather than $>0$ in \cite{King1994} to agree with physics conventions once we identify $\theta=\xi$ below.}: Given a vector $\bm{\theta} \in \bZ^G$, a quiver representation of dimension vector $\bm{N}$ is $\theta$-stable (resp. semi-stable) if and only if $\bm{\theta}\bm{N}=0$ and for any proper subrepresentation of dimension $\bm{N}'$ we have $\bm{\theta}\bm{N}'<0$ (resp. $\bm{\theta}\bm{N}'\leq 0$).

Consider a quiver with gauge group (\ref{generic gauge group of Q}) and Fayet-Iliopoulos (FI) terms $\bm{\xi}$. The moduli space of a supersymmetric quiver $(\cQ, \bm{N})$ is usually described by a  K\"ahler quotient
\be\label{KQ description of VMS}
\cM(\cQ, \bm{N}; \bm{\xi})_{K} \, =\, (X_a | \partial W=0) //_{\bm{\xi}} \; \cG\, ,
\ee
where the moments maps at level $\bm{\xi}$ correspond to the (4-dimensional) D-terms equations
\be\label{D terms equs in general}
\sum_{X_a | s(a)=i} X_a^{\dagger}X_a - \sum_{X_b | t(b)=i} X_b X_b^{\dagger} \, = \, \xi_i\, .
\ee
We can describe the same moduli space algebraically as a GIT (Geometric Invariant Theory) quotient, which is a quotient by the complexified gauge group $\cG_{\bC}$ \cite{King1994, Martelli:2008cm}. To perform a GIT quotient we need to pick a $\bm{\theta}\in \bZ^G$ (the $\bm{\theta}=0$ case is well-known to physicists, the general case a bit less so). Let $\cZ= \{X_a |\partial W=0 \} = \text{Spec}\, \bC[X_a]/(\partial W)$ be the set of solutions of the F-term equations, now viewed as an affine space, and let $z_a$ be some affine coordinates on $\cZ$. Let us also introduce a trivial fiber $\bC$ with coordinate $t$. The choice of $\bm{\theta}$ determines a one-dimensional representation $\chi_{\bm{\theta}}$ (also known as character) of $(\bC^{*})^G\subset \cG_{\bC}$ on the trivial line bundle $\cZ\times \bC$,
\be\label{action of character theta}
(z_a, \, t)\mapsto  (\lambda\cdot z_a,\, \chi_{\bm{\theta}}(\lambda)\, t)\, , \qquad \mathrm{with}\qquad \chi_{\bm{\theta}} = \prod\nolimits_{i=1}^{G} \lambda_i^{\theta_i}\, ,
\ee
for  $\lambda=(\lambda_1, \cdots, \lambda_G) \, \in (\bC^{*})^G$. Let $\cG_{\bC}(\bm{\theta})$ denote the action of the gauge group $\cG_{\bC}$ on $\cZ\times \bC$ where the torus $(\bC^*)^G\subset \cG_{\bC}$ acts as (\ref{action of character theta}). The GIT quotient is given by
\be\label{GIT construction 00}
\cM(\bm{\theta})_{GIT} = \text{Proj} \; \bC[\cZ\times \bC]^{\cG_{\bC}(\bm{\theta})} \, ,
\ee
which means that we take the $\text{Proj}$ of the space of $\cG_{\bC}(\bm{\theta})$-invariant regular functions on $\cZ\times \bC$, where the scaling weights are the magnetic charges (see for instance section 2.5 of \cite{Martelli:2008cm} for more background on this). The crucial result of \cite{King1994} is that the space of $\theta$-(semi)stable representations of dimension $\bm{N}$ is given by (\ref{GIT construction 00}). Moreover, GIT quotient and K\"ahler quotient are directly related (by the Kempf-Ness theorem):
\be
\cM(\cQ, \bm{N}; \bm{\theta})_{GIT} \cong \cM(\cQ, \bm{N}; \bm{\xi})_{K} \, ,\qquad \mathrm{with}\quad \bm{\theta} = \bm{\xi}\, .
\ee
The parameters $\bm{\theta}$ are \emph{discretized} FI parameters ---and thus they determine discretized K\"ahler classes of the underlying $CY_3$ cone for a $CY_3$ quiver.

\begin{figure}[t]
\begin{center}
\subfigure[\small The $\bF_0$ quiver (Phase I)]{
\includegraphics[height=4.3cm]{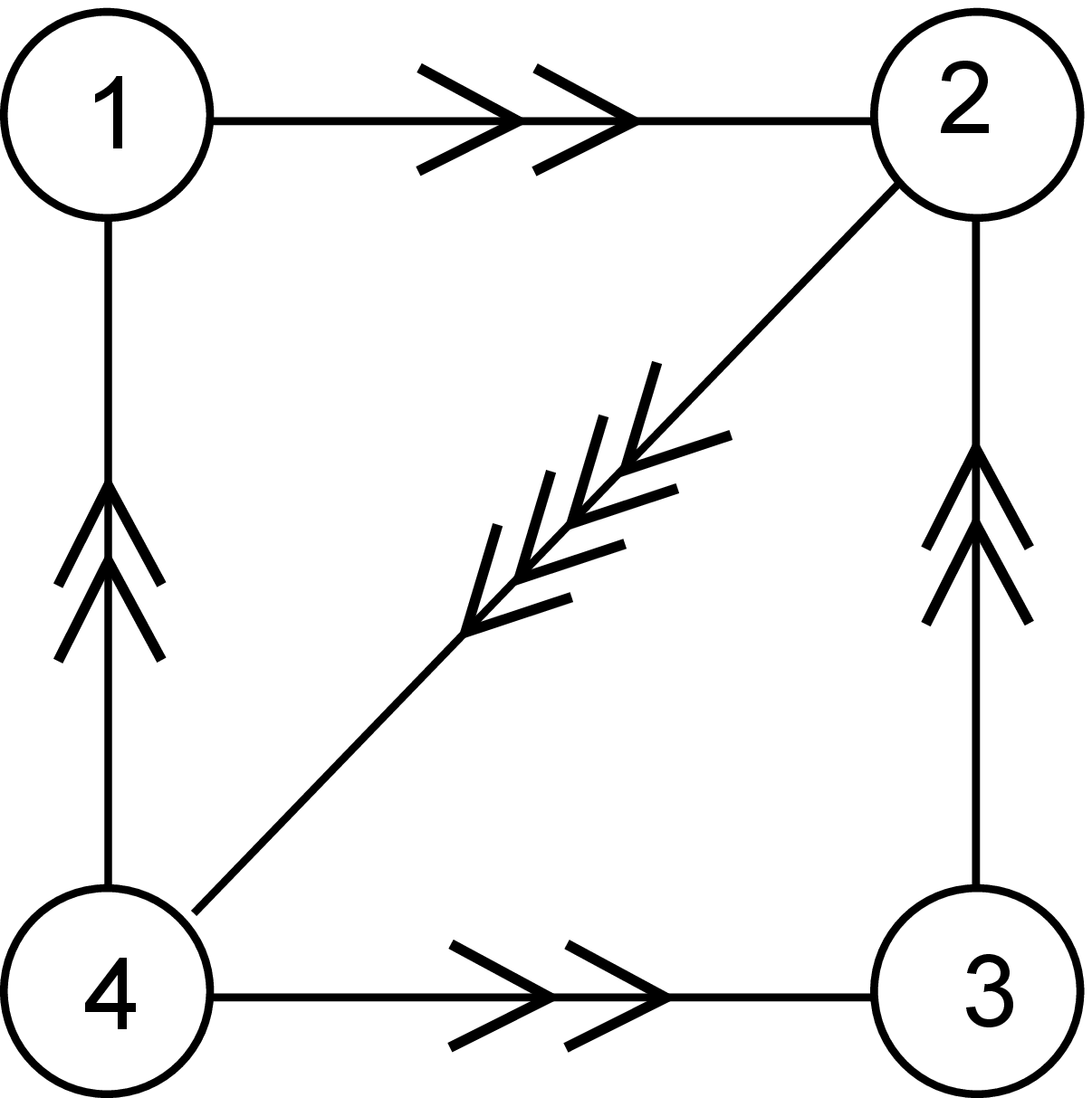}
\label{fig:quiver F0 I}}
\qquad \qquad \quad
\subfigure[\small Beilinson quivers]{
\includegraphics[height=6cm]{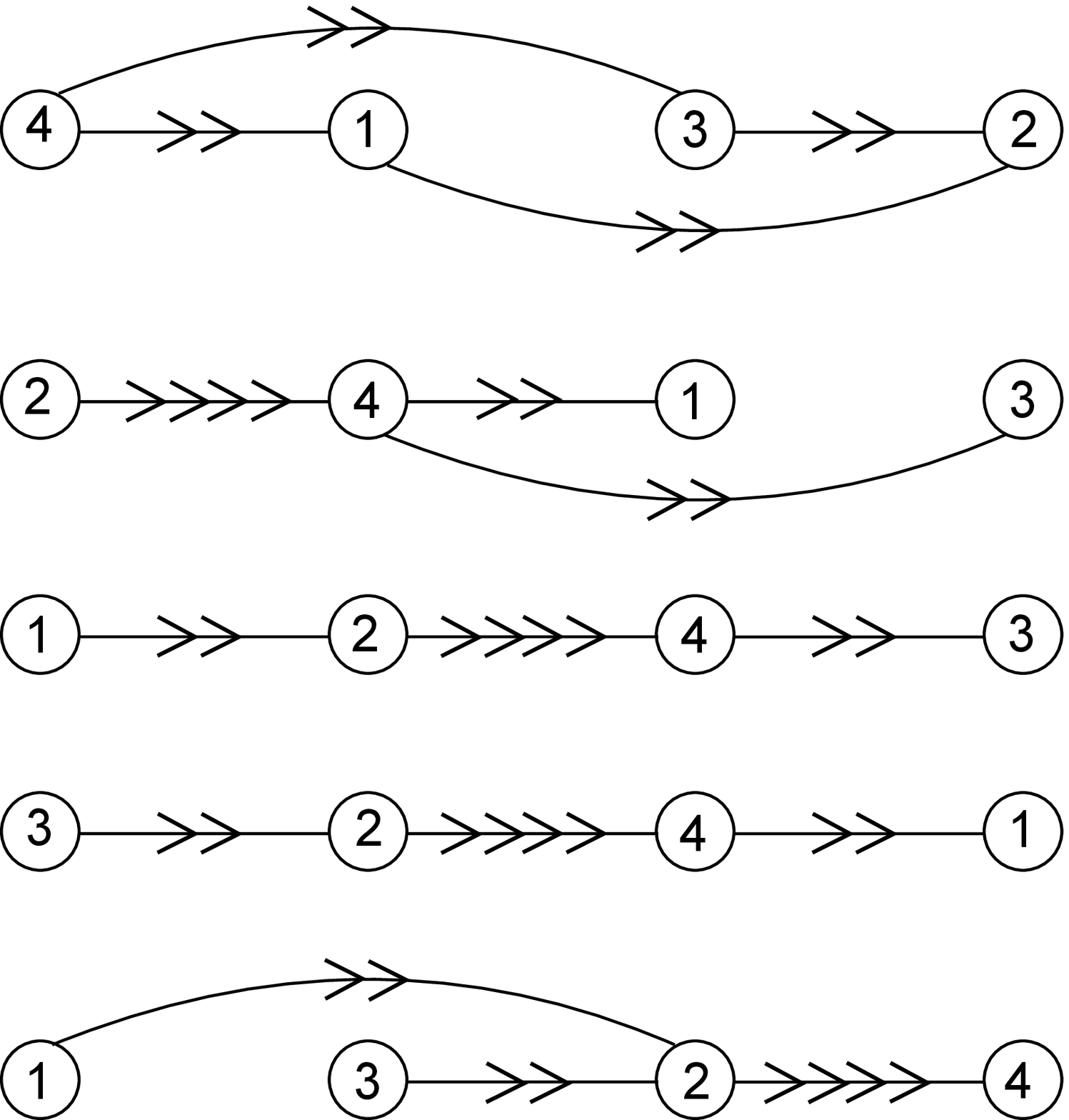}
\label{fig:Beil quiver F0}
}
\caption{\small Examples of Beilison quivers associated to a $CY_3$ quiver for $Y= \cO_{B_4}(K)$. In this case $B_4$ is $\bF_0= \CP^1\times \CP^1$. There are five Beilinson quivers, corresponding to five K\"ahler chambers for this quiver and geometry. As an example of how $\theta$-stability determines the K\"ahler chambers, consider the first Beilison quiver. The $\cO_p$ representation has dimension vector $\bm{r}=(1,1,1,1)$, and one can show that the proper subrepresentations have dimensions $(0,1,0,0)$, $(1,1,0,0)$, $(0,1,1,0)$ and $(1,1,1,0)$. $\theta$-stability then requires $\xi_2 <0$, $\xi_1+\xi_2<0$, $\xi_2+\xi_3<0$, $\xi_1+\xi_2+\xi_3<0$, which defines the K\"ahler chambers. We can go on for the other Beilison quivers, and these results can also be checked independently by toric methods.
}\label{figs: examples Bel quivers}
\end{center}
\end{figure}
Consider now a $CY_3$ quiver $\cQ$ and a supersymmetric theory $(\cQ,  \bm{r})$ corresponding to the point-like D-brane $\cO_p$ on a cone $Y$ of the type considered in section \ref{subsec: from Dbrane to quivers} ---see (\ref{definition of ri}). The FI parameters $\bm{\xi}$ of $(\cQ, \bm{r})$ (satisfying $\bm{\xi}\,\bm{r}=0$) span the K\"ahler moduli space of $Y$ near (a particular point in) the quiver locus. At some real codimension one \emph{K\"ahler walls} in the FI parameter space $V_{FI}\cong \bR^{G-1}$ the variety $\tilde{Y}$ can change to a birationally equivalent one $\tilde{Y}'$. The space $V_{FI}$ is thus subdivided into so-called K\"ahler chambers, separated by K\"ahler walls%
\footnote{The K\"ahler walls we refer to here are walls of marginal stability (where the central charges of two BPS states align, allowing for decay or recombination) also called ``walls of the first kind'' in \cite{2008arXiv0811.2435K}. The walls of the second kind of \cite{2008arXiv0811.2435K} correspond to Seiberg dualities in the quiver language \cite{Aganagic:2010qr}.}.
If we reconstruct $\tilde{Y}$ from the quiver as the K\"ahler quotient
\be
\tilde{Y} = \cM(\cQ, \bm{r}; \bm{\xi})_{K}, ,
\ee
the boundary of the K\"ahler walls are walls where the way we solve the D-term equations (\ref{D terms equs in general}) changes, although it is hard to make it very precise in general%
\footnote{In the \emph{toric} case however this can be made very precise using perfect matching variables. Similarly, the Beilinson quiver that is discussed next can be found explicitly using the results of \cite{Hanany:2006nm, 2009arXiv0909.2013B}. This  is discussed in detail in \cite{Closset:2012ep}.
}.
In the GIT description, we should fix some $\bm{\theta}$ and look for the subspace of $\tilde{Y}= \cM(\cQ, \bm{r}; \bm{\theta})_{GIT}$ corresponding to $\cO_p$ localized on the exceptional divisor $B_4$. Such quiver representations are all representations of a (generalized) Beilinson quiver $\cQ_{B_4}$, which is obtained from $\cQ$ by removing some arrows. The quiver $\cQ_{B_4}$ describes sheaves on $B_4$, including the skyscraper sheaf $\cO_p$, in agreement with the discussion of section \ref{subsec: from Dbrane to quivers}. See Figure \ref{figs: examples Bel quivers} for some concrete example.
Because a $CY_3$ quiver is always strongly connected (meaning that we can go from any node to any other nodes followings the arrows), one can see that the simple representations of dimension $\bm{r}$, corresponding to $\cO_p$, are $\theta$-stable for any $\bm{\theta}$. On the other hand $\cQ_{B_4}$ has no closed loops and consequently the $\theta$-stability of such representations is non-trivial. The requirement of $\theta$-stability of the $\cO_p$ representation determines the boundary of the  K\"ahler chamber in $V_{FI}$ (using $\bm{\xi}\cong \bm{\theta}$). We spelled out an example in the comments below Figure \ref{figs: examples Bel quivers}.

Before concluding this section let us review the role of the FI parameters in the brane picture of Seiberg duality. In the quiver language of subsection \ref{subsec: quiver mutations} the FI parameters are what selects a particular type of quiver representations, distinguishing between left and right mutations \cite{Berenstein:2002fi}: Due to (\ref{D terms equs in general}), taking $\xi_k \ll 0$ forces the fields $X_{jk}$ (arrows ending on node $k$) to take maximal VEVs while taking $\xi_k \gg 0$ forces to turn on the fields $X_{kj}$ (arrows starting at $k$). This distinction in allowed quiver representations forces to use the left or right mutation of (\ref{L and R mutation on Q reps}), respectively \cite{2009arXiv0906.0761K}.
From the fractional brane point of view this is understood similarly. Let $\bm{\theta}\sim \bm{\xi}$ be the stability parameter determining the resolved cone $\tilde{Y}$. Near the quiver locus the fractional brane $E_k^{\vee}$ has central charge
\be
Z(E_k^{\vee}) \sim \frac{1}{g^2_k} +i \theta_k\, ,
\ee
where $g^2_k$ is the (appropriately normalized) gauge coupling. Seiberg duality on node $k$ corresponds to continuing $g^2_k$ to negative value; see \emph{e.g.} \cite{Aganagic:2010qr} for a recent discussion of this. As we do that the anti-brane $\overline{E_k^{\vee}}$ acquires positive tension, and we should use it as part of the new basis of the Seiberg dual quiver. Having $\theta_k \neq 0$ we can go around the singular origin of the $Z(E^{\vee}_k)$ plane smoothly, either in a clockwise or anti-clockwise manner. The difference between left and right mutation ---see (\ref{L and R mutation on sheaves})--- comes from this choice%
\footnote{A more precise explanation uses the derived category structure and the grading of \cite{Douglas:2000ah}: A left mutation sends the B-brane $E_k^{\vee}$ to $E_k^{\vee}[-1]$, the complex shifted one position to the left, while a right mutation sends $E_k^{\vee}$ to $E_k^{\vee}[1]$, the complex shifted to the right. In the derived category $E_k^{\vee}[-1]$ and $E_k^{\vee}[1]$ are two different objects, even though they are both superficially the same ``anti-brane'' $\overline{E_k^{\vee}}$.}, which is forced on us by the sign of $\theta_{k}$
:
\be\label{how to choose L or R mut}
\mu_k^L \;\longleftrightarrow\; \theta_k <0\, , \qquad \quad \mu_k^R \;\longleftrightarrow\;\theta_k >0\, .
\ee
This distinction will be crucial in the following. The FI parameters $\bm{\theta}$ transform as $\bm{\theta}' = M_k^T \bm{\theta}$ under mutation.

\section{Seiberg duality for Chern-Simons quivers}\label{sec: SD for 3d CS quivers}
A $\cN=2$ Chern-Simons quiver theory is a three-dimensional $\cN=2$ supersymmetric field theory whose gauge fields appear with Chern-Simons interactions. In recent years they have been shown to be able to describe the low energy dynamics of M2-branes.
In this paper we only discuss CS quivers whose underlying quiver $\cQ$ is a maximally chiral $CY_3$ quiver. We denote by $(\cQ, \bm{N}, \bm{k})$ a CS quiver theory with ranks $\bm{N}$ and CS levels $\bm{k}= (k_1, \cdots, k_G)$. The Chern-Simons levels are half-integer quantized such that $k_i + \frac12 \sum_j A_{ij} N_j \in \bZ$.

We are interested in CS quiver theories whose CS levels satisfy
\be\label{condition on k to have CY4}
\bm{r} \bm{k} = \sum\nolimits_{i=1}^G  r_i  k_i =0\, ,
\ee
where $\bm{r}$ is the dimension vector of the $\cO_p$ representation discussed in (\ref{definition of ri}). This constraint will be assumed in the following.

\subsection{Moduli space of CS quivers and GIT quotient}\label{subsec: One loop VMS CS quivers}
At the semi-classical level, the crucial difference between a 4d quiver theory moduli space and a 3d CS quiver theory moduli space is that the 3d theory has additional real adjoint scalars $\sigma_i$ in the $\cN=2$ vector multiplets, which might take VEVs as well. The classical vacuum equations of $(\cQ, \bm{N}, \bm{k})$ are
\bea
\partial_{X_a} W  \,& = \,  0\, ,\\
\sigma_i X_{ij}- X_{ij} \sigma_j  \,& = \, 0 \, ,\\
\sum_{X_a | s(a)=i} X_a^{\dagger}X_a - \sum_{X_b | t(b)=i} X_b X_b^{\dagger} \,& = \, \sigma_i k_i\, ,
\eea
and the general solution can be rather intricate. For $\cQ$ a $CY_3$ quiver, let us write the ranks as
\be
N_i = r_i \tilde{N}+N_i\, ,
\ee
with $\tilde{N}= \min(N_i)$. We will focus on the \emph{geometric branch}, which we define by setting
\be
\sigma_i = \text{diag}( \underbrace{\sigma_{1}, \cdots,\sigma_{1}}_{r_i \; \text{times}},\cdots, \underbrace{\sigma_{\tilde{N}},\cdots,  \sigma_{\tilde{N}}}_{r_i \; \text{times}}, 0, \cdots, 0)\, ,\qquad  \forall\, i\, .
\ee
In the brane language this branch will correspond to $\tilde{N}$ D2-branes $\cO_p$ moving on a $CY_3$ fibered on a line $\{\sigma\}\cong \bR$. The semi-classical moduli space of a single $\cO_p$ is given by
\bea\label{CS VMS for Op}
\partial_{X_a} W  \,& = \,  0\, ,\\
\sum_{X_a | s(a)=i} X_a^{\dagger}X_a - \sum_{X_b | t(b)=i} X_b X_b^{\dagger} \,& = \, \sigma_i k_i^{eff}(\sigma)\, ,
\eea
with the effective CS levels $\bm{k}^{eff}$  \cite{Benini:2011cma}
\be\label{kpm defined 00}
\bm{k}^{eff}(\sigma) = \begin{cases}
   \bm{k}_+ \, & \qquad \text{if}\quad \sigma >0 \\
   \bm{k}_- \, & \qquad \text{if}\quad \sigma <0 \\
\end{cases}\, , \qquad \quad \mathrm{with}\quad \bm{k}_{\pm} = \bm{k} \pm \frac12 A\, \bm{N}\, ,
\ee
written in term of the adjacency matrix $A$. Remark that $\bm{k}_{\pm}$ are always integers. The condition (\ref{condition on k to have CY4}) insures that we can have solutions of the D-term equations at $\sigma \neq 0$. At any fixed $\sigma \neq 0$, the equations (\ref{CS VMS for Op}) lead to the  K\"ahler quotient description (\ref{KQ description of VMS}) of a resolved CY$_3$ cone $\tilde{Y}$. Therefore for $\tilde{N}=1$ the full geometric branch is a resolved cone $\tilde{Y}$ fibered on a line $\bR\cong \{\sigma \}$ according to (\ref{CS VMS for Op})-(\ref{kpm defined 00}). More precisely we have \emph{two} resolved cones $\tilde{Y}_{\pm}$ depending on the sign of $\sigma$. We can describe them as efficiently by the GIT quotient reviewed in section \ref{sec: stability}:
\be\label{GIT quotient in CS quivers}
\tilde{Y}_{\pm} \, = \, \cM(\cQ, \bm{r}; \bm{\theta}_{\pm})_{GIT} \, , \qquad \text{with}\quad  \bm{\theta}_{\pm}= \pm \bm{k}_{\pm} \, .
\ee
For $\tilde{N}>1$ one can show that we have a $\tilde{N}$-symmetric product of the $\tilde{N}=1$ case.
Our main point here is that the geometric branch of a CS quiver theory can be recast in purely algebraic language, similarly to what happens in 4d quiver gauge theories describing D3-branes. Interestingly, the $\theta$-stability parameters are given by the effective Chern-Simons levels according to (\ref{GIT quotient in CS quivers}).

Moreover, the coordinates $t_{\pm}$ on the stabilizing line bundle of the GIT construction (\ref{action of character theta})-(\ref{GIT construction 00}) for $\tilde{Y}_{\pm}$ are naturally identified with ``bare'' diagonal monopole operators \cite{Closset:2012ep}. It is conjectured that the full CY$_4$ conical geometry probed by the M2-branes is recovered algebraically from the quantum chiral ring of the $\tilde{N}=1$ Chern-Simons quiver, including the gauge invariant diagonal monopole operators $f_{n\bm{\theta}_{\pm}} t^{n}_{\pm}$ of flux $\pm n$, according to
\be\label{M2 geometric branch algebraically}
\cM_{\text{M2-branes}} = \text{Spec}\; \frac{\bigoplus_{n\geq 0}\, \bC[f_{n\bm{\theta}_{-}}\,  t^{n}_{-},f_{n\bm{\theta}_{+}} t^{n}_{+}]}{\cI_{QR}}  \, \, ,
\ee
which is the ring of all holomorphic gauge invariant operators modulo the relations coming from the superpotential (classical F-term relations), graded by the magnetic charge $n$, an further divided by an ideal $\cI_{QR}$ generated by so-called ``quantum F-term relations'' \cite{Closset:2012ep}.  We refer to \cite{Closset:2012ep} for more comments on this conjecture; see also \cite{Gaiotto:2009tk, Benini:2009qs, Jafferis:2009th}.

\subsection{From M2-brane on CY fourfold to CS quiver}\label{subsec: from M2 to CS quiver}
In this section we explain how to derive a CS quiver for M2-branes at CY$_4$ singularity, in the cases of interest for this paper; we refer to \cite{Closset:2012ep} for a fuller explanation of this algorithm.

Consider $N$ M2-branes on a conical CY fourfold%
\footnote{The CY$_4$ is a cone over a Sasaki-Einstein 7-fold, and this SE$_7$ should have at least one $U(1)$ isometry beyond the (generally non-compact) action generated by the Reeb vector.}, which preserves $\cN=2$ supersymmetry on the M2-branes worldvolume, and choose a type IIA reduction
\be\label{M-theory fibration}
U(1)_M \rightarrow CY_4  \rightarrow X_7 \, ,
\ee
where $U(1)_M$ is the M-theory circle and $X_7$ is the type IIA geometric background. We choose the $U(1)_M$-fibration such that $X_7$ can itself be described as the fibration%
\footnote{Technically this is not a fibration but rather only a foliation, since the ``fiber'' changes topology at we cross $r_0=0$.}
 of a CY$_3$ (resolved) cone $\tilde{Y}$ over a line $\bR\cong \{r_0\}$ (in particular for a toric CY$_4$  this is very easy to do),
\be\label{Y fibered on R}
\tilde{Y} \rightarrow X_7 \rightarrow \bR\, .
\ee
M2-branes become \emph{D2-branes} located at $r_0=0$ and at the tip of $Y$, and we can use our deeper knowledge of D-branes on Calabi-Yau threefolds to learn more about the M2-branes on the $CY_4$, through the type IIA/M-theory duality  \cite{Aganagic:2009zk}; this proposal was fruitfully developed in \cite{Benini:2009qs, Jafferis:2009th, Benini:2011cma, Closset:2012ep}.

For a generic fibration (\ref{M-theory fibration}), however, the IIA background $X_7$ can have all kinds of singularities which are not readily manageable%
\footnote{See \cite{Jafferis:2009th} for a proposal of how to deal with some of these ``bad'' cases.}. In favorable cases, the $U(1)_M$ fiber degenerates over $X_7$ in a way we understand well, leading to \emph{D6-branes} in type IIA, which are localized at $r_0=0$ and can be wrapped on non-compact \cite{Benini:2009qs, Jafferis:2009th} or compact \cite{Benini:2011cma, Closset:2012ep} 4-cycles in $\tilde{Y}$.

Here as in \cite{Benini:2011cma, Closset:2012ep} we consider exclusively the case where the IIA reductions leads to D6-branes wrapped on compact 4-cycles. We take $\tilde{Y}= \cO_{B_4}(K)$ like in section \ref{subsec: from Dbrane to quivers}. The M-theory CY$_4$ can often have $G_4$ torsion flux, and this leads to IIA backgrounds with further D4-branes sources (on 2-cycles in $B_4$) and rather generic compactly supported RR fluxes. The RR fluxes induce the Chern-Simons interactions on the fractional D-branes worldvolume.

We can refer to \cite{Benini:2011cma, Closset:2012ep} for explicit examples of the reduction to type IIA, and consider a generic IIA background characterized by the fluxes and sources
\bea\label{fluxes and sources}
\bm{Q}_{\text{flux}, \pm} \, & \equiv\, \left( -Q_{4;\,  \pm} \,|\, Q_{6;\, \alpha \pm} \,|\, 0 \right)\, , \\
\bm{Q}_\text{source} \, & \equiv\, \left( Q_{D6} \,|\, (\cI^{-1})^{\alpha\beta} Q_{D4;\, \beta} \,|\, Q_{D2} \right)\, .
\eea
The vectors $\bm{Q}_{\text{flux}, \pm}$ encode the RR fluxes (more precisely the Page charges ---see section \ref{section: brane charges}) that are collected through the 4- and 2-cycles $B_4$ and $\cC_{\alpha}$, at $r_0>0$ or $r_0<0$. The fluxes jump at $r_0=0$ due to explicit D6- and D4-branes sources there (encoded in $\bm{Q}_\text{source}$), according to
\be\label{rels between sources and fluxes}
 Q_{4;\,  +} - Q_{4;\,  -}  = \cI_{0\alpha} (\cI^{-1})^{\alpha\beta} Q_{D4;\, \beta}\, ,\qquad\qquad
 Q_{6;\, \alpha +}-Q_{6;\, \alpha -}  = \cI_{0\alpha} Q_{D6}    \, ,
\ee
with $\cI_{\alpha\beta}$ as in (\ref{def 2 cycles and cI}) and $\cI_{0\alpha}$ the intersection number between $B_4$ and $\cC_{\alpha}$ in $\tilde{Y}$. The Chern-Simons quiver theory is then constructed from the data (\ref{fluxes and sources}) and from the D-brane machinery reviewed in section \ref{section: Dbranes and SD, general}. One can show that the geometric background (\ref{Y fibered on R}) is characterized by two resolved cones $\tilde{Y}_{\pm}$ with K\"ahler parameters
\be\label{chi of IIA geom}
\bm{\chi}_{\pm} =\, \begin{cases} \bm{Q}_{\text{flux}, +}\, r_0\quad &\text{for}\quad r_0>0 \\
                            \bm{Q}_{\text{flux}, -}\, r_0\quad &\text{for}\quad r_0<0   \end{cases},
\ee
where $\bm{\chi}$ was defined in section \ref{subsection: KMS and quiver locus}. The cones $\tilde{Y}_{\pm}$ can in general lie in two different K\"ahler chambers of the underlying $CY_3$ quiver $\cQ$. Assuming that we know the correct dictionaries $Q^{\vee}_{\pm}$ ---see (\ref{definition Q Qv})---  associated to these K\"ahler chambers, (\ref{dico xi to chi 0}) gives the effective FI parameters of the quiver describing a D2-brane $\cO_p$ moving at $r_0>0$ or $r_0<0$, respectively:
\be
\bm{\xi}_{-} = Q^{\vee}_{-}\, \bm{\chi}_{-}\, , \qquad \qquad \bm{\xi}_{+} = Q^{\vee}_{+}\, \bm{\chi}_{+}\, .
\ee
These effective FI parameters come from a CS quiver of the type studied in the last subsection, with
\be
\bm{k}_{-} = Q^{\vee}_{-} \bm{Q}_{\text{flux}, -}\, , \qquad \bm{k}_{+} = Q^{\vee}_{+} \bm{Q}_{\text{flux}, +}\, , \qquad \text{and}\quad \sigma = r_0\, ,
\ee
which follows from the Wess-Zumino action of the fractional branes.
The  underlying CS quiver theory has Chern-Simons levels
\be
\bm{k}= \frac12 (k_- + k_+)\, .
\ee
The ranks $\bm{N}$ of the CS quiver follow from the brane sources and the flux in a more subtle way. In order to use the dictionnaries and read the quiver ranks from the branes, we need to split these D-brane sources to the left and right of $r_0=0$:
\be
\bm{Q}_{\text{source}, -}= \bm{\delta Q}_\text{source}\, , \qquad \qquad
\bm{Q}_{\text{source},  +}= \bm{Q}_\text{source} -\bm{\delta Q}_\text{source}\, ,
\ee
in such a way that the bunches $\bm{Q}_{\text{source}, \pm}$ still lie inside the K\"ahler chambers where $Q^{\vee}_{\pm}$ are respectively valid; since these branes change the background flux, this is a non-trivial constraint. In practice we can take an arbitrary splitting $ \bm{\delta Q}_\text{source}$, and compute
\be
\bm{N}_{\text{trial}} = \bm{Q}_{\text{source},-} (Q^{\vee}_-)^{-1} +\bm{Q}_{\text{source},+} (Q^{\vee}_+)^{-1}\, ,
\ee
which depends on some of the unknowns in the arbitrary splitting $ \bm{\delta Q}_\text{source}$ (it only depends on the so-called anomalous D-branes, which wrap cycles dual to compact cycles and therefore source the fluxes $\bm{Q}_{\text{flux}, +}$). The correct $\bm{N}$ is found by requiring that
\be
 A\bm{N}_{\text{trial}}= \bm{k}_+\, -  \, \bm{k}_- \, ,
\ee
with $A$ the adjacency matrix of $\cQ$.

This algorithm  relates the string theory data (\ref{fluxes and sources}) to a Chern-Simons quiver $(\cQ, \bm{N}, \bm{k})$, which by construction has the IIA geometry (\ref{Y fibered on R})-(\ref{chi of IIA geom}) as its semi-classical geometric branch. Moreover, one can show in many examples (although we yet lack a general proof of that fact) that the algebraic construction (\ref{M2 geometric branch algebraically}) in term of monopole operators reproduces the CY$_4$ geometry of M-theory.

\subsection{D-brane mutations and 3d Seiberg duality}
Consider a IIA background like in section \ref{subsec: from M2 to CS quiver} and its associated CS quiver theory $(\cQ, \bm{N}, \bm{k})$:
\bea
\tilde{Y}_{\pm} \quad  & \longleftrightarrow \quad \cQ\, \\
\bm{Q}_\text{source} \, , \bm{Q}_{\text{flux}, \pm}\quad  & \stackrel{Q^{\vee}_{\pm}}{\longleftrightarrow} \quad \bm{N}\, , \bm{k}\, .
\eea
The pair of dictionaries $Q^{\vee}_-$, $Q^{\vee}_+$ relate the two descriptions. This setup is essentially a ``doubling'' of the relation between branes on $Y$ and quiver theory $(\cQ, \bm{N})$ reviewed in section \ref{section: Dbranes and SD, general}, with the effective CS levels $\bm{k}_{\pm}$ playing the role of stability parameters for a D2-brane on $B_4 \subset \tilde{Y}_{\pm}$.

Consider doing a Seiberg duality on the node $i_0\in Q_0$. This is realized as a quiver mutation $\mu_{i_0}$ (\ref{quiver mutation on A}), or equivalently as a (right or left) mutation of the underlying B-branes:
\be\label{mutated quiver for CS quiver theory}
\cQ' = \mu_{i_0}( \cQ )\, , \qquad \quad A' = M_{i_0}^T\, A\, M_{i_0}\, .
\ee
The dictionaries $Q^{\vee}_{\pm}$ are mutated according to (\ref{mutation of dictionary Qv}), giving us the fractional brane dictionaries for the new quiver (\ref{mutated quiver for CS quiver theory}). The only thing we need to understand is whether we should apply a \emph{left} or \emph{right} mutation to the dictionaries. We explained around (\ref{how to choose L or R mut}) that the choice between left or right mutation of B-branes corresponds to whether the parameter $\theta_{i_0}$ of $\mathsf{E}_{i_0}^{\vee}$ is positive or negative.
Thus the dictionaries $Q^{\vee}_{\pm}$ are mutated according to:
\bea\label{change of Qv: 4 cases}
Q_-^{\vee}\phantom{}' \,=\, M^T_{i_0 -} Q^{\vee}_- \, \qquad &\text{with}\quad
&\begin{cases} M_{i_0 -} = M_{(i_0;\, L)} \qquad & \text{if}\; \theta_{i_0\, -} \leq 0 \\
                M_{i_0 -} = M_{(i_0;\, R)} \qquad & \text{if}\; \theta_{i_0\, -} \geq 0  \end{cases}\, , \\
Q_+^{\vee}\phantom{}' \,=\, M^T_{i_0 +} Q^{\vee}_+ \, \qquad &\text{with}\quad
&\begin{cases} M_{i_0 +} = M_{(i_0;\, L)} \qquad & \text{if}\; \theta_{i_0\, +} \leq 0 \\
                M_{i_0 +} = M_{(i_0;\, R)} \qquad & \text{if}\; \theta_{i_0\, +} \geq 0  \end{cases}\, ,
\eea
There are four different possibilities, but only two qualitatively different behaviors: Either $\theta_{i_0\, -}$ and $\theta_{i_0\, +}$ have the same sign, or they have opposite signs. The theory $(\cQ', \bm{N}', \bm{k}')$ is then again obtained from the same type IIA background:
\bea
\tilde{Y}_{\pm} \quad  & \longleftrightarrow \quad \cQ'\, \\
\bm{Q}_\text{source} \, , \bm{Q}_{\text{flux}, \pm}\quad  & \stackrel{Q^{\vee}_{\pm}\phantom{}'}{\longleftrightarrow} \quad \bm{N}'\, , \bm{k}'\, .
\eea
By construction, the Chern-Simons quiver theories $(\cQ, \bm{N}, \bm{k})$ and $(\cQ', \bm{N}', \bm{k}')$ share the same ``geometric branch'' as their semi-classical moduli space.
In section \ref{examples: dP0 Seiberg duals}  we will check in some examples that the quantum chiral ring matches as well%
\footnote{Up to possible subtleties that we will explain.}.

What we have shown here is that the understanding of Seiberg duality from \emph{brane mutations} in maximally chiral quivers carries almost verbatim to three dimensional Chern-Simons quivers: 3d Seiberg duality is a brane mutation.

\subsection{Seiberg duality for chiral Chern-Simons quivers}
Seiberg duality on quivers is a  ``local'' operation, acting on a single node $i_0\in Q_0$ and only affecting the structure of the quiver in the neighborood of $\cQ$ directly connected to $i_0$. Therefore we should be able to understand Seiberg duality for any CS quiver theory $(\cQ,\bm{N}, \bm{k})$ as an operation on a $\cN=2$  $U(N_{i_0})$ theory with $s_1$ and $s_2$ chiral superfields in the antifundamental and fundamental representation, respectively, with
\be\label{def s1 s2}
s_1 = \sum_{j \in I_L} N_j A_{j i_0}\, , \qquad\quad s_2 = \sum_{j\in I_R}  A_{i_0 j}N_j\, .
\ee
For a purely chiral quiver the $A_{ij}$ in (\ref{def s1 s2}) is the antisymmetric adjacency matrix we used above, and more generally it would be the total number of arrows from $i$ to $j$. In the case $s_1=s_2$ Seiberg dualities have been proposed in \cite{Aharony:1997gp,Giveon:2008zn}; see also \cite{Aharony:2008gk, Amariti:2009rb}. The case $s_1 \neq s_2$ has been studied more recently in \cite{Benini:2011mf}, where new Seiberg dualities relevant to that case where derived
 from the ones of \cite{Aharony:1997gp, Giveon:2008zn} using straightforward field theory arguments.

It was found in \cite{Benini:2011mf} (see also \cite{Cremonesi:2010ae}) that Seiberg duality for such a theory depends on the signs of the effective CS levels $k_{i_0\pm}$, giving us four cases, in perfect agreement with (\ref{change of Qv: 4 cases}). The dual theory has gauge group $U(N_{i_0}')$ with \be\label{rules for change of ranks under SD}
N_{i_0}'\,=\, \begin{cases}  \frac12(s_1+s_2) +k_{i_0} - N_{i_0} &\quad \text{if}\;\; k_{i_0\,-} \geq 0,\; k_{i_0\,+}\geq 0 \\
 \frac12(s_1+s_2) -k_{i_0} - N_{i_0} &\quad \text{if}\;\; k_{i_0\,-}\leq 0,\; k_{i_0\,+}\leq 0 \\
   s_1 - N_{i_0} &\quad \text{if}\;\; k_{i_0\,-}\geq 0,\; k_{i_0\,+}\leq 0 \\
   s_2- N_{i_0} &\quad \text{if}\;\; k_{i_0\,-}\leq 0,\; k_{i_0\,+}\geq 0
\end{cases}
\ee
and CS level
\be
k_{i_0}'= -k_{i_0}\,.
\ee
The first two and last two cases in (\ref{rules for change of ranks under SD}) were dubbed ``minimally chiral'' and ``maximally chiral'' 3d Seiberg duality, respectively. The minimally chiral case is a natural generalization of the rule $N_{i_0}'=n_f+|k_{i_0}|-N_{i_0}$ for $s_1=s_2=n_f$ \cite{Giveon:2008zn, Amariti:2009rb}, while the maximally chiral case might look more exotic at first sight.

 Crucially, the CS levels of the flavor symmetry group $SU(s_1)\times SU(s_2)$ are affected by the duality, and when we embed this theory into a quiver (breaking $SU(s_1)\times SU(s_2)$ into subgroups and gauging them) this results in particular shifts of the Chern-Simons levels of the neighboring nodes. Let $\tilde{k}_i$ denote the CS levels of the nodes $i\in  I_L$ connected to $i_0$ by incoming arrows $a: i\rightarrow i_0$, and $k_j$ the levels of the nodes $j\in I_R$ connected to $i_0$ by outgoing arrows $a: i_0\rightarrow j$. Those CS levels transform under 3d Seiberg duality according to \cite{Benini:2011mf}
\be\label{rules for changein CS under SD}
\tilde{k}_i'\,=\, \begin{cases}  \tilde{k}_i+\frac12 A_{i i_0} k_{i_0\, -}  \\
                                     \tilde{k}_i+\frac12 A_{i i_0} k_{i_0\, +}   \\
                                    \tilde{k}_i+ A_{i i_0} k_{i_0}  \\
                                  \tilde{k}_i \end{cases}\, , \qquad
k_j'\,=\, \begin{cases}     k_j+\frac12 A_{i_0j}k_{i_0\, +}            &\qquad \text{if}\;\; k_{i_0\,-} \geq 0,\; k_{i_0\,+}\geq 0 \\
                            k_j+\frac12 A_{i_0j}k_{i_0\, -}   &\qquad \text{if}\;\; k_{i_0\,-}\leq 0,\; k_{i_0\,+}\leq 0 \\
                           k_j            &\qquad \text{if}\;\; k_{i_0\,-}\geq 0,\; k_{i_0\,+}\leq 0 \\
                           k_j+A_{i_0 j}k_{i_0}              &\qquad \text{if}\;\; k_{i_0\,-}\leq 0,\; k_{i_0\,+}\geq 0
\end{cases}
\ee
It is easy to show that these rules are precisely realized by our D-brane argument. Dualizing on node $i_0$, the dual CS levels are given by
\be
\bm{k}' = \frac12 (M_{i_0-}^T \bm{k}_- + M_{i_0+}^T \bm{k}_+) =\frac12(M_{i_0-}^T+M_{i_0+}^T) \bm{k} +\frac12(M_{i_0-}^T-M_{i_0+}^T) \frac12 A\bm{N}\, ,
\ee
and a short computation shows that it reproduces the rules (\ref{rules for changein CS under SD}).

It is more tricky to prove that the change of brane dictionary (\ref{change of Qv: 4 cases}) always reproduces the change of rank (\ref{rules for change of ranks under SD}), but it is the case in all the many examples we checked.

\section{$Y^{p,q}(\CP^2)$ geometry, $dP_0$  quivers and Seiberg duality}\label{examples: dP0 Seiberg duals}
\begin{figure}[t]
\begin{center}
\subfigure[\small The toric $dP_0$ quiver $\cQ_{(0)}$.]{
\includegraphics[height=3cm]{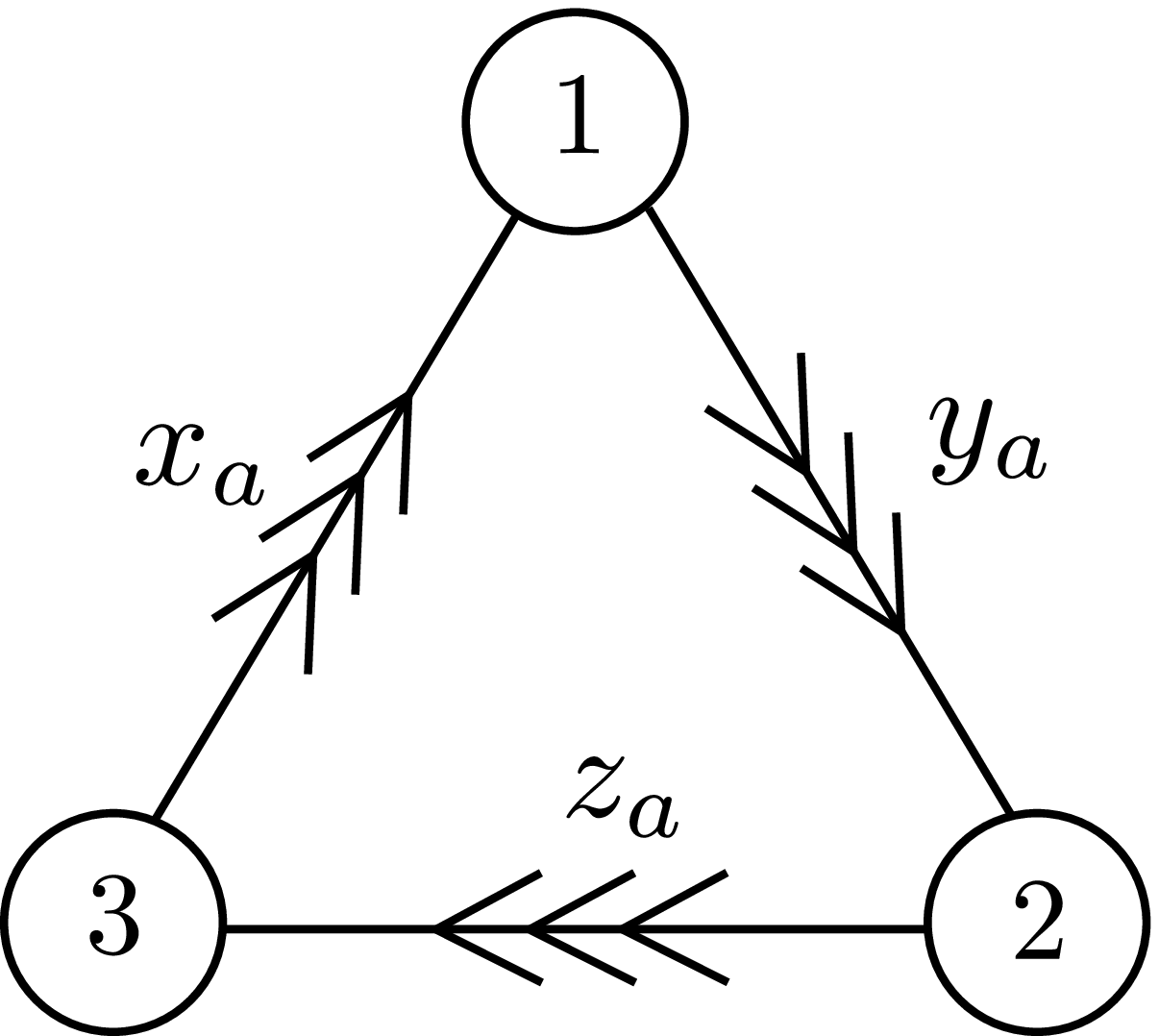}
\label{fig:quiver dP0 toric}}
\qquad \qquad \quad
\subfigure[\small Generic $dP_0$ quiver]{
\includegraphics[height=3cm]{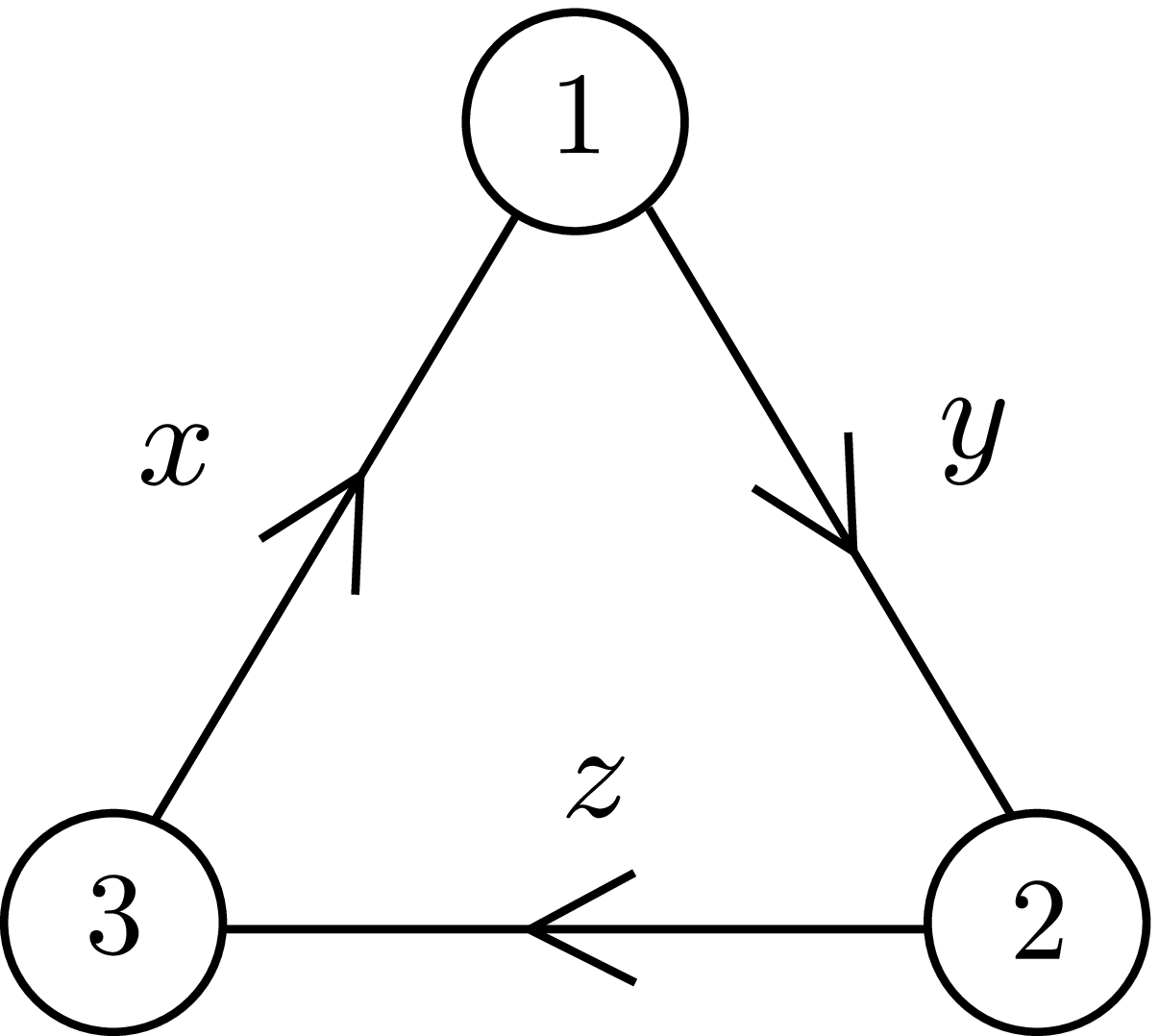}
\label{fig: general dP0 quiver}
}
\caption{\small Quivers for D-branes on $\cO_{\CP^2}(-3)$. The quiver $\cQ_{(0)}$ on the left is the simplest one, while the one on the right is the generic case. In the latter, the labels $(x,y,z)$ stand for the number of arrows, and the labels $(r_1, r_2, r_3)$ are the ranks of the gauge group for a single point like D2-brane.}
\end{center}
\end{figure}

\begin{figure}[h!]
\begin{center}
\subfigure[\small Windows for Seiberg duality on node 1.]{
\includegraphics[height=4cm]{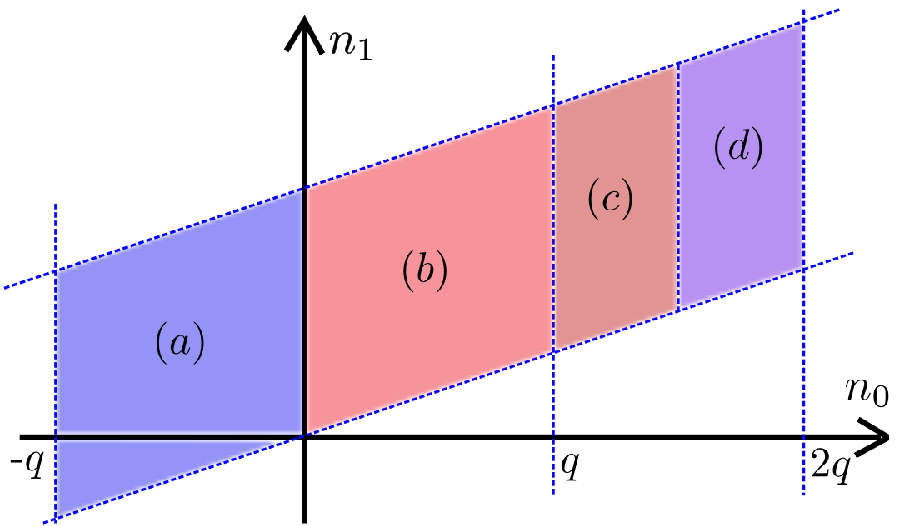}
\label{fig:torsion domain SD1}}
\qquad
\subfigure[\small $\cQ_{(1; 1)}$]{
\includegraphics[height=3cm]{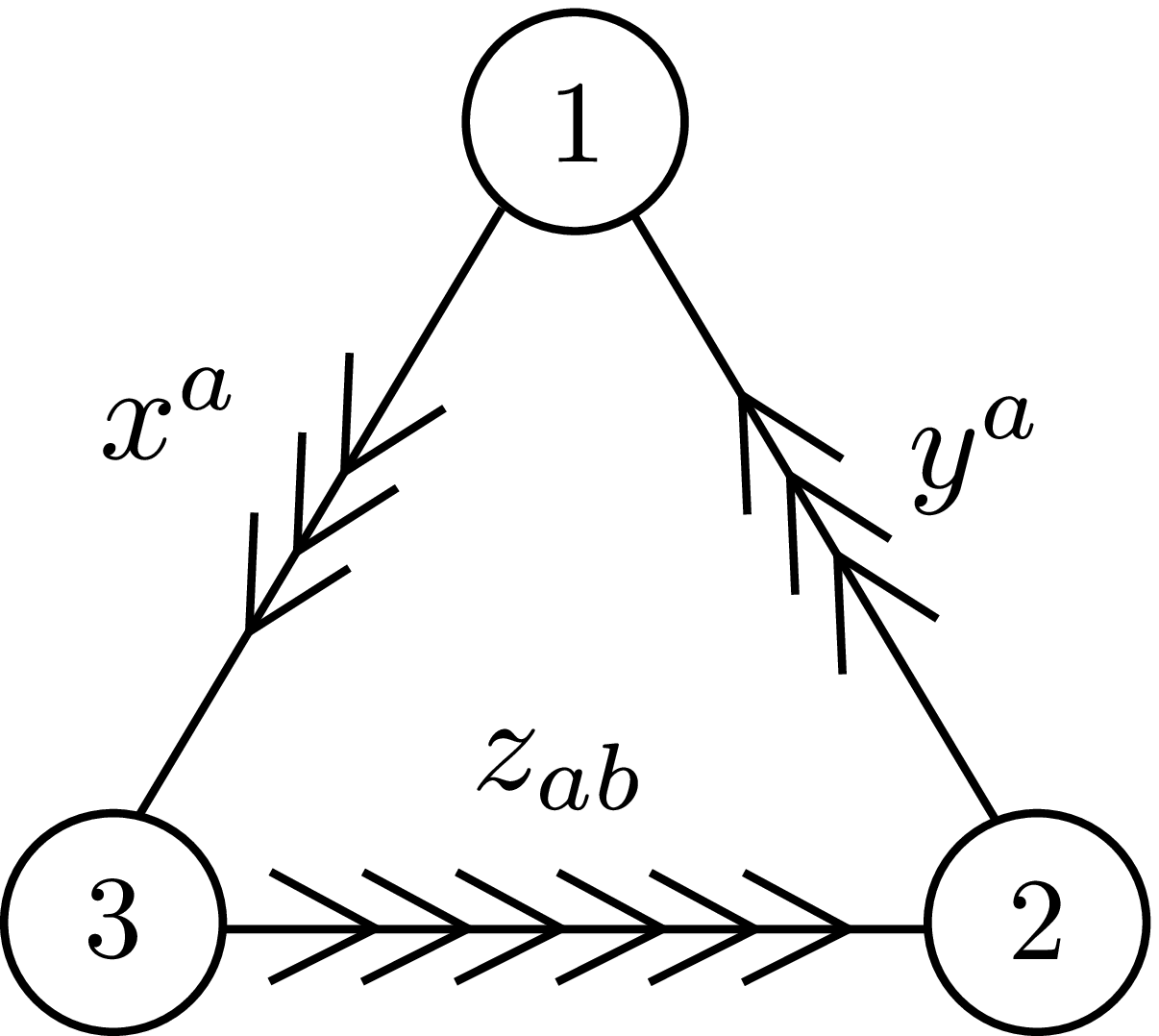}
\label{fig:SD1on1 quiver}}
\qquad
\subfigure[\small Windows for Seiberg duality on node 2.]{
\includegraphics[height=4cm]{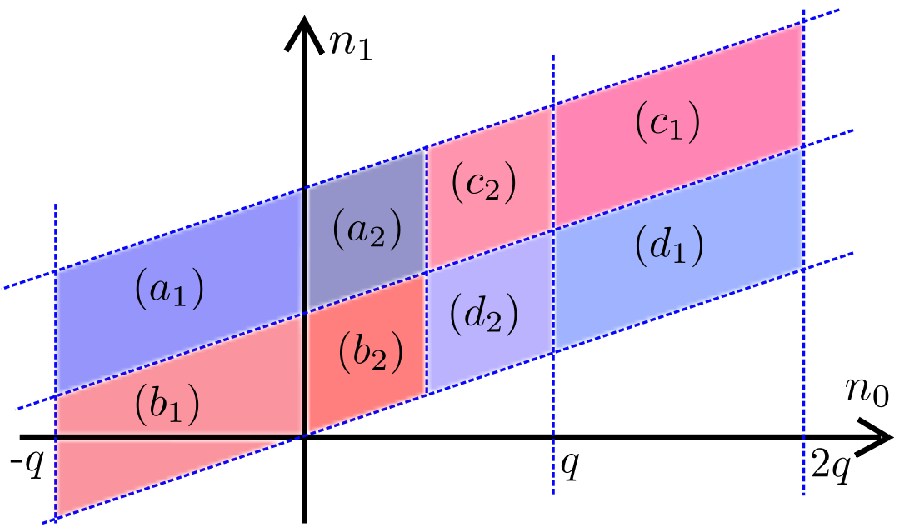}
\label{fig:torsion domain SD2}}
\qquad
\subfigure[\small $\cQ_{(1; 2)}$]{
\includegraphics[height=3cm]{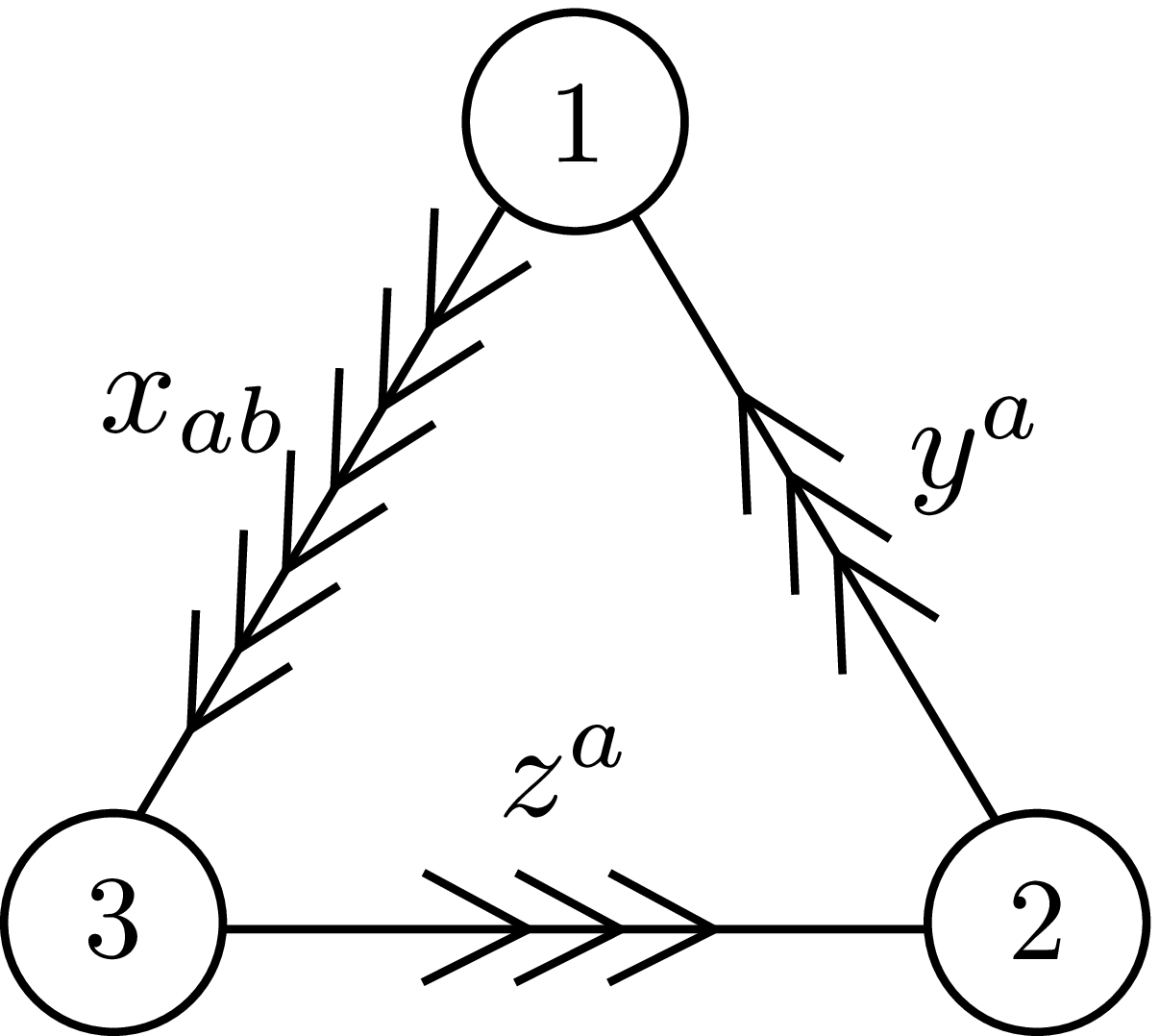}
\label{fig:SD1on2 quiver}}
\qquad
\subfigure[\small Windows for Seiberg duality on node 3.]{
\includegraphics[height=4cm]{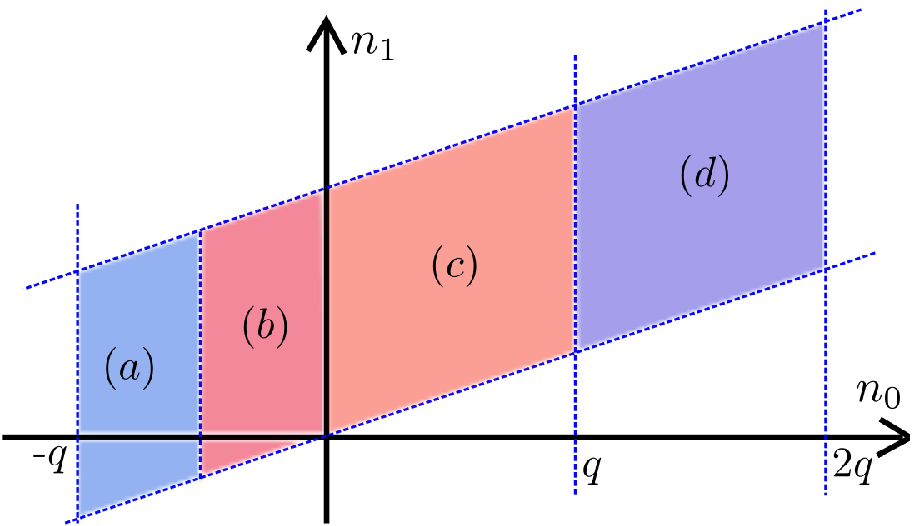}
\label{fig:torsion domain SD3}}
\qquad
\subfigure[\small $\cQ_{(1; 3)}$]{
\includegraphics[height=3cm]{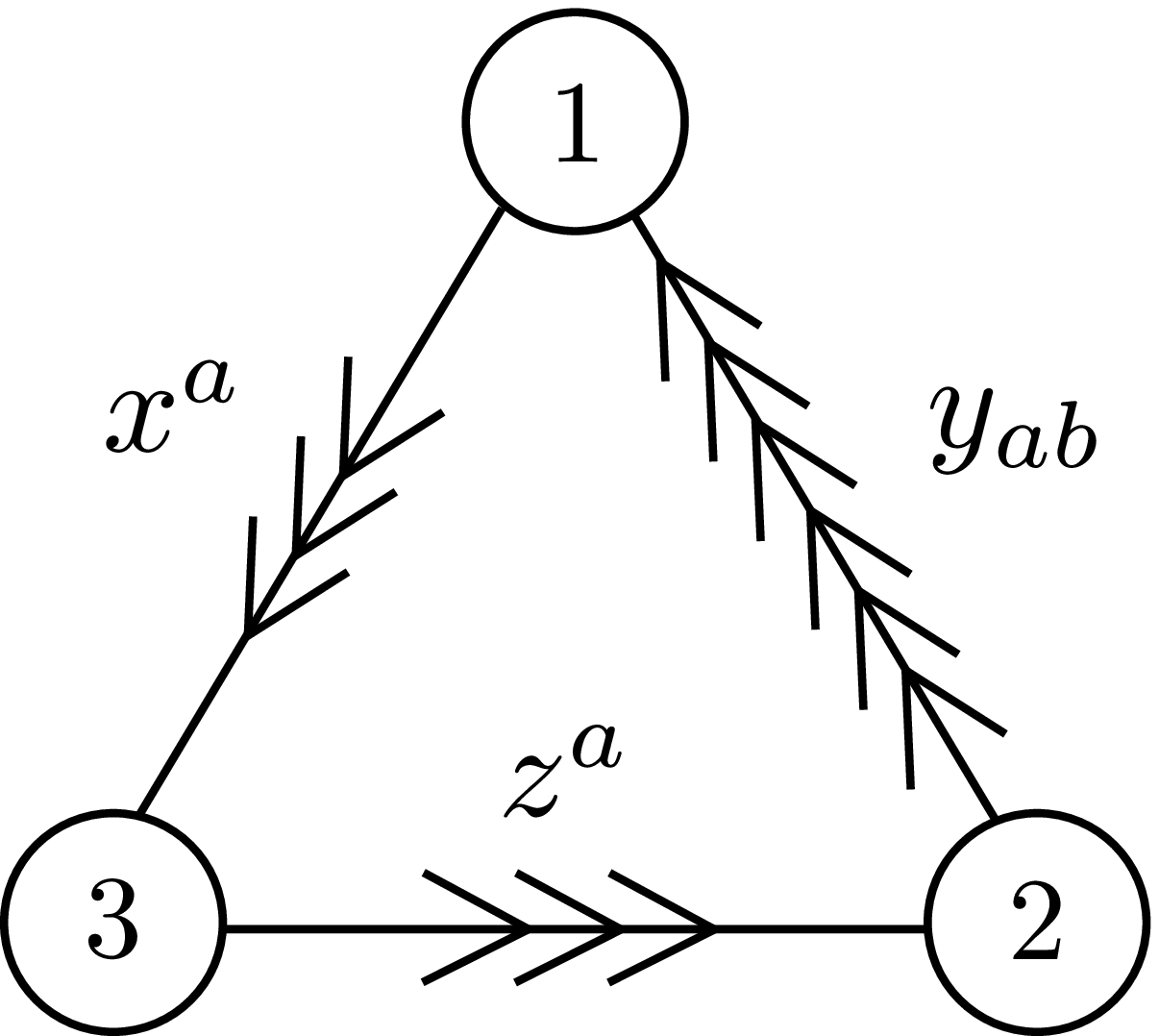}
\label{fig:SD1on3 quiver}}
\caption{\small Left: Windows in the fundamental domain on the $(n_0,n_1)$ plane, depending on which kind of mutation one should perform on the CS theories $(\cQ_{(0)}, \bm{N}, \bm{k})$ at a given node $i_0$. The windows are bluish when $\theta_{i_0-}$ and $\theta_{i_0+}$ have opposite signs, and reddish when they have the same sign. Right: Corresponding dual quivers $\cQ_{(1; i_0)}$. \label{Figs: 14 windows and 3 quivers}}
\end{center}
\end{figure}

Consider D-branes on $Y=\bC^3/\bZ_3$, also known as complex cone over $dP_0= \CP^2$: this singularity is the blow down of $\tilde{Y}=\cO_{\CP^2}(-3)$. The simplest of the quivers associated to $\tilde{Y}$ (denoted $\cQ_{(0)}$ in the following) is shown in Figure \ref{fig:quiver dP0 toric}. It has three nodes and nine arrows $x_a$, $y_b$, $z_c$ ($a,b,c= 1, \cdots, 3$), as indicated on the Figure. There is a superpotential
\be
W= \epsilon^{abc}\, x_a\, y_b\, z_c\,
\ee
which reduces the global symmetry group of the quiver to $SU(3)$. The arrows $x$, $y$, $z$ each transform in the $\bm{3}$ of this $SU(3)$. In three dimensions we can consider a CS theory $(\cQ_{(0)}, \bm{N}, \bm{k})$ with generic gauge group
\be\label{gauge groups of dP0 quiver}
U(N_1)\times U(N_2)\times U(N_3)\, ,
\ee
and Chern-Simons levels $\bm{k}=(k_1, k_2, k_2)$. The only constraint is from the cancelation of  $\bZ_2$ anomalies%
\footnote{This is to cancel the anomaly in the non-Abelian part of the gauge group. There are still some $\bZ_2$ anomalies in the Abelian sector whose cancelation requires to introduce off-diagonal CS levels \cite{Benini:2011cma}; in this paper we neglect this subtlety.} ,
$k_i + \frac12 \sum_j A_{ij} N_i \in \bZ$ .
We will also impose $\sum_i k_i=0$ in order to have a M-theory dual.
The quiver representation $\cO_p$ (for D2-branes) has dimension $\bm{r}= (1,1,1)$. Successive quiver mutations lead to an infinite tree of Seiberg dual quivers of the type shown in Figure \ref{fig: general dP0 quiver} \cite{Cachazo:2001sg, Feng:2002kk}. The $\cO_p$ representation of a generic quiver in the duality tree has dimension vector $\bm{r}= (r_1,r_2,r_3)$ with the $r_i$'s satisfying a Markov equation \cite{Feng:2002kk}
\be\label{Markov eq for dP0}
r_1^2+r_2^2+r_3^2 = 3r_1r_2r_3\, .
\ee
The numbers of arrows $(n_x,n_y,n_z)$ are given by
\be
|n_x|= 3r_2\, , \qquad |n_y| = 3r_3\, , \qquad |n_z|= 3r_1\, ,
\ee
The superpotential of any of these theories $W$ can be found by following the rules of quiver mutations.
We will concentrate on the first Seiberg dual quiver in this infinite family:
\be
\cQ_{(1; i_0)} \equiv \mu_{i_0}(\cQ_{(0)})\, ,
\ee
obtained by a quiver mutation on node $i_0$. Due to the $\bZ_3$ symmetry of $\cQ_{(0)}$ we get the same quiver for any $i_0$, but this $\bZ_3$ symmetry is broken by generic quiver representations, and therefore it is better to think of $\cQ_{(1; i_0)}$, $i_0=1,2,3$ as three distinct quivers ---see Figures \ref{fig:SD1on1 quiver}, \ref{fig:SD1on2 quiver}, \ref{fig:SD1on3 quiver}.
Choosing for instance $i_0=1$, the (left and right) mutation matrices defined in section \ref{sec: mutation, branes and Qs} are
\be
M_{(1;\, L)}= \mat{ -1 & 0 & 3\\ 0&1&0\\0&0&1 }\, ,\qquad M_{(1;\, R)}= \mat{ -1 & 3 & 0\\ 0&1&0\\0&0&1 }\, .
\ee
We have
\be
\bm{r}= (1,1,1) M_{(1;\, L)}^T = (1,1,1) M_{(1;\, R)}^T  = (2,1,1)\, ,
\ee
and $(n_x,n_y,n_z)= (-3,-3,-6)$.  The minus corresponds to the fact that one should flip the orientation of the arrows in Fig.\ref{fig: general dP0 quiver}. The arrows $x$, $y$ and $z$ transform in the $\bm{\bar{3}}$, $\bm{\bar{3}}$ and $\bm{6}$ of $SU(3)$, respectively, and we have a superpotential
\be\label{W of first SD for dP0 quiver}
W = y^a\, x^b\, z_{(ab)}\, .
\ee
Similar considerations apply for $i_0=2$ and $3$.

\subsection{CS quivers for M2-branes on the cone over $Y^{p,q}(\CP^2)$}\label{subsec: results of BCC}
\begin{figure}[t]
\begin{center}
\subfigure[\small Beilinson quivers for $\CP^2$ from $\cQ_{(0)}$.]{
\includegraphics[height=3.5cm]{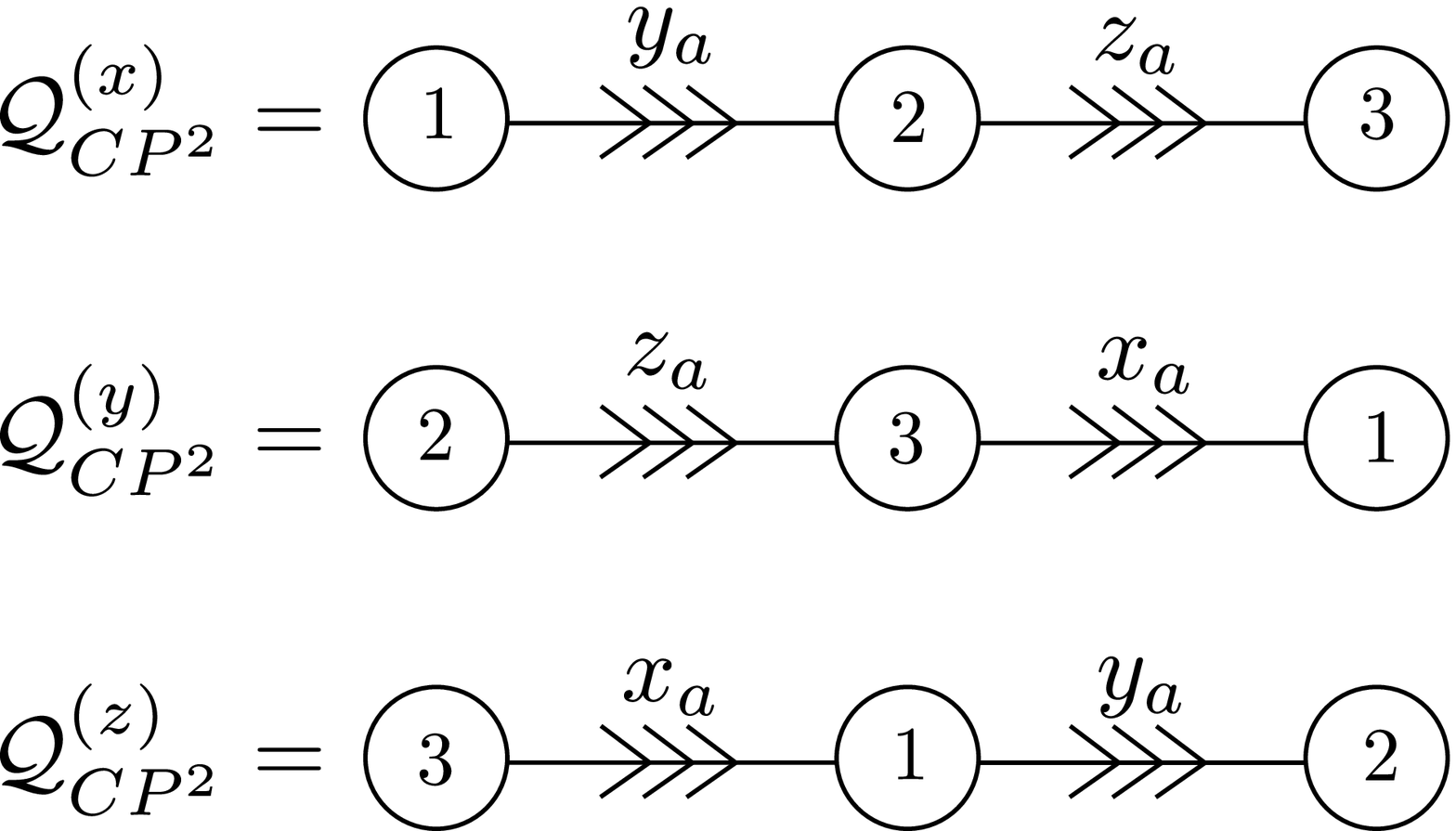}
\label{fig:Bel quivers dP0 toric}}
\,
\subfigure[\small  K\"ahler chambers in FI space of $\cQ_{(0)}$.]{
\includegraphics[height=4cm]{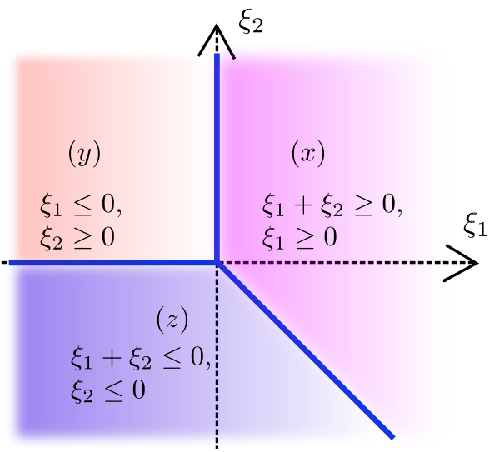}
\label{fig: KCs toric dP0}}
\,\\
\subfigure[\small Torsion group $\Gamma$.]{
\includegraphics[height=4.5cm]{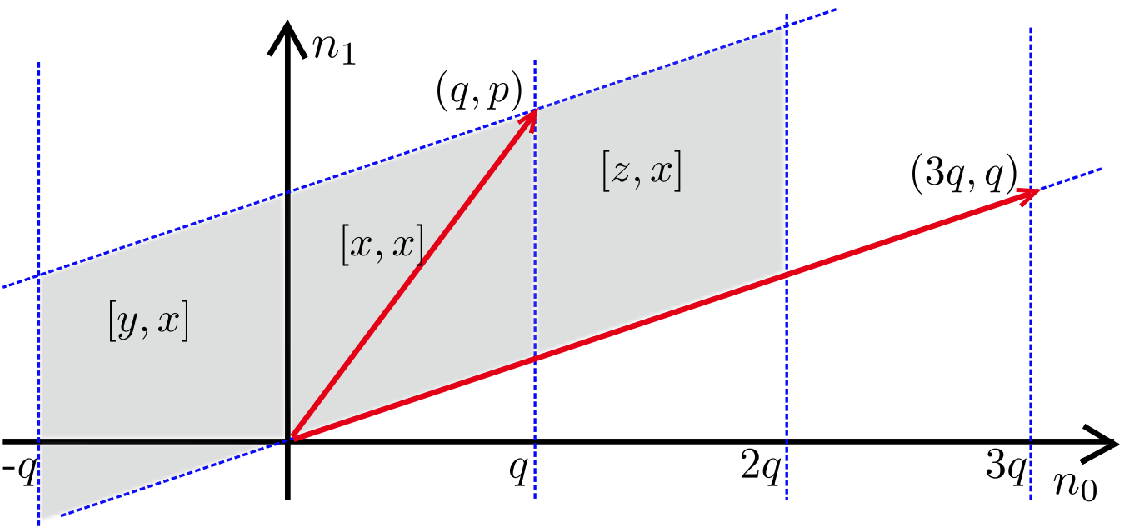}
\label{fig: torsion group Gamma}}
\caption{\small  Upper left:  Beilinson quivers for the three K\"ahler chambers of the toric $dP_0$ quiver. Upper right: K\"ahler chambers in FI parameter space. Below: The  $(n_0, n_1)$ plane parametrizing M-theory torsion flux, with a choice of fundamental domain. Points related by $(q,p)$ or $(3q,q)$ (red vectors) are identified.}
\end{center}
\end{figure}
In \cite{Benini:2011cma} it was shown that the generic Chern-Simons quiver (\ref{gauge groups of dP0 quiver}) describes M2-branes on a cone over the manifold  $Y^{p,q}(\CP^2)$ of \cite{Martelli:2008rt}, with generic value $(n_0,n_1)$ of the torsion $G_4$ flux in
\be\label{torsion group Gamma}
 H^4(Y^{p,q}(\CP^2), \bZ) \,=\, \bZ^2 \,/\, \langle (3q,q) \,,\, (q,p) \rangle  \, \equiv\, \Gamma \, .
\ee
Let us review this correspondence in the notation of this paper.
The dual  type IIA background is of the form (\ref{Y fibered on R}) with $\tilde{Y}=\cO_{\CP^2}(-3)$,
and with \cite{Benini:2011cma}
\bea\label{fluxes and sources for YpqCP2}
\bm{Q}_{\text{flux}, -} \, & \equiv\, \left( -n_0+\tfrac12 q \,|\, -q \,|\, 0 \right)\, , \\
\bm{Q}_{\text{flux}, +} \, & \equiv\, \left( -n_0+3n_1 +  \tfrac12 q -\tfrac32 p \,|\, -q+ 3 p \,|\, 0 \right)\, , \\
\bm{Q}_\text{source} \, & \equiv\, \left( -p \,|\, n_1- \tfrac12 p \,|\, N-\tfrac18 p \right)\, .
\eea
The quiver $\cQ_{(0)}$ has three K\"ahler chambers in FI parameter space. The corresponding Beilinson quivers are shown in Figure \ref{fig:Bel quivers dP0 toric}; since they are obtained by deleting the arrows $x$, $y$ or $z$, respectively, we denote them by $\cQ_{\CP^2}^{(x)}$, $\cQ_{\CP^2}^{(y)}$ and $\cQ_{\CP^2}^{(z)}$. From the Beilinson quivers we find the three K\"ahler chambers shown in Figure \ref{fig: KCs toric dP0}. The corresponding dictionaries between brane charge and quiver ranks are
\be\label{3 dicos for BCC paper}
Q^{\vee}_{(x)} = \mat{1 & \frac12 & \frac18 \\ -2 & 0 & \frac34 \\ 1 & -\frac12 & \frac18} \, , \quad
Q^{\vee}_{(y)} = \mat{ 1 & \frac12 & \frac18 \\ 1 & \frac32 & \frac98 \\ -2 & -2 & - \frac14} \, , \quad
Q^{\vee}_{(z)} =\mat{ -2 & 2 & - \frac14 \\ 1 & - \frac32 & \frac98 \\ 1 & - \frac12 & \frac18}\, .
\ee
One can cover the torsion group (\ref{torsion group Gamma}) by three windows shown in Fig.\ref{fig: torsion group Gamma}. In window $[y,x]$ we should use $Q^{\vee}_{(y)}$ at $r_0<0$ and $Q^{\vee}_{(x)}$ at $r_0>0$, and similarly for the other two windows $[x,x]$ and $[z,x]$.
By running the algorithm of section \ref{subsec: from M2 to CS quiver} one reproduces the result of \cite{Benini:2011cma}, shown in Table \ref{tab: three theories dP0 toric}.
\begin{table}[t]
\bea\nn
&\qquad  \text{Conditions}: \qquad -q \leq n_0 \leq q\, , \qquad\qquad 0 \leq 3n_1 - n_0 \leq 3p - q \\
&\boxed{ \text{Window $[y,x]$:} \, \begin{cases}
&\bm{N} \,=\, (N-n_ 0 + n_ 1  - p, N, N - n_ 1 ) \\
&\bm{k} \,=\, ( \frac {3 n_ 1} {2} - n_ 0, \frac {n_ 0} {2} - 3 n_ 1 + \frac {3 p}{2} - q, \frac {n_ 0}{2} + \frac {3 n_ 1}{2} - \frac {3 p}{2} + q)
\end{cases}}\\
\\
&\qquad \text{Conditions}: \qquad 0\leq n_0 \leq q \, ,\qquad\qquad 0 \leq 3n_1 - n_0 \leq 3p-q \\
& \boxed{ \text{Window $[x,x]$:} \, \begin{cases}
&\bm{N} \,=\, (N+ n_ 1  - p, N, N - n_ 1) \\
&\bm{k} \,=\, ( \frac {3 n_ 1} {2} - n_ 0,
2 n_ 0 - 3 n_ 1 + \frac {3 p} {2} - q, -n_ 0 + \frac {3 n_ 1} {2} - \frac{3 p} {2} + q)
\end{cases}}\\
\\
& \qquad \text{Conditions} : \qquad q  \leq n_0 \leq 2q \, , \qquad\qquad 0 \leq 3n_1 - n_0 \leq 3p - q \\
& \boxed{\text{Window $[z,x]$:} \, \begin{cases}
&\bm{N} \,=\, (  N+n_ 1  - p, N, N+n_ 0 - n_ 1  - q) \\
&\bm{k} \,=\, (\frac {n_ 0} {2} + \frac {3 n_ 1} {2} - \frac {3 q} {2}, \frac {n_ 0}{2} - 3 n_ 1 + \frac {3 p} {2} + \frac {q} {2}, -n_ 0 + \frac {3 n_1} {2} - \frac {3 p} {2} + q )
\end{cases}}
\eea
\caption{\small Theories for $Y^{p,q}(\CP^2)$ based on the toric quiver $\cQ_{(0)}$, for any value of $(n_0, n_1)$ in the fundamental domain (spanning $\Gamma$).  \label{tab: three theories dP0 toric}}
\end{table}

\subsection{Dual theories from mutations}
Given the technology we have introduced, it is quite easy (with some help from \emph{Mathematica}) to work out the Seiberg dual Chern-Simons theories. For definiteness we present the results for the theories based on the quivers $\cQ_{(1; i_0)}$, at the first level in the duality tree%
\footnote{If we go on dualizing there is more and more ways to mutate on the various nodes and the duality tree ramifies more and more, but this adds nothing new to the story}. In that case the quiver representation $\cO_p$ for a D2-brane corresponds to a gauge group $U(2)\times U(1)\times U(1)$ and cyclic permutations thereof.

\subsubsection*{Duality on node 1}
Consider mutating the quiver $\cQ_{(0)}$ of Figure \ref{fig:quiver dP0 toric} on node $1$, giving rise to the quiver $\cQ_{(1; 1)}$ of Figure \ref{fig:SD1on1 quiver}. The $\theta$ parameters of the fractional D-branes described by the field theories of Table \ref{tab: three theories dP0 toric} are:
\bea
\text{Window $[y,x]$:} &\qquad \theta_{1 -} = n_0\, &,& \quad \theta_{1 +}= 3 n_1-n_0\,  \\
\text{Window $[x,x]$:} &\qquad \theta_{1 -} = n_0\, &,& \quad \theta_{1 +}= 3 n_1-n_0\,  \\
\text{Window $[z,x]$:} &\qquad \theta_{1 -} = -2 n_0+ 3q\, &,& \quad \theta_{1 +}= 3 n_1-n_0\,  \\
\eea
This gives us four windows, denoted $(a)$, $(b)$, $(c)$, $(d)$ in Figure \ref{fig:torsion domain SD1}:
\bea
(a)\, : &\quad \theta_{1 -} \leq 0\, , \quad \theta_{1 +} \geq 0 : &\quad
              &  -q \leq n_0 \leq 0 \, , &\quad 0 \leq 3n_1 - n_0 \leq 3p -q  \\
(b)\, : &\quad \theta_{1 -} \geq 0\, , \quad \theta_{1 +} \geq 0 : &\quad
             &   0 \leq n_0 \leq q\, , &\quad 0 \leq 3n_1 - n_0 \leq 3p -q  \\
(c)\, : &\quad \theta_{1 -} \geq 0\, , \quad \theta_{1 +} \geq 0 : &\quad
              &     q \leq n_0 \leq \tfrac32 q\, , &\quad 0 \leq 3n_1 - n_0 \leq 3p -q  \\
(d)\, : &\quad \theta_{1 -} \leq 0\, , \quad \theta_{1 +} \geq 0 : &\quad
              &    \tfrac32 q \leq n_0 \leq 2q\, , &\quad 0 \leq 3n_1 - n_0 \leq 3p -q  \\
\eea
The dual $\cQ_{(1; 1)}$ quiver theory has ranks and CS levels obtained from the mutated dictionaries according to (\ref{change of Qv: 4 cases}). In window $(a)$ we have
\be\label{SD1 (a) dicos}
Q^{\vee}_{-} = M_{1; L}^T Q^{\vee}_{(y)} =
 \mat{  -1 & -\frac{1}{2} & -\frac{1}{8} \\
 1 & \frac{3}{2} & \frac{9}{8} \\
 1 & -\frac{1}{2} & \frac{1}{8}  }\, ,\qquad
Q^{\vee}_{+} = M_{1; R}^T Q^{\vee}_{(x)} =
\mat{-1 & -\frac{1}{2} & -\frac{1}{8} \\
 1 & \frac{3}{2} & \frac{9}{8} \\
 1 & -\frac{1}{2} & \frac{1}{8}}\, .
\ee
We see that $Q^{\vee}_- = Q^{\vee}_+$ in this case. In window $(b)$ we have
\be
Q^{\vee}_{-}= M_{1; R}^T Q^{\vee}_{(x)}\, , \qquad Q^{\vee}_{+}= M_{1; R}^T Q^{\vee}_{(x)}
\ee
so that  $Q^{\vee}_- = Q^{\vee}_+$ again, and it turns out that this dictionary is the same as (\ref{SD1 (a) dicos}) for window $(a)$. Therefore the CS theories in window $(a)$ and $(b)$ are the same.
Similarly we find the dictionaries for windows $(c)$
\be\label{SD1 (c) dicos}
Q^{\vee}_{-} = M_{1; R}^T Q^{\vee}_{(z)} =
 \mat{2 & -2 & \frac{1}{4} \\
 -5 & \frac{9}{2} & \frac{3}{8} \\
 1 & -\frac{1}{2} & \frac{1}{8}  }\, ,\qquad
Q^{\vee}_{+} = M_{1; R}^T Q^{\vee}_{(x)} =
\mat{  -1 & -\frac{1}{2} & -\frac{1}{8} \\
 1 & \frac{3}{2} & \frac{9}{8} \\
 1 & -\frac{1}{2} & \frac{1}{8}   }\, ,
\ee
and for window $(d)$
\be\label{SD1 (d) dicos}
Q^{\vee}_{-} = M_{1; R}^T Q^{\vee}_{(z)} =
 \mat{ 2 & -2 & \frac{1}{4} \\
 1 & -\frac{3}{2} & \frac{9}{8} \\
 -5 & \frac{11}{2} & -\frac{5}{8} }\, ,\qquad
Q^{\vee}_{+} = M_{1; R}^T Q^{\vee}_{(x)} =
\mat{  -1 & -\frac{1}{2} & -\frac{1}{8} \\
 1 & \frac{3}{2} & \frac{9}{8} \\
 1 & -\frac{1}{2} & \frac{1}{8} }\, ,
\ee
Using the mutated dictionaries in their respective windows of validity, we find the CS quivers theories $(\cQ_{(1;1)}, \bm{N}, \bm{k} )$ of Table \ref{tab: CS theories dP0 SD1}.
\begin{table}[t]
\bea\nn
&\qquad  \text{Conditions}: \qquad -q \leq n_0 \leq q\, , \qquad\qquad 0 \leq 3n_1 - n_0 \leq 3p - q \\
&\boxed{ \text{SD}_1 (a)\&(b) \, : \, \begin{cases}
&\bm{N} \,=\, ( 2 N+ p-n_ 1, N, N - n_ 1) \\
&\bm{k} \,=\, (n_ 0 - \frac {3 n_ 1} {2}, -n_ 0 + \frac {3 n_ 1} {2} + \frac {3 p}{2} - q, -n_ 0 + \frac {3 n_ 1} {2} - \frac {3 p} {2} + q)
\end{cases}}\\
\\
&\qquad \text{Conditions}: \qquad q \leq n_0 \leq \tfrac32 q \, ,\qquad\qquad 0 \leq 3n_1 - n_0 \leq 3p-q \\
& \boxed{ \text{SD}_1 (c) \, : \, \begin{cases}
&\bm{N} \,=\, ( 2 N + p-n_ 1 , N, N - n_ 1 +n_ 0 - q) \\
&\bm{k} \,=\, (-\frac {n_ 0} {2} - \frac {3 n_ 1} {2} + \frac {3 q} {2},
2 n_ 0 + \frac {3 n_ 1} {2} + \frac {3 p} {2} -
 4 q, -n_ 0 + \frac {3 n_ 1} {2} - \frac {3 p} {2} + q)
\end{cases}}\\
\\
& \qquad \text{Conditions} : \qquad \tfrac32 q \leq n_0 \leq 2q \, , \qquad\qquad 0 \leq 3n_1 - n_0 \leq 3p - q \\
& \boxed{ \text{SD}_1 (d) \, : \, \begin{cases}
&\bm{N} \,=\, ( 2 N + p - n_ 1   + 2 n_ 0 - 3 q, N, N - n_ 1 +n_ 0 - q) \\
&\bm{k} \,=\, (-\frac {n_ 0} {2} - \frac {3 n_ 1} {2} + \frac {3 q} {2}, -n_ 0 + \
\frac {3 n_ 1} {2} + \frac {3 p} {2} + \frac {q} {2},
2 n_ 0 + \frac {3 n_ 1} {2} - \frac {3 p} {2} - \frac {7 q} {2})
\end{cases}}
\eea
\caption{\small Seiberg dual CS theories ``SD$_1$'' for the quiver $\cQ_{(1;1)}$, for any value of $(n_0, n_1) \in \Gamma$. The notation SD$_{i_0}$ means that the duality was performed on node $i_0$ of the $\cQ_{(0)}$ quiver. \label{tab: CS theories dP0 SD1}}
\end{table}

\subsubsection*{Duality on node 2}
We should play the same game for each of the nodes of $\cQ_{(0)}$. Dualizing on $i_0= 2$ leads to a subdivision of the fundamental domain into 8 windows, shown in Figure \ref{fig:torsion domain SD2}.  The resulting CS quiver theories based on $\cQ_{(1,2)}$ are listed in Tables \ref{tab: CS theories dP0 SD2} and \ref{tab: CS theories dP0 SD2 bis}. In this case each of the 8 windows lead to distinct ranks and CS levels.
\begin{table}[h]
\bea\nn
&\qquad  \text{Conditions}: \qquad -q \leq n_0 \leq 0\, , \qquad\qquad  \tfrac{1 }{2}(3p - q) \leq 3n_1 - n_0 \leq 3p - q \\
&\boxed{ \text{SD}_2 (a_1) \, : \, \begin{cases}
&\bm{N} \,=\, (N-n_ 0 + n_ 1  - p, 2N -2 n_ 0 + 3 n_ 1  - 3 p + q, N - n_1 ) \\
&\bm{k} \,=\, ( 2 n_ 0 - \frac {15 n_ 1} {2} + \frac {9 p} {2} - \frac {3 q}{2}, -\frac {n_ 0} {2} + 3 n_ 1 - \frac {3 p} {2} + q, -n_ 0 + \frac {3 n_ 1} {2} - \frac {3 \
p} {2} - \frac {q} {2} )
\end{cases}}\\
\\
&\qquad  \text{Conditions}: \qquad -q \leq n_0 \leq 0\, , \qquad\qquad  0 \leq 3n_1 - n_0 \leq \tfrac{1 }{2}(3p - q) \\
&\boxed{ \text{SD}_2 (b_1) \, : \, \begin{cases}
&\bm{N} \,=\, (N-n_ 0 + n_ 1  - p, 2 N - 3 n_ 1, N - n_ 1 ) \\
&\bm{k} \,=\, (\frac {3 n_ 1} {2} - n_ 0, -\frac {n_ 0} {2} +
 3 n_ 1 - \frac {3 p} {2} + q, 2 n_ 0 - \frac {15 n_ 1} {2} + 3 p - 2 q )
\end{cases}}\\
\\
&\qquad  \text{Conditions}: \qquad 0 \leq n_0 \leq \tfrac12 q\, , \qquad\qquad  \tfrac{1 }{2}(3p - q) \leq 3n_1 - n_0 \leq 3p - q \\
&\boxed{ \text{SD}_2 (a_2) \, : \, \begin{cases}
&\bm{N} \,=\, (N+ n_ 1  - p, 2N -2 n_ 0 + 3 n_ 1 - 3 p + q, N - n_ 1 ) \\
&\bm{k} \,=\, (2 n_ 0 - \frac {15 n_ 1} {2} + \frac {9 p} {2} - \frac {3 q} {2}, -2 n_ 0 + 3 n_ 1 - \frac {3 p} {2} + q, 2 n_ 0 + \frac {3 n_ 1} {2} - \frac {3 p} {2} - \frac {q} {2} )
\end{cases}}\\
\\
&\qquad  \text{Conditions}: \qquad 0 \leq n_0 \leq \tfrac12 q\, , \qquad\qquad 0 \leq 3n_1 - n_0 \leq \tfrac{1 }{2}(3p - q) \\
&\boxed{ \text{SD}_2 (b_2) \, : \, \begin{cases}
&\bm{N} \,=\, (N+n_ 1  - p, 2 N - 3 n_ 1, N - n_ 1 ) \\
&\bm{k} \,=\, (\frac {3 n_ 1} {2} - n_ 0, -2 n_ 0 + 3 n_ 1 - \frac {3 p} {2} + q,5 n_ 0 - \frac {15 n_ 1} {2} + 3 p - 2 q )
\end{cases}}\\
\eea
\caption{\small Seiberg dual CS theories ``SD$_2$'' for the quiver $\cQ_{(1;2)}$.  \label{tab: CS theories dP0 SD2}}
\end{table}
\begin{table}[h]
\bea\nn
&\qquad  \text{Conditions}: \qquad \tfrac12 q \leq n_0 \leq q\, , \qquad\qquad  \tfrac{1 }{2}(3p - q) \leq 3n_1 - n_0 \leq 3p - q \\
&\boxed{ \text{SD}_2 (c_2) \, : \, \begin{cases}
&\bm{N} \,=\, (N+ n_ 1  - p, 2N+ 3 n_ 1  - 3 p, N - n_ 1 ) \\
&\bm{k} \,=\, (5 n_ 0 - \frac {15 n_ 1} {2} + \frac {9 p} {2} - 3 q, -2 n_ 0 + 3 n_ 1 - \frac {3 p} {2} + q, -n_ 0 + \frac {3 n_ 1} {2} - \frac {3p}{2} + q )
\end{cases}}\\
\\
&\qquad  \text{Conditions}: \qquad \tfrac12 q \leq n_0 \leq  q\, , \qquad\qquad 0 \leq 3n_1 - n_0 \leq \tfrac{1 }{2}(3p - q) \\
&\boxed{ \text{SD}_2 (d_2) \, : \, \begin{cases}
&\bm{N} \,=\, (N+n_ 1  - p,2N+ 2 n_ 0 - 3 n_ 1  - q, N - n_ 1 ) \\
&\bm{k} \,=\, (2 n_ 0 + \frac {3 n_ 1} {2} - \frac {3 q} {2}, -2 n_ 0 + 3 n_ 1 - \frac {3 p} {2} + q,2 n_ 0 - \frac {15 n_ 1} {2} + 3 p - \frac{q}{2} )
\end{cases}}\\
\\
&\qquad  \text{Conditions}: \qquad q \leq n_0 \leq 2 q\, , \qquad\qquad \tfrac{1 }{2}(3p - q)  \leq 3n_1 - n_0 \leq  3p - q\\
&\boxed{ \text{SD}_2 (c_1) \, : \, \begin{cases}
&\bm{N} \,=\, ( N+ n_ 1  - p, 2 N+ 3 n_ 1  - 3 p,N+ n_ 0 - n_ 1  - q ) \\
&\bm{k} \,=\, ( 2 n_ 0 - \frac {15 n_ 1} {2} + \frac {9 p} {2}, -\frac {n_ 0} {2} + 3 n_ 1 - \frac {3 p} {2} - \frac {q} {2}, -n_ 0 + \frac{3 n_1}{2}- \frac {3 p} {2} + q)
\end{cases}}\\
\\
&\qquad  \text{Conditions}: \qquad  q \leq n_0 \leq  2q\, , \qquad\qquad 0 \leq 3n_1 - n_0 \leq \tfrac{1 }{2}(3p - q) \\
&\boxed{ \text{SD}_2 (d_1) \, : \, \begin{cases}
&\bm{N} \,=\, ( N+ n_ 1  - p,2N+ 2 n_ 0 - 3 n_ 1 - q, N+ n_ 0 - n_ 1  - q) \\
&\bm{k} \,=\, (-n_ 0 + \frac {3 n_ 1} {2} + \frac {3 q} {2}, -\frac {n_ 0} {2} + 3 n_ 1 - \frac {3 p} {2} - \frac {q} {2},2 n_ 0 - \frac {15 n_ 1} {2} + 3 p - \frac {q} {2} )
\end{cases}}\\
\eea
\caption{\small Seiberg dual CS theories ``SD$_2$'', continued.  \label{tab: CS theories dP0 SD2 bis}}
\end{table}

\subsubsection*{Duality on node 3}
Dualizing on node 3 we find a pattern similar to the node 1 case, to which it is related by a $\bZ_2$ operation (which is CP in the original $\cQ_{(0)}$ quiver). We have four windows in Figure \ref{fig:torsion domain SD3} and it turns out there are only three distinct pair of dictionaries. The dual theories are the $\cQ_{(1; 3)}$ quiver with the ranks and CS levels of Table \ref{tab: CS theories dP0 SD3}.
\begin{table}[h]
\bea\nn
&\qquad  \text{Conditions}: \qquad -q \leq n_0 \leq -\tfrac12 q\, , \qquad\qquad 0 \leq 3n_1 - n_0 \leq 3p - q \\
&\boxed{ \text{SD}_3 (a) \, : \, \begin{cases}
&\bm{N} \,=\, (N -n_ 0 + n_ 1  - p, N,2 N  -2 n_ 0 + n_ 1  - q ) \\
&\bm{k} \,=\, (2 n_ 0 + \frac {3 n_ 1} {2} + \frac {3 q} {2}, -n_ 0 + \frac {3 n_ 1}{2} - 3 p + \frac {q} {2}, -\frac {n_ 0} {2} - \frac {3 n_1} {2} + \frac {3 p} {2} - q )
\end{cases}}\\
\\
&\qquad \text{Conditions}: \qquad -\tfrac12 q \leq n_0 \leq 0 \, ,\qquad\qquad 0 \leq 3n_1 - n_0 \leq 3p-q \\
& \boxed{ \text{SD}_3 (b) \, : \, \begin{cases}
&\bm{N} \,=\, (N -n_ 0 + n_ 1  - p, N,2N+ n_ 1 ) \\
&\bm{k} \,=\, ( \frac {3 n_ 1} {2} - n_ 0,2 n_ 0 + \frac {3 n_ 1} {2} - 3p + 2 q, -\frac {n_ 0} {2} - \frac {3 n_ 1} {2} + \frac {3 p} {2} - q)
\end{cases}}\\
\\
& \qquad \text{Conditions} : \qquad 0 \leq n_0 \leq 2q \, , \qquad\qquad 0 \leq 3n_1 - n_0 \leq 3p - q \\
& \boxed{ \text{SD}_3 (c)\&(d) \, : \, \begin{cases}
&\bm{N} \,=\, (N+n_ 1  - p, N, 2 N+n_ 1  ) \\
&\bm{k} \,=\, ( \frac {3 n_ 1} {2} - n_ 0, -n_ 0 + \frac {3 n_ 1} {2} - 3 p + 2 q,n_ 0 - \frac {3 n_ 1} {2} + \frac {3 p} {2} - q)
\end{cases}}
\eea
\caption{\small Seiberg dual CS theories ``SD$_3$'' for the quiver $\cQ_{(1;3)}$.  \label{tab: CS theories dP0 SD3}}
\end{table}

\subsection{Consistency checks and remarks}
We have just presented the complete list of the Chern-Simons quiver theories $(\cQ_{(1;i_0)}, \bm{N}, \bm{k})$ obtained by doing one Seiberg duality on the CS theories of Table \ref{tab: three theories dP0 toric}. At any fixed $(n_0, n_1)$ the corresponding $\cQ_{(0)}$-quiver theory has three distinct Seiberg dual theories. For instance, the dual theories to the torsionless point $(n_0, n_1)=(0,0)$ are
\bea\label{SD for torsionless point}
\cQ_{(1; 1)}\, &: \,  \qquad U(2N +p)_0\times U(N)_{\frac32 p-q}\times U(N)_{-\frac32 p+q}\, , \\
\cQ_{(1; 2)}\, &: \,  \qquad U(N -p)_0\times U(2N)_{-\frac32 p+q}\times U(N)_{3 p-2q}\, , \\
\cQ_{(1; 3)}\, &: \,  \qquad U(N -p)_0\times U(N)_{-3 p+ 2q}\times U(2N)_{\frac32 p-q}\, . \\
\eea
The semi-classical moduli space can be analyzed similarly to \cite{Benini:2011cma}. Remark that the first theory in (\ref{SD for torsionless point}) has naively a larger Coulomb branch than the one of the original theory, but in this case the semi-classical brane derivation we performed actually fails: It was shown by Aharony \cite{Aharony:1997gp} that the Seiberg dual in that case with $k_{i_0}=0$ involves extra singlets and additional couplings to monopoles, which should conspire to give the correct matching of moduli spaces.

For generic $(n_0, n_1)$ it is obvious that the geometric branch (see section \ref{subsec: One loop VMS CS quivers}) of the semi-classical moduli space of $(\cQ, \bm{N}, \bm{k})$ reproduces the type IIA geometry. This is true by construction: We should compute the parameters $\bm{k_{\pm}}= \pm \bm{\theta}_{\pm}$ from the quiver data and use the dictionaries $Q^{\vee}_{\pm}$ to find the IIA resolutions parameters (\ref{chi of IIA geom}), but this is just inverting the steps we followed to find the field theories in the first place.

A better consistency check is that the Seiberg dual field theories reproduce the periodicities $(n_0, n_1)\sim (n_0+3q, n_1+q)$ and $(n_0, n_1)\sim (n_0+q, n_1+p)$ of the torsion group (\ref{torsion group Gamma}). This is indeed the case. The field theories on the external boundaries of the various windows in Figures \ref{fig:torsion domain SD1}, \ref{fig:torsion domain SD2}, \ref{fig:torsion domain SD3} are identified according to%
\footnote{They are identified up to the expected shift of the D2-brane charge $N$ which arises because the translation along a periodicity vector is a large gauge transformation in type IIA, which shifts the Page charges. See equation (5.12) of \cite{Benini:2011cma}.}
\bea
\delta{(n_0, n_1)}= (3q,q) \, & :\quad \text{SD}_{i_0}\rightarrow \text{SD}_{i_0}\, , \\
\delta{(n_0, n_1)}= (q,p) \, & :\quad \text{SD}_{i_0}\rightarrow \text{SD}_{i_0-1}\, , \\
\eea
meaning that the periodicity $(3q,q)$ sends a theory in the family $\text{SD}_{1}$ to another identical one in the same family, while the periodicity $(q,p)$ sends a theory in $\text{SD}_{i_0}$ to one in another family $\text{SD}_{i_0-1}$. In other words, the first periodicity is apparent inside any of the three  families of $\cQ_{(1; i_0)}$-quivers while the second one holds because of Seiberg dualities.

\subsection{Monopole operators and matching the chiral ring}
Beyond the semi-classical analysis of the moduli space, which reproduces the IIA geometry, we can check our Seiberg dual quivers by seeing whether the quantum chiral rings of dual theories coincide, as they should. In particular the chiral ring should reproduce the CY$_4$ cone probed by the M2-branes according to the construction (\ref{M2 geometric branch algebraically}). The analysis we perform here is ``pseudo-Abelian'': the monopole operators we construct are best thought of as functions on the Coulomb branch. A comprehensive analysis of the monopoles in the non-Abelian theory (for instance along the lines of \cite{Bashkirov:2011vy}) would require more sophisticated tools, and we leave it for future work.

Some useful facts about the relevant ``diagonal'' monopole operators are collected in Appendix \ref{App: monopole operator charges}.

\subsubsection*{Toric quiver monopoles}
The gauge invariant operators generating the chiral ring (or rather the subspace of it corresponding to (\ref{M2 geometric branch algebraically})) of the toric CS quiver theories of \cite{Benini:2011cma} are:
\bea\label{gauge_invariants toric dP0}
 [y,x]\;: &\quad\bm{a}_{(3)} = XYZ \;,\quad \bm{b}_{(q)} = TX^{-n_0}Z^{q+n_0} \;,\quad \bm{c}_{(3p-q)} = \tilde TY^{3n_1 - n_0}Z^{3p-q-3n_1 + n_0}\, , \\
 [x,x]\;: &\quad \bm{a}_{(3)} = XYZ \;,\quad \bm{b}_{(q)} = TY^{n_0}Z^{q-n_0} \;,\quad \bm{c}_{(3p-q)} = \tilde TY^{3n_1 - n_0}Z^{3p-q-3n_1 + n_0}\, ,\\
 [z,x]\;: &\quad \bm{a}_{(3)} = XYZ \;,\quad \bm{b}_{(q)} = TX^{-q+n_0}Y^{2q-n_0} \;,\quad \bm{c}_{(3p-q)} = \tilde TY^{3n_1 - n_0}Z^{3p-q-3n_1 + n_0} \, .
\eea
The subscript corresponds to the number of symmetrized $SU(3)$ indices: The operator $\bm{a}_3$ transforms in the $\text{Sym}^3\, \bm{3}= \bm{10}$, etc. We stress again that the operators (\ref{gauge_invariants toric dP0}) should be thought of as regular functions on the Coulomb branch. On the other hand at the origin of the Coulomb branch (where the CFT lives) we should contract the indices according to the non-Abelian representations of $T$, $\tilde{T}$ (see Appendix \ref{App: monopole operator charges}), and looking at all possible gauge invariant operators in the CFT one finds that the diagonal monopole operators fall in larger $SU(3)$ representations than the ones apparent in (\ref{gauge_invariants toric dP0}).%
\footnote{This result was communicated to me by Stefano Cremonesi.}. We are left to assume that these extra operators fail to be chiral due to non-perturbative effects, but this issue would deserve a more serious study.
In any case the operators (\ref{gauge_invariants toric dP0}) together with the quantum relation
\be\label{quantum rel for YpqCP2}
\bm{b}_{(q)}\, \bm{c}_{(3p-q)} \, \sim \, (\bm{a}_{(3)})^p
\ee
reproduce the coordinate ring of the CY$_4$ cone $C(Y^{p,q}(\CP^2))$.

\subsubsection*{Monopoles in Seiberg dual quivers}
For definiteness let us focus on the first of the 14 Seiberg dual theories listed in Tables \ref{tab: CS theories dP0 SD1} to \ref{tab: CS theories dP0 SD3}, which we called ``$\text{SD}_1(a)\&(b)$''. This is the quiver of Figure \ref{fig:SD1on1 quiver} with gauge group and CS levels
\be
 U(2N+p-n_1)_{n_0-\frac32 n_1}\times U(N)_{-n_0+\frac32 n_1+\frac32 p-q} \times U(N-n_1)_{-n_0+\frac32 n_1-\frac32p +q}
\ee
Let us denote by $g_i$ the electric charge under the diagonal $U(1)_i \subset U(N_i)$. The electric and global charges of the bifundamental fields and of the bare monopoles $T$ and $\tilde T$ are:
\be\label{table charges SD1a}
\begin{array}{c|ccc|c|c}
& X^a & Y^a & Z_{ab} & T & \tilde T \\
\hline
g_1 & -1 & 1 & 0 & 2[n_0] & 2[-n_0 + 3n_1] \\
g_2 & 0 & -1 & 1 & -n_0 - q & n_0 - 3n_1 - 3p+q \\
g_3 & 1 & 0 & -1 & -n_0 + q & n_0 - 3n_1 +3p - q \\
\hline
\tabs R & 2-R_Y - R_Z & R_Y & R_Z & R_T & R_T \\
SU(3) & \rep{\bar{3}} & \rep{\bar{3}} & \rep{6} & \rep{1} & \rep{1} \\
U(1)_M & 0 & 0 & 0 & 1 & -1
\end{array}
\ee
where we denoted $R_X=R[X]$ the R-charge of any field $X$, and
\be
R_T= - 2 p +3n_1 -3 n_1 R_Y +\frac32 (p-n_1) R_Z\, .
\ee
From (\ref{table charges SD1a}) we find the gauge invariant operators
\be\label{gauge_invariants SD1(a)}
\bm{a}_{(3)} = XYZ \;,\qquad \bm{b}_{(q)} = TX^{q- n_0} Y^{q+n_0} \;,\qquad \bm{c}_{(3p-q)} = \tilde T X^{3p-q + n_0-3n_1} Z^{3p-q+3n_1 - n_0}\, .
\ee
The operators $\bm{b}$ and $\bm{c}$  are regular functions (all the powers in the expressions (\ref{gauge_invariants SD1(a)}) are positives) if and only if
\be
-q \leq n_0 \leq q\, , \qquad  -(3p-q) \leq 3n_1 - n_0 \leq 3p - q
\ee
This agrees with the condition for the $\text{SD}_1(a)\&(b)$ theory to be valid, and in fact it cuts out a window in the $(n_0, n_1)$ plane which is twice larger than windows $(a)\&(c)$ of Figure \ref{fig:torsion domain SD1}.

The operators (\ref{gauge_invariants SD1(a)}) have the correct R-charges to allow the quantum relation (\ref{quantum rel for YpqCP2}). They also transform in the same $SU(3)$ representation as the operators (\ref{gauge_invariants toric dP0}) of the original theory, namely $\text{Sym}^q\bm{3}$ and $\text{Sym}^{3p-q}\bm{3}$. To understand this we have to deal with the fact that the theory on the Coulomb branch is not Abelian anymore. The ``pseudo-Abelian'' theory (one single D2-brane) is a $U(2)\times U(1)\times U(1)$ theory with CS levels $\bm{k}_{-}$ or $\bm{k}_+$ for $\sigma<0$ or $\sigma >0$, respectively. Consider the $\sigma<0$ case, where the bare monopole is $T$ transforming in the $[n_0, n_0]$ of $U(2)$ and with charge $-n_0-q$ and $-n_0+q$ under the $U(1)$'s of the second and third node. Let $Y\cdot X$ denote the contraction of $Y$ and $X$ on their $U(2)$ indices. Due to the superpotential (\ref{W of first SD for dP0 quiver}) $Y\cdot X$ transforms in the $\rep{3}$ of $SU(3)$. We have
\be\label{3 cases for pseudoAbelian monopoles}
 \bm{b}_{(q)} = \begin{cases} T (Y\cdot X)^q\, , &\qquad \text{if}\; n_0= 0\\
  T \cdot Y^{2n_0} (Y\cdot X)^{q-n_0}  \, ,&\qquad \text{if}\; n_0> 0 \\
 T\cdot X^{-2n_0} (Y\cdot X)^{q+n_0}   \, ,&\qquad \text{if}\; n_0< 0
 \end{cases}
\ee
In the first case we obviously have the correct $SU(3)$ representation. In the second case the contraction $T\cdot Y^{2n_0}$ gives the representation $\text{Sym}^{n_0}\bm{3}$  (using $\bm{\bar{3}}\otimes_A \bm{\bar{3}}= \rep{3}$ and so forth), and thus the operator $T \cdot Y^{2n_0} (Y\cdot X)^{q-n_0}$ lies in the
\be\label{finding the SU3 irrep, expl 00}
\text{Sym}^{n_0}\bm{3} \, \otimes \,   \text{Sym}^{q- n_0}\bm{3}\, \supset \,  \text{Sym}^{q}\bm{3}, ,
\ee
which contains the expected representation $\text{Sym}^{q}\bm{3}$.
We can  similarly analyze the third case in (\ref{3 cases for pseudoAbelian monopoles}), and  the monopoles $\bm{c}_{(3p-q)}$ at $\sigma>0$.
The extra $SU(3)$ representations in (\ref{finding the SU3 irrep, expl 00}) are puzzling. For lack of better tools, we are left to conjecture that the corresponding operators are set to zero in the quantum chiral ring. At the torsionless point $n_0=n_1=0$ the monopole operators match across Seiberg duality without extra assumption.

A similar analysis can be performed in any of the 14 windows of Figure \ref{Figs: 14 windows and 3 quivers}, with similar conclusions.

\section{Remark on brane charge and CS quivers dual to massive IIA geometries}\label{section: brane charges}
This section lies outside the main line of development of the paper. It aims to clarify a point in the treatment of brane charge of \cite{Benini:2011cma}, and it allows to generalize the construction of \cite{Benini:2011cma} to CS quivers with $\sum_i k_i \neq 0$, corresponding to the presence of D8-brane charge in type IIA.

The quantities that we loosely referred to as ``brane charges'' in this work are quantized and conserved charges corresponding to either quantized RR flux or explicit D-brane sources (the latter sourcing the former). In \cite{Benini:2011cma} we identified such brane charges with Page charges \cite{Marolf:2000cb}, which is a concept mostly relevant to the supergravity limit of string theory. Consider a type IIA Dp-brane $\mathsf{E}^{\vee}$ wrapped on some $(p-2)$-cycle $S\subset \tilde{Y}$ with some worlvolume flux $F_{wv}$ turned on. The brane charges of such a source can be read from its Wess-Zumino action,
\be\label{d brane charge in WZ}
S_{WZ} = \tau_p \int_{\bR^{2,1} \times S} C^{(P)} Q(\mathsf{E}^{\vee})\, ,
\ee
where $C^{(P)}$ is defined in term of the ordinary RR potential $C$ (a polyform) as
$C^{(P)}= C\wedge e^B$
with $B$ the background $B$-field pulled-back on the D-brane \cite{Benini:2007gx}. We restrict ourselves to cases where $B$ is flat, and in such a situation this $C^{(P)}$ is the potential for the Page currents
\be
F^{(P)}= F e^B\, .
\ee
The point of this distinction is that the Page charge is sourced only by the D-brane and its worlvolume flux, and therefore they are properly quantized%
\footnote{We should also remark that the Freed-Witten anomaly \cite{Freed:1999vc} (when $S$ is spin$^{c}$ but not spin) leads to half-integer quantized worlvolume fluxes in general \cite{Benini:2011cma}.}; on the other hand they are not invariant under large gauge tranformations $B \rightarrow B+ \Lambda$, $F_{wv}\rightarrow F_{wv}-\Lambda$. The quantity $Q(\mathsf{E}^{\vee})$ in (\ref{d brane charge in WZ}) is by definition the (Page) brane charge. It is given by \cite{Minasian:1997mm}
\be\label{brane charge in general}
Q(E^{\vee})=   ch(E^{\vee})\sqrt{\frac{\hat{A}(TS)}{\hat{A}(NS)}}\, ,
\ee
with $\hat{A}$ the A-roof characteristic class, here for the tangent and normal bundles to $S$. This brane charge is a K-theory class, and indeed K-theory is the natural concept of charge for the B-brane category \cite{Sharpe:1999qz, Aspinwall:2004jr}. In \cite{Benini:2011cma} and in this work, we just took $Q(E^{\vee})=   ch(E^{\vee})$ as our working definition of D-brane charge, but in general we should use the exact formula (\ref{brane charge in general}). The gravitational correction in (\ref{brane charge in general}) translates to small corrections to the brane charge dictionaries  $Q^{\vee}$. This does not affect the results of \cite{Benini:2011cma}, where these corrections where ignored in a consistent way.

These considerations are however important if we want to generalize the string theory derivations of CS quivers of \cite{Benini:2011cma} to cases with $\sum_i r_i k_i \neq 0$. This corresponds to having D8-brane charge (in the guise of $F_0$ flux) in the type IIA dual \cite{Gaiotto:2009mv, Fujita:2009kw}: the fluxes in (\ref{fluxes and sources}) become
\be
\bm{Q}_{\text{flux}} \,  =\, \left( -Q_{4} \,|\, Q_{6;\, \alpha} \,|\, Q_8 \right)\, , \\
\ee
It is possible to work out the effect of $Q_8= F_0 \neq 0$ on the derivation of \cite{Benini:2011cma}, reviewed in section \ref{subsec: results of BCC}. Accounting for the gravitational correction in (\ref{brane charge in general}), the brane charges dictionaries (\ref{3 dicos for BCC paper}) become
\be
Q^{\vee}_{(x)} = \mat{1 & \frac12 & \frac14 \\ -2 & 0 & \frac12 \\ 1 & -\frac12 & \frac14} \;,\quad
Q^{\vee}_{(y)}= \mat{ 1 & \frac12 & \frac14 \\ 1 & \frac32 & \frac54 \\ -2 & -2 & - \frac12} \;,\quad
Q^{\vee}_{(z)} = \mat{ -2 & 2 & - \frac12 \\ 1 & - \frac32 & \frac54 \\ 1 & - \frac12 & \frac14} \;,
\ee
and the sources and fluxes (\ref{fluxes and sources for YpqCP2}) are shifted to
\bea
\bm{Q}_{\text{flux}, -} \, & \equiv\, \left( -n_0+\tfrac12 q -\frac14 F_0\,|\, -q \,|\, F_0 \right)\, , \\
\bm{Q}_\text{source} \, & \equiv\, \left( -p \,|\, n_1- \tfrac12 p \,|\, N-\tfrac14 p \right)\, .
\eea
Running our algorithm, this gives the same results as in Table \ref{tab: three theories dP0 toric}, except that the Chern-Simons levels are shifted to
\be
\bm{k} \rightarrow \bm{k}_{\, (\text{Table \ref{tab: three theories dP0 toric})}} + (0, F_0, 0)\, ,
\ee
and therefore $\sum_i k_i = F_0$.

These Chern-Simons quiver theories should be dual to the massive type IIA $AdS_4$ solutions of \cite{Petrini:2009ur}. We can thus provide a precise map between the six field theory parameters $(\bm{N}, \bm{k})$ and the supergravity parameters of that paper. It would be interesting to see whether one can perform non-trivial tests of this duality that would be sensitive to the finer details of this map.

\section{Conclusions}
We have shown that for Chern-Simons quivers realized as fractional brane quivers in type IIA string theory, Seiberg-like dualities are easily found by changing the fractional brane basis, much like what happens for D3-brane quiver theories. In particular, we concentrated on maximally chiral quivers for D-branes at CY$_3$ cones with a single exceptional divisor. It would be worthwhile but technically challenging  to generalize the argument of this paper to completely general $CY_3$ quivers. In any case the subclass we analyzed contains the newly discovered ``chiral'' Seiberg-like dualities of \cite{Benini:2011mf}.

We believe our approach to be quite powerful, as demonstrated in the example of section \ref{examples: dP0 Seiberg duals}: All the computations of mutated dictionaries and dual CS quiver gauge theories are easily done on a computer, for arbitrarily complicated examples. Indeed some more general examples of mutated dictionaries have already appeared in the related work \cite{Closset:2012ep}.

There are a number of issues and open questions that we did not address. We should stress that the B-brane pictures is not sensible to the finer details of 3d Seiberg duality%
\footnote{Much like it does not know the difference between $U(N)$ and $SU(N)$ in 4d, for instance.}. It does not see the parity anomaly in the Abelian sector of chiral quivers, which so far has to be fixed ``by hand'' \cite{Benini:2011cma}. Moreover, it cannot account for the extra singlet fields (dual to monopole operators) which appear in 3d Seiberg duality when some effective CS level vanishes \cite{Aharony:1997gp, Benini:2011mf}. These shortcoming are expected since the B-model is a $g_s=0$ approximation of string theory, and we should not expect too much from B-branes. Seiberg duality for chiral quivers should thus also be further studied with field theory methods.

Originally, this work was motivated by a desire to understand 3d Seiberg duality in a manner conceptually similar to the Berenstein-Douglas approach to Seiberg duality \cite{Berenstein:2002fi}. Such point of view was also recently taken in \cite{Berenstein:2011dr}. In that respect we did not reach our goal, because we could not abstract the discussion from the underlying D-branes. It would be interesting to pursue this avenue. A possibility would be to formalize $\cN=2$ Chern-Simons quiver theories as families of ``decorated'' quiver representations, similar (but different) from those studied in \cite{2007arXiv0704.0649D}, and try to understand how quiver mutation induces an action on such objects.

Finally, we believe mutations of fractional branes should be studied again in their own right by physicists. Previous work on D-branes on Fano varieties \cite{Herzog:2003zc,Herzog:2004qw,  Herzog:2005sy, Hanany:2006nm}  focussed on complete strongly exceptional collections%
\footnote{As understood in \cite{Herzog:2004qw} in particular, it is important that the strongly exceptional collection be complete, which means that it generates the B-brane category on the corresponding space. }, but it seems that the more general notion one should use is the one of tilting collection, which is slightly weaker.
We conjectured in section \ref{sec: mutation, branes and Qs} that there exists a good notion of mutation of a tilting collection which gives another tilting collection, but this certainly calls for a proof. Indeed we did not even define the correct mutation operation in term of sheaves, but only in term of their charges.
One should also study more seriously the relation between tilting collections on $\tilde{B}_4$ and on its canonical bundle $\tilde{Y}= \cO_{\tilde{B}_4}(K)$.
We hope that continuing progress in the mathematical literature on the subject will allow to address those questions rigorously.

\section*{Acknowledgments}
I am grateful to Ofer Aharony and Mauricio Romo for interesting discussions and feedback, and especially to Stefano Cremonesi for continous discussion and exchanges related to this work.
This work is supported by a  Feinberg Postdoctoral Fellowship at the Weizmann Institute of Sciences.

\appendix

\section{More on quivers and Seiberg duality}\label{appendix: on quiver and such}
A category is a class of objects and a class of map between them, called morphisms, which can be composed naturally. There are all sorts of categories satisfying more specialized axioms. A simple kind of categories are Abelian categories, which are such that every morphism has a kernel and cokernel (in the appropriate abstract sense of homological algebra). For every Abelian category $\mathsf{A}$ we can construct a derived category $D(\mathsf{A})$ whose objects are chain complexes of the objects in $\mathsf{A}$ (up to some equivalence relations); see \cite{Aspinwall:2004jr} for a physics-oriented introduction.

In the following we review some elementary facts about the category of quiver representations for $CY_3$ quivers, and we review the Berenstein-Douglas \cite{Berenstein:2002fi} understanding of Seiberg duality as an equivalence of derived category of quiver representations. These facts are reviewed for completeness: they are not essential to the main flow of ideas of the paper, but they provide its conceptual context.

 We always work over the field $\bC$.

\subsection{Quivers, path algebra and $\cA$-modules}

\paragraph{Quivers and their representations.} Formally, a quiver $\cQ= (Q_0, Q_1, s, t)$ is a set of nodes $Q_0$, a set of arrows $Q_1$ between the nodes, and two functions
\be
s: Q_1\rightarrow Q_0\, , \qquad t: Q_1\rightarrow Q_0
\ee
such that $s(a_{ij})= i$ (the ``source'' node) and $t(a_{ij})=j$ (the ``target'' node). A path from $i$ to $j$, $p: i\rightarrow j$, is a sequence of arrows\footnote{Remark that we write a paths  $a\cdots b$  from $s(a)$ to $t(b)$ in the opposite order with respect to composition of maps, because it agrees better with usual conventions in supersymmetric quiver gauge theories.}
\be
p=a_1 a_2\cdots a_p \quad\qquad \text{such that} \quad  t(a_1)= s(a_2)\, , \; \cdots, \; t(a_{p-1})= s(a_p)\, .
\ee
and with $s(a_1)=i$, $t(a_p)=j$. A quiver representation $X=(V_i, X_a)$ is a choice of vector space $V_i$ for each node and a linear map $X_a$ for each arrow. A morphism $\phi$ between two quiver representations $X$, $X'$ is a set of linear maps $\phi_i: V_i\rightarrow V'_i$ such that $\phi_i X'_a = X_a \phi_j$ for any $a:i\rightarrow j$. For a quiver with relations the maps $X_a$ must also satisfy the relations. Two representations are identified if there exist an isomorphism (invertible morphism) between them. Quiver representations and their morphisms form the category of quiver representations, $\cQ-\mathrm{rep}$ for short. Moreover $\cQ-\mathrm{rep}$ is an Abelian category, rather obviously since the morphisms are linear maps.

\paragraph{Path algebra.} Recall that an algebra is a vector space equipped with a multiplication (not necessarily commutative). The unconstrained path algebra $\bC\cQ$ of a quiver $\cQ$ is the algebra generated by all the possible paths in the quiver, with the multiplication given by the concatenation of paths, namely $a\cdot b= ab$ if $t(a)=s(b)$, and zero otherwise%
\footnote{As a simple example, take a quiver with one single node and one arrow $a:1\rightarrow 1$. In that particular case $\bC\cQ= \bC[a]$, the (commutative) ring of polynomials in $a$. The next simplest example is a single node with $r$ arrows $a_1, \cdots, a_r :1\rightarrow 1$. In that case $\bC\cQ= \bC\langle a_1, \cdots, a_r\rangle$, the free associative algebra of words in an alphabet of $r$ letters.}. For each vertex we define a trivial path $e_i$ satisfying $e_i e_j = \delta_{ij} e_i$. They act as a projectors and provide an identity element $1=\sum_i e_i$ in $\bC\cQ$.
For a quiver with superpotential, the path algebra $\cA$ is the algebra $\bC\cQ$ modulo the relations (\ref{F-term rels}),
\be
\cA \equiv \bC\cQ/(\partial W)\, ,
\ee
with $(\partial W)$ the ideal of $\bC\cQ$ generated by (\ref{F-term rels}).

\paragraph{Modules.} Given any ring $R$, a (right) R-module $M$ is an abelian group $M$ together with a (right) action of $R$, $R: M\times R\rightarrow M$ acting just like a scalar multiplication on a vector space. Morphisms between modules are homomorphisms, $\phi: M \rightarrow M'$ such that $\forall m,n \in M$ and  $r,s\in R$, $\phi(m r +n s )= \phi(m)r+\phi(n)s\in M'$. The modules and morphism over a given ring $R$ form the category $R-\mathrm{mod}$.

A module $M$ is a submodule of $M'$ if there exist an injective morphism $\phi: M \rightarrow M'$. The dimension of a module $M$ is its dimension as an Abelian group.

A \textbf{free $R$-module} is a module which has a basis (a generating linearly independent set); free modules are what come closest to vector spaces.

A \textbf{projective $R$-module} is a module which is the summand of a free module. In other words, the $R$-module $M$ is projective if there exists an $R$-module $N$ such that $M\oplus N$ is free.

A \textbf{simple $R$-module} is a module which has no  proper submodule (no submodule other than itself and the trivial one).

\paragraph{$\cA$-modules and $\cQ$-representations.} In particular, we can consider $\cA$-modules. A standard result is that there exist an equivalence of categories
\be\label{equiv categories Amod Qrep}
\cA-\mathrm{mod}\cong  \cQ-\mathrm{rep}\, .
\ee
The equivalence is given by the following maps. For any quiver representation $X=(V_i, X_a)$, we define a (right) module%
\footnote{The fact that is is a right-module instead of a left-module is just due to our particular notation for the paths. Thus we will keep the unfortunate notation $xa=y$ for $x\in V_i$, $y\in V_j$ and $a:V_i\rightarrow V_j$.}
$\cX$ as the abelian group $\cX= \oplus_i V_i$, with the (right) action of the paths in $\cA$ on $\cX$ given by
\bea
x\, a_1\cdots a_p & \;=\; x\, \pi_{s(a_1)}X_{a_1}\cdots X_{a_p} \iota_{t(p)}  \, ,\\
x\, e_i & \;=\; x\, \pi_i \iota_i\, ,
\eea
$ \forall\, x\in \cX$, with  $\iota_i$ and $\pi_i$ the inclusion and projection maps $V_i \stackrel{\iota_i}{\rightarrow} \cX \stackrel{\pi_i}{\rightarrow} V_i$. In the other direction, for any $\cA$-module $\cX$, we have a representation $X=(V_i, X_a)$ with
\bea
V_i& \;=\; \cX e_i\, ,\\
xX_a & \;=\; x a \, ,
\eea
according to the way $a\in \cA$ was represented on $\cX$. Hence we can talk interchangeably about quiver representation or $\cA$-modules, but sometimes one or the other language is more convenient. The dimension vector $\bm{N}$ of a $\cA$-module is defined as the $\bm{N}$ of the associated $\cQ$-representation. Note that two modules $M$, $N$ are the same if there is an isomorphism between them, by definition, and thus an $\cA$-module of dimension $\bm{N}$ corresponds to the gauge orbit of a particular F-term solution in a supersymmetric quiver theory $(\cQ, \bm{N})$.

\subsection{Seiberg duality as a tilting equivalence}\label{app:SD and derived equiv}
Consider a $CY_3$ quiver $\cQ$ and its path algebra $\cA$. Of particular interest are the simple $\cA$-modules, because they correspond to ``single brane'' states. To each node of the quiver we associate a simple module $e_i$ and a projective module $P_i$.  The module $e_i$ is the quiver representation with $V_j = \delta_{i j} \bC$, and correspond to a fractional brane. On the other hand $P_i$ is the $\cA$-module consisting of all paths ending at node $i$:
\be
P_i = \cA e_i\, .
\ee
These right $\cA$-modules are projective since $\oplus_i P_i = \cA$, and we have the identity
\be
\cA= \text{End} (\oplus_i P_i)^{op}\, ,
\ee
where $^{op}$ corresponds to reversing all the arrows in a path algebra. Now, consider going to the derived category of $\cA$-modules, $D(\cA-\text{mod})$; the image of $P_i$ in $D(\cA-\text{mod})$ is the single entry complex $P_i \equiv 0\rightarrow P_i\rightarrow 0 $ in position $0$. The objects $P_i$ form a tilting collection in the derived category, meaning that they satisfy
\be\label{condition tilting coll 00}
\text{Ext}^q(P_i, P_j)= 0\, \qquad  \qquad\forall q>0\,
\ee
and generate the full category $D(\cA-\text{mod})$.
If we have \emph{any} tilting collection $\{P_i'\}$ in $D(\cA-\text{mod})$, we can define a path algebra
\be
\cA' = \text{End} (\oplus_i P'_i)^{op}\, ,
\ee
and therefore a quiver $\cQ'$. One says that there exists a tilting equivalence between $\cQ$ and $\cQ'$. By Rickard's theorem \cite{Rickard1989}, any tilting equivalence is an equivalence of derived categories
\be
D(\cA) \cong D(\cA')\,
\ee
(and conversely any derived category equivalence can be realized as a tilting equivalence).
The proposal of \cite{Berenstein:2002fi} is that Seiberg duality for quivers is such a tilting equivalence. In \cite{Mukhopadhyay:2003ky} it was shown that the notion of ``doing Seiberg duality at node $i_0 \in Q_0$'' is realized by taking the following tilting collections:
\be
P'_i = P_i \, \quad \text{if}\; i\neq i_0\, ,    \qquad P_i' = \left(0\longrightarrow \oplus_{j \in I_L} P_j \stackrel{(a_{ji_0})}{\longrightarrow} P_{i_0}\longrightarrow 0 \right)\, ,
\ee
where $I_L\subset Q_0$ is the set of nodes connected to $i_0$ by ingoing arrows, and the first non-trivial entry in the complex $P_i'$ is at position $0$. This construction corresponds to the left mutation of quiver representations discussed in the main text, while a similar tilting collection can be constructed for right mutations. Indeed, this construction is also completely parallel to the mutations on the sheaves $E_i$ of (\ref{L and R mutation on sheaves}), with $P_i \sim E_i$.

\section{More on B-branes and quivers}\label{sec:App:sheaves and quivers}
The relation between a conical CY$_3$ $Y$ and a quiver $\cQ$ can be summarized in a single line,
\be\label{rel der cat}
D(\text{Coh}\,\tilde{Y}) \cong D(\cA_{\cQ}\text{-mod})\, ,
\ee
namely it is a derived category equivalence similar to the one for quivers reviewed in the last Appendix.
$\text{Coh}\,\tilde{Y}$ is the Abelian category of coherent sheaves on $\tilde{Y}$. In more physical terms, $D(\text{Coh}\,\tilde{Y})$ is the category of B-branes, the boundary states in the topological B-model. In the following we provide some more context for this relation, at an intuitive level.%
\footnote{ We refer to \cite{Aspinwall:2004jr,Sharpe:2003dr} for a thorough introduction to the subject.}

We consider a crepant resolution $\tilde{Y}$ of the singularity $Y$. In the large volume limit (and for $g_s=0$), D-branes are objects that wrap some cycles in $\tilde{Y}$, and carry some Chan-Paton bundle. We are ultimately interested in BPS D-branes (preserving half of the 8 supercharges of the $\tilde{Y}$ background), which must wrap holomorphic cycles and carry some holomorphic vector bundle (amongst other conditions).
Mathematically they are called coherent sheaves. To obtain the most general brane, we should allow for branes and anti-branes at the same time; for instance we can think of a stack of brane/anti-branes wrapping $\tilde{Y}$ and hope that tachyon condensation can give us the most general BPS D-brane as bound states, in the spirit of Sen's conjecture \cite{Sen:1998sm}. One can formalize this by considering \emph{complexes} of coherent sheaves%
\footnote{A more rigorous justification for introducing complexes involves the BRST operator of the topological B-model \cite{Douglas:2000gi}.},
\be
\label{generic element of D(Y)}
\cdots \rightarrow \mathsf{E}_{(-2)} \rightarrow \mathsf{E}_{(-1)} \rightarrow \mathsf{E}_{(0)} \rightarrow \mathsf{E}_{(1)} \rightarrow \cdots \;,
\ee
where the sheaf $\mathsf{E}_{(n)}$ can be thought of as a brane for $n$ even and as an anti-brane for $n$ odd. Considering the physically relevant equivalence classes of objects leads to the picture of D-branes as objects in the derived category of coherent sheaves  \cite{1994alg.geom.11018K, Douglas:2000gi}. A single coherent sheaf $\mathsf{E}$ lifts to the trivial complex
\be
\label{trivial complex for E}
 0 \rightarrow \mathsf{E}_{(0)} \rightarrow 0  \;
\ee
in $D(\text{Coh}\, \tilde{Y})$, which we simply write as $\mathsf{E}$.

The category $D(\text{Coh}\, \tilde{Y})$ contains many more objects than BPS D-branes. Indeed, already at the level of a brane wrapping some cycle the BPS condition is more restrictive than just picking up some holomorphic representative \cite{Harvey:1996gc}. The objects of $D(\text{Coh}\, \tilde{Y})$ are \emph{B-branes}, the branes in the topological B-model, and as such they do not depend on the K\"ahler moduli of the CY$_3$ background. On the other hand the set of BPS D-branes very much depend on the K\"ahler structure. Given a K\"ahler structure, a B-brane is called \emph{stable} if it corresponds to a BPS D-brane. This separation of the problem into an ``holomorphic part'' and and ``real'' (and harder) part is a traditional theme in supersymmetric theories. In the supersymmetric quiver description  this corresponds to the familiar distinction between F-terms and D-terms.

Suppose that we find a finite collection of sheaves $\cE=\{\mathsf{P}_i\}$  which form a tilting collection, namely they generates the full $D(\text{Coh}\, \tilde{Y})$ and  satisfy the conditions
\be\label{condition tilting coll}
\text{Ext}^q(\mathsf{P}_i, \mathsf{P}_j)= 0\, \qquad  \qquad\forall q>0\, .
\ee
These $\mathsf{P}_i$'s provide a so-called tilting sheaf
\be\label{tilting sheaf}
T= \mathsf{P}_1 \oplus \mathsf{P}_2\oplus \cdots \oplus \mathsf{P}_G\, ,
\ee
and they define an algebra%
\be
\cA= \mathrm{End} (T)^{\text{op}} = \oplus_{i, j}\,\text{Hom}(\mathsf{P}_i, \mathsf{P}_j)^{\text{op}}
\ee
which is a $CY_3$ quiver algebra. The object (\ref{tilting sheaf}) gives the isomorphism (\ref{rel der cat}):
\bea
\text{Hom}(T,-)\; &:\; D(\text{Coh}Y)\rightarrow D(\cA\text{-mod})\, , \\
T\otimes_{\cA}-\; &:\;  D(\cA\text{-mod})\rightarrow D(\text{Coh}Y)\, .
\eea
See for instance \cite{Aspinwall:2004vm, Aspinwall:2008jk}. Under this isomorphism the sheaves $\mathsf{P}_i$ map to the projective $\cA$-modules $P_i$ of Appendix \ref{app:SD and derived equiv}.

\section{Map between FI and K\"ahler parameters}\label{App: rel FI and Kparam}
It is well known that fractional branes couple to the K\"ahler parameters of the string theory background through Fayet-Iliopoulos parameters \cite{Douglas:1996sw}, but the explicit map between the two is not often explicitly given. Here we provide such a map in the case of a resolved CY$_3$ cone of the type studied in the main text, $\tilde{Y}= \cO_{B_4}(K)$.

The complexified K\"ahler parameters of $\tilde{Y}$ seen by the type II string are
\be
t_{\alpha}\,\equiv \, \int_{\cC_{\alpha}}  (B+ i\, J)  \, \equiv \, b_{\alpha} + i\, \chi_{\alpha}\, ,
\ee
where the 2-cycles $\cC_{\alpha}$, $\alpha= 1, \cdots, m$, where defined in (\ref{def 2 cycles and cI}), and thus $\tilde{Y}$ has a K\"ahler moduli space $\cM_K$ of real dimension $2m$.

Let $\mathsf{E}^{\vee}$ be a D-brane state with $ch(\mathsf{E}^{\vee}) = (r(\mathsf{E}^{\vee}),\,  c_1^{\alpha}(\mathsf{E}^{\vee}) ,\, ch_2( \mathsf{E}^{\vee} ) )$. The \emph{central charge} of $\mathsf{E}^{\vee}$ is the complex number
\be\label{exact central charge Z}
Z(\mathsf{E}) \, = \, r(\mathsf{E}^{\vee})\, \Pi_6 + c_1^{\alpha}(\mathsf{E}^{\vee}) \,\Pi_{4, \alpha} + ch_2(\mathsf{E}^{\vee}) \,\Pi_2\, ,
\ee
where $\Pi_4$, $\Pi_{2, \alpha}$ and $\Pi_0$ are so-called \emph{periods} associated to the states with Chern characters $(1,0,0)$, $(0,1,0)$ and $(0,0,1)$, respectively.
At large volume ($\chi_{\alpha} \rightarrow \infty$, $\forall \alpha$), the central charge is given explicitly by
\be\label{central charge Z at LV}
Z(\mathsf{E}^{\vee}) = \int_{\tilde{B}_4} e^{B+ iJ} ch(\mathsf{E}^{\vee}) \sqrt{\frac{\mathrm{\hat A}(T \tilde{B}_4)}{\mathrm{\hat A}(N \tilde{B}_4)}} \, \; + \; o(e^{2\pi i t_{\alpha}})\, ,
\ee
and this gives
\bea
&\Pi_6 & &= \quad \frac{1}{2}\int_{\tilde{B}_4} (B+iJ)^2 \, + \frac{1}{24}\chi(\tilde{B}_4) \\
&\Pi_{4, \alpha}  & &=   \quad t_{\alpha}    \\
&\Pi_2 & &=  \quad 1
\eea
The term $\chi(\tilde{B}_4)/24$ comes from the gravitational term in (\ref{central charge Z at LV}), with $\chi(\tilde{B}_4)= m+2$ the Euler character of $\tilde{B}_4$ \cite{Cheung:1997az}. The central charge is invariant under the large gauge transformation $B \rightarrow B + D$, $F\rightarrow F-D$ with $D\in H^2(\tilde{B}_4, \bZ)$. This is obvious for the large volume expression (\ref{central charge Z at LV}), while imposing that the exact central charge (\ref{exact central charge Z}) is also invariant imposes non-trivial constraints on the exact periods. We can show in that way that the relations  $\Pi_{4, \alpha}  =  t_{\alpha}$ and $\Pi_2 = 1$ are in fact exact. $\Pi_2$ corresponds to the point-like object on $\tilde{Y}$, denoted $\cO_p$.

Near the quiver locus (when all the central charges $Z(\mathsf{E}_i^{\vee})$ are almost aligned, the FI parameters $\xi_i$ on the fractional branes are simply \cite{Douglas:2000ah}
\be\label{FI eq ImZ}
\xi_i = \text{Im} Z(\mathsf{E}_i^{\vee})\, .
\ee
Let us denote $\bm{\xi}= (\xi_i)$ the vector of $m+2$ FI parameters of the quiver, and introduce
\be\label{def of chi generalized}
\bm{\chi} \equiv (\chi_0, \chi_{\alpha}, 0)\, ,
\ee
corresponding to the $m+1$ K\"ahler parameters transverse to $\cM_Q$. From the expression (\ref{exact central charge Z}) and the fact that $\Pi_{4, \alpha}= t_{\alpha}$ exactly, we can rewrite (\ref{FI eq ImZ}) as:
\be\label{dico xi to chi}
\bm{\xi} \, = \, Q^{\vee}\, \bm{\chi}\, ,
\ee
where $Q^{\vee}$ is the brane charge dictionary introduced in section \ref{subsec: from Dbrane to quivers}. One can easily show that $\sum_i r_i \xi_i=0$ by construction, with the $r_i$ defined in (\ref{definition of ri}). We also have the identification
\be\label{def of chi0}
\chi_0 \,\equiv \, \text{Im} \Pi_6 \, = \,  \int_{\tilde{B}_4} B\wedge J \, =\,  \,  b_{\alpha} (\cI^{-1})^{\alpha\beta} \chi_{\beta}.
\ee
The second equality is the large volume result, which can be modified by $\alpha'$ corrections.

\section{Charges of diagonal monopoles in $\cN=2$ CS quivers}\label{App: monopole operator charges}
The so-called diagonal monopole operators in $\cN=2$ CS quiver theories are monopole operators which insert the same flux for each gauge group $U(N_i)$ ---see the precise definition below. They correspond to gravitons along the M-theory circle in the M-theory $AdS_4$ dual, and to D0-branes in type IIA.
In \cite{Benini:2011cma} these operators were discussed in detail in the case of \emph{toric} quivers. In this Appendix we present the appropriate generalization of the analysis of \cite{Benini:2011cma} to the more general case.

Consider an CS quiver theory $(\cQ, \bm{N}, \bm{k})$. By assumption, $\cQ$ describes D-branes at a $CY_3$ singularity, and the point-like D-brane $\cO_p$ corresponds to a quiver with gauge group
\be\label{quiver for point-like D-brane}
\cG= \prod_{i=1}^G U(r_i)\, ,
\ee
with $r_i\geq 1$ (in a toric quiver $r_i=1$). The bare \emph{diagonal monopole operator} of flux $n$ is the operator that creates the fluxes
\be
H_{i} = (\underbrace{n, \, n, \cdots, \, n}_{r_i \; \text{times}},\, 0, \cdots, \, 0)\, , \qquad \, i= 1, \cdots, G\, ,
\ee
in the Cartan of $\cG$.
We denote such an operator by $T^{(n)}$ (and by $T$, $\tilde{T}$ the special cases $n = \pm 1$). It transforms under the gauge group $U(N_i)$ according to the irreducible representation of highest weight
\be
\label{full hw}
w_i = (g_{i,1}, g_{i,2}, \cdots, g_{i,N_i}) \;,
\ee
with
\be
g_{i,\; l}[T^{(n)}] =  \begin{cases} k_i n -\frac{|n|}{2} \sum_{i=1}^G A_{ij} N_j & \quad \text{if} \quad l=1, \cdots, r_i\, ,\\
0 & \quad \text{if} \quad l> r_i\, ,\end{cases}
\ee
where $A_{ij}$ is the adjacency matrix of $\cQ$.

Assuming that
\be\label{assumption Tr eq 0}
 \sum_{X_{ij}} ( R[X_{ij}] - 1 ) r_i r_j + \sum_i^G  r_i^2 =0\, ,
\ee
we can write the R-charge of the bare monopole as:
\be\label{R charge in general}
R[T^{(n)}] = -\frac{|n|}{2} \sum_{X_{ij}} ( R[X_{ij}] - 1 ) (r_i N_j+r_jN_i) \, -\frac{|n|}{2} \sum_i^G 2 r_i N_i\, .
\ee
When $N_i= r_i \tilde{N}$, the R-charge vanishes, due to (\ref{assumption Tr eq 0}). Therefore, if we consider generic ranks $N_i= r_i \tilde{N} + M_i$, we can replace $N_i$ by $M_i$ in the last formula.
The assumption (\ref{assumption Tr eq 0}) corresponds to the cancelation of the (global) gravitational anomaly $\Tr(R)$ in 4 dimensions; it does not need to hold in general, but it does for any D3-brane quiver with gauge group (\ref{quiver for point-like D-brane}) that we know of%
\footnote{At large $N$ this follows from having an $AdS_5$ supergravity dual \cite{Henningson:1998gx}.}
 ---in particular it holds for toric quivers \cite{Franco:2005rj}.

Remark also that for the family of $dP_0$ quivers of Figure \ref{fig: general dP0 quiver} in section \ref{examples: dP0 Seiberg duals} the constraint (\ref{assumption Tr eq 0}) is the Markov equation (\ref{Markov eq for dP0}), once we impose that $R[XYZ]=2$ from the superpotential.

\bibliography{bib3dSeibergDual}{}

\providecommand{\href}[2]{#2}\begingroup\raggedright\begin{thebibliography}{10}

\bibitem{Seiberg:1994pq}
N.~Seiberg, ``{Electric - magnetic duality in supersymmetric nonAbelian gauge
  theories},'' \href{http://dx.doi.org/10.1016/0550-3213(94)00023-8}{{\em
  Nucl.Phys.} {\bfseries B435} (1995) 129--146},
  \href{http://arxiv.org/abs/hep-th/9411149}{{\ttfamily arXiv:hep-th/9411149
  [hep-th]}}.

\bibitem{Aharony:2008gk}
O.~Aharony, O.~Bergman, and D.~L. Jafferis, ``{Fractional M2-branes},''
  \href{http://dx.doi.org/10.1088/1126-6708/2008/11/043}{{\em JHEP} {\bfseries
  0811} (2008) 043}, \href{http://arxiv.org/abs/0807.4924}{{\ttfamily
  arXiv:0807.4924 [hep-th]}}.

\bibitem{Giveon:2008zn}
A.~Giveon and D.~Kutasov, ``{Seiberg Duality in Chern-Simons Theory},''
  \href{http://dx.doi.org/10.1016/j.nuclphysb.2008.09.045}{{\em Nucl. Phys.}
  {\bfseries B812} (2009) 1--11},
\href{http://arxiv.org/abs/0808.0360}{{\ttfamily arXiv:0808.0360 [hep-th]}}.

\bibitem{Amariti:2009rb}
A.~Amariti, D.~Forcella, L.~Girardello, and A.~Mariotti, ``{3D Seiberg-like
  Dualities and M2 Branes},''
  \href{http://dx.doi.org/10.1007/JHEP05(2010)025}{{\em JHEP} {\bfseries 05}
  (2010) 025},
\href{http://arxiv.org/abs/0903.3222}{{\ttfamily arXiv:0903.3222 [hep-th]}}.

\bibitem{Niarchos:2009aa}
V.~Niarchos, ``{R-charges, Chiral Rings and RG Flows in Supersymmetric
  Chern-Simons-Matter Theories},''
  \href{http://dx.doi.org/10.1088/1126-6708/2009/05/054}{{\em JHEP} {\bfseries
  05} (2009) 054},
\href{http://arxiv.org/abs/0903.0435}{{\ttfamily arXiv:0903.0435 [hep-th]}}.

\bibitem{Hanany:1996ie}
A.~Hanany and E.~Witten, ``{Type IIB superstrings, BPS monopoles, and
  three-dimensional gauge dynamics},''
  \href{http://dx.doi.org/10.1016/S0550-3213(97)00157-0}{{\em Nucl.Phys.}
  {\bfseries B492} (1997) 152--190},
  \href{http://arxiv.org/abs/hep-th/9611230}{{\ttfamily arXiv:hep-th/9611230
  [hep-th]}}.

\bibitem{Cremonesi:2010ae}
S.~Cremonesi, ``{Type IIB construction of flavoured ABJ(M) and fractional M2
  branes},'' \href{http://dx.doi.org/10.1007/JHEP01(2011)076}{{\em JHEP}
  {\bfseries 01} (2011) 076},
\href{http://arxiv.org/abs/1007.4562}{{\ttfamily arXiv:1007.4562 [hep-th]}}.

\bibitem{Aganagic:2009zk}
M.~Aganagic, ``{A Stringy Origin of M2 Brane Chern-Simons Theories},''
\href{http://arxiv.org/abs/0905.3415}{{\ttfamily arXiv:0905.3415 [hep-th]}}.

\bibitem{Benini:2009qs}
F.~Benini, C.~Closset, and S.~Cremonesi, ``{Chiral flavors and M2-branes at
  toric CY4 singularities},''
  \href{http://dx.doi.org/10.1007/JHEP02(2010)036}{{\em JHEP} {\bfseries 1002}
  (2010) 036}, \href{http://arxiv.org/abs/0911.4127}{{\ttfamily arXiv:0911.4127
  [hep-th]}}.

\bibitem{Jafferis:2009th}
D.~L. Jafferis, ``{Quantum corrections to N=2 Chern-Simons theories with flavor
  and their AdS(4) duals},'' \href{http://arxiv.org/abs/0911.4324}{{\ttfamily
  arXiv:0911.4324 [hep-th]}}.

\bibitem{Benini:2011cma}
F.~Benini, C.~Closset, and S.~Cremonesi, ``{Quantum moduli space of
  Chern-Simons quivers, wrapped D6-branes and AdS4/CFT3},''
  \href{http://dx.doi.org/10.1007/JHEP09(2011)005}{{\em JHEP} {\bfseries 1109}
  (2011) 005}, \href{http://arxiv.org/abs/1105.2299}{{\ttfamily arXiv:1105.2299
  [hep-th]}}.

\bibitem{Closset:2012ep}
C.~Closset and S.~Cremonesi, ``{Toric Fano varieties and Chern-Simons
  quivers},''
\href{http://arxiv.org/abs/1201.2431}{{\ttfamily arXiv:1201.2431 [hep-th]}}.

\bibitem{Aharony:2008ug}
O.~Aharony, O.~Bergman, D.~L. Jafferis, and J.~Maldacena, ``{N=6 superconformal
  Chern-Simons-matter theories, M2-branes and their gravity duals},''
  \href{http://dx.doi.org/10.1088/1126-6708/2008/10/091}{{\em JHEP} {\bfseries
  10} (2008) 091},
\href{http://arxiv.org/abs/0806.1218}{{\ttfamily arXiv:0806.1218 [hep-th]}}.

\bibitem{Benini:2011mf}
F.~Benini, C.~Closset, and S.~Cremonesi, ``{Comments on 3d Seiberg-like
  dualities},'' \href{http://dx.doi.org/10.1007/JHEP10(2011)075}{{\em JHEP}
  {\bfseries 1110} (2011) 075},
  \href{http://arxiv.org/abs/1108.5373}{{\ttfamily arXiv:1108.5373 [hep-th]}}.
  * Temporary entry *.

\bibitem{Aspinwall:2004vm}
P.~S. Aspinwall and I.~V. Melnikov, ``{D-branes on vanishing del Pezzo
  surfaces},'' \href{http://dx.doi.org/10.1088/1126-6708/2004/12/042}{{\em
  JHEP} {\bfseries 0412} (2004) 042},
  \href{http://arxiv.org/abs/hep-th/0405134}{{\ttfamily arXiv:hep-th/0405134
  [hep-th]}}.

\bibitem{Herzog:2004qw}
C.~P. Herzog, ``{Seiberg duality is an exceptional mutation},''
  \href{http://dx.doi.org/10.1088/1126-6708/2004/08/064}{{\em JHEP} {\bfseries
  0408} (2004) 064}, \href{http://arxiv.org/abs/hep-th/0405118}{{\ttfamily
  arXiv:hep-th/0405118 [hep-th]}}.

\bibitem{Brandhuber:1997ta}
A.~Brandhuber, J.~Sonnenschein, S.~Theisen, and S.~Yankielowicz, ``{Brane
  configurations and 4-D field theory dualities},''
  \href{http://dx.doi.org/10.1016/S0550-3213(97)00414-8}{{\em Nucl.Phys.}
  {\bfseries B502} (1997) 125--148},
\href{http://arxiv.org/abs/hep-th/9704044}{{\ttfamily arXiv:hep-th/9704044
  [hep-th]}}.

\bibitem{Elitzur:1997hc}
S.~Elitzur, A.~Giveon, D.~Kutasov, E.~Rabinovici, and A.~Schwimmer, ``{Brane
  dynamics and N=1 supersymmetric gauge theory},''
  \href{http://dx.doi.org/10.1016/S0550-3213(97)00446-X}{{\em Nucl.Phys.}
  {\bfseries B505} (1997) 202--250},
\href{http://arxiv.org/abs/hep-th/9704104}{{\ttfamily arXiv:hep-th/9704104
  [hep-th]}}.

\bibitem{Cachazo:2001sg}
F.~Cachazo, B.~Fiol, K.~A. Intriligator, S.~Katz, and C.~Vafa, ``{A Geometric
  unification of dualities},''
  \href{http://dx.doi.org/10.1016/S0550-3213(02)00078-0}{{\em Nucl.Phys.}
  {\bfseries B628} (2002) 3--78},
  \href{http://arxiv.org/abs/hep-th/0110028}{{\ttfamily arXiv:hep-th/0110028
  [hep-th]}}.

\bibitem{Feng:2002kk}
B.~Feng, A.~Hanany, Y.~H. He, and A.~Iqbal, ``{Quiver theories, soliton spectra
  and Picard-Lefschetz transformations},'' {\em JHEP} {\bfseries 02} (2003)
  056,
\href{http://arxiv.org/abs/hep-th/0206152}{{\ttfamily arXiv:hep-th/0206152}}.

\bibitem{Aharony:2010af}
O.~Aharony, D.~Jafferis, A.~Tomasiello, and A.~Zaffaroni, ``{Massive type IIA
  string theory cannot be strongly coupled},''
  \href{http://dx.doi.org/10.1007/JHEP11(2010)047}{{\em JHEP} {\bfseries 1011}
  (2010) 047},
\href{http://arxiv.org/abs/1007.2451}{{\ttfamily arXiv:1007.2451 [hep-th]}}.

\bibitem{2006math.....12139G}
V.~{Ginzburg}, ``{Calabi-Yau algebras},'' {\em ArXiv Mathematics e-prints}
  (Dec., 2006) , \href{http://arxiv.org/abs/arXiv:math/0612139}{{\ttfamily
  arXiv:math/0612139}}.

\bibitem{Douglas:2000qw}
M.~R. Douglas, B.~Fiol, and C.~Romelsberger, ``{The Spectrum of BPS branes on a
  noncompact Calabi-Yau},''
  \href{http://dx.doi.org/10.1088/1126-6708/2005/09/057}{{\em JHEP} {\bfseries
  0509} (2005) 057}, \href{http://arxiv.org/abs/hep-th/0003263}{{\ttfamily
  arXiv:hep-th/0003263 [hep-th]}}.

\bibitem{Berenstein:2002fi}
D.~Berenstein and M.~R. Douglas, ``{Seiberg duality for quiver gauge
  theories},''
\href{http://arxiv.org/abs/hep-th/0207027}{{\ttfamily arXiv:hep-th/0207027}}.

\bibitem{Luty:1995sd}
M.~A. Luty and W.~Taylor, ``{Varieties of vacua in classical supersymmetric
  gauge theories},'' \href{http://dx.doi.org/10.1103/PhysRevD.53.3399}{{\em
  Phys. Rev.} {\bfseries D53} (1996) 3399--3405},
\href{http://arxiv.org/abs/hep-th/9506098}{{\ttfamily arXiv:hep-th/9506098}}.

\bibitem{2007arXiv0704.0649D}
H.~{Derksen}, J.~{Weyman}, and A.~{Zelevinsky}, ``{Quivers with potentials and
  their representations I: Mutations},'' {\em ArXiv e-prints} (Apr., 2007) ,
  \href{http://arxiv.org/abs/0704.0649}{{\ttfamily arXiv:0704.0649 [math.RA]}}.

\bibitem{2009arXiv0906.0761K}
B.~{Keller} and D.~{Yang}, ``{Derived equivalences from mutations of quivers
  with potential},'' {\em ArXiv e-prints} (June, 2009) ,
  \href{http://arxiv.org/abs/0906.0761}{{\ttfamily arXiv:0906.0761 [math.RT]}}.

\bibitem{Diaconescu:1997br}
D.-E. Diaconescu, M.~R. Douglas, and J.~Gomis, ``{Fractional branes and wrapped
  branes},'' {\em JHEP} {\bfseries 9802} (1998) 013,
  \href{http://arxiv.org/abs/hep-th/9712230}{{\ttfamily arXiv:hep-th/9712230
  [hep-th]}}.

\bibitem{Douglas:1996sw}
M.~R. Douglas and G.~W. Moore, ``{D-branes, quivers, and ALE instantons},''
  \href{http://arxiv.org/abs/hep-th/9603167}{{\ttfamily arXiv:hep-th/9603167
  [hep-th]}}.

\bibitem{Wijnholt:2002qz}
M.~Wijnholt, ``{Large volume perspective on branes at singularities},'' {\em
  Adv. Theor. Math. Phys.} {\bfseries 7} (2004) 1117--1153,
\href{http://arxiv.org/abs/hep-th/0212021}{{\ttfamily arXiv:hep-th/0212021}}.

\bibitem{Herzog:2003zc}
C.~P. Herzog, ``{Exceptional collections and del Pezzo gauge theories},''
  \href{http://dx.doi.org/10.1088/1126-6708/2004/04/069}{{\em JHEP} {\bfseries
  04} (2004) 069},
\href{http://arxiv.org/abs/hep-th/0310262}{{\ttfamily arXiv:hep-th/0310262}}.

\bibitem{Herzog:2005sy}
C.~P. Herzog and R.~L. Karp, ``{Exceptional collections and D-branes probing
  toric singularities},''
  \href{http://dx.doi.org/10.1088/1126-6708/2006/02/061}{{\em JHEP} {\bfseries
  0602} (2006) 061}, \href{http://arxiv.org/abs/hep-th/0507175}{{\ttfamily
  arXiv:hep-th/0507175 [hep-th]}}.

\bibitem{Franco:2005sm}
S.~Franco {\em et al.}, ``{Gauge theories from toric geometry and brane
  tilings},'' \href{http://dx.doi.org/10.1088/1126-6708/2006/01/128}{{\em JHEP}
  {\bfseries 01} (2006) 128},
\href{http://arxiv.org/abs/hep-th/0505211}{{\ttfamily arXiv:hep-th/0505211}}.

\bibitem{Hanany:2005ss}
A.~Hanany and D.~Vegh, ``{Quivers, tilings, branes and rhombi},''
  \href{http://dx.doi.org/10.1088/1126-6708/2007/10/029}{{\em JHEP} {\bfseries
  10} (2007) 029},
\href{http://arxiv.org/abs/hep-th/0511063}{{\ttfamily arXiv:hep-th/0511063}}.

\bibitem{Kennaway:2007tq}
K.~D. Kennaway, ``{Brane Tilings},''
  \href{http://dx.doi.org/10.1142/S0217751X07036877}{{\em Int. J. Mod. Phys.}
  {\bfseries A22} (2007) 2977--3038},
\href{http://arxiv.org/abs/0706.1660}{{\ttfamily arXiv:0706.1660 [hep-th]}}.

\bibitem{Gulotta:2008ef}
D.~R. Gulotta, ``{Properly ordered dimers, R-charges, and an efficient inverse
  algorithm},'' \href{http://dx.doi.org/10.1088/1126-6708/2008/10/014}{{\em
  JHEP} {\bfseries 0810} (2008) 014},
  \href{http://arxiv.org/abs/0807.3012}{{\ttfamily arXiv:0807.3012 [hep-th]}}.

\bibitem{Aspinwall:2004jr}
P.~S. Aspinwall, ``{D-branes on Calabi-Yau manifolds},''
  \href{http://arxiv.org/abs/hep-th/0403166}{{\ttfamily arXiv:hep-th/0403166
  [hep-th]}}.

\bibitem{Herzog:2006bu}
C.~P. Herzog and R.~L. Karp, ``{On the geometry of quiver gauge theories
  (Stacking exceptional collections)},''
  \href{http://arxiv.org/abs/hep-th/0605177}{{\ttfamily arXiv:hep-th/0605177
  [hep-th]}}.

\bibitem{Aspinwall:2005ur}
P.~S. Aspinwall and L.~M. Fidkowski, ``{Superpotentials for quiver gauge
  theories},'' \href{http://dx.doi.org/10.1088/1126-6708/2006/10/047}{{\em
  JHEP} {\bfseries 0610} (2006) 047},
  \href{http://arxiv.org/abs/hep-th/0506041}{{\ttfamily arXiv:hep-th/0506041
  [hep-th]}}.

\bibitem{Hanany:2006nm}
A.~Hanany, C.~P. Herzog, and D.~Vegh, ``{Brane tilings and exceptional
  collections},'' \href{http://dx.doi.org/10.1088/1126-6708/2006/07/001}{{\em
  JHEP} {\bfseries 0607} (2006) 001},
  \href{http://arxiv.org/abs/hep-th/0602041}{{\ttfamily arXiv:hep-th/0602041
  [hep-th]}}.

\bibitem{2009arXiv0909.2013B}
M.~{Bender} and S.~{Mozgovoy}, ``{Crepant resolutions and brane tilings II:
  Tilting bundles},'' {\em ArXiv e-prints} (Sept., 2009) ,
  \href{http://arxiv.org/abs/0909.2013}{{\ttfamily arXiv:0909.2013 [math.AG]}}.

\bibitem{Diaconescu:1999dt}
D.-E. Diaconescu and J.~Gomis, ``{Fractional branes and boundary states in
  orbifold theories},'' {\em JHEP} {\bfseries 0010} (2000) 001,
  \href{http://arxiv.org/abs/hep-th/9906242}{{\ttfamily arXiv:hep-th/9906242
  [hep-th]}}.

\bibitem{Douglas:2000ah}
M.~R. Douglas, B.~Fiol, and C.~Romelsberger, ``{Stability and BPS branes},''
  \href{http://dx.doi.org/10.1088/1126-6708/2005/09/006}{{\em JHEP} {\bfseries
  0509} (2005) 006}, \href{http://arxiv.org/abs/hep-th/0002037}{{\ttfamily
  arXiv:hep-th/0002037 [hep-th]}}.

\bibitem{King1994}
A.~D. King, ``{Moduli of representations of finite-dimensional algebras},''
  {\em Quart. J. Math. Oxford Ser. (2), Vol. 45, No. 180.} (1994) .

\bibitem{Aspinwall:2004mb}
P.~S. Aspinwall, ``{D-branes, Pi-stability and theta-stability},''
  \href{http://arxiv.org/abs/hep-th/0407123}{{\ttfamily arXiv:hep-th/0407123
  [hep-th]}}.

\bibitem{Martelli:2008cm}
D.~Martelli and J.~Sparks, ``{Symmetry-breaking vacua and baryon condensates in
  AdS/CFT},'' \href{http://dx.doi.org/10.1103/PhysRevD.79.065009}{{\em
  Phys.Rev.} {\bfseries D79} (2009) 065009},
  \href{http://arxiv.org/abs/0804.3999}{{\ttfamily arXiv:0804.3999 [hep-th]}}.

\bibitem{2008arXiv0811.2435K}
M.~{Kontsevich} and Y.~{Soibelman}, ``{Stability structures, motivic
  Donaldson-Thomas invariants and cluster transformations},'' {\em ArXiv
  e-prints} (Nov., 2008) , \href{http://arxiv.org/abs/0811.2435}{{\ttfamily
  arXiv:0811.2435 [math.AG]}}.

\bibitem{Aganagic:2010qr}
M.~Aganagic and K.~Schaeffer, ``{Wall Crossing, Quivers and Crystals},''
  \href{http://arxiv.org/abs/1006.2113}{{\ttfamily arXiv:1006.2113 [hep-th]}}.

\bibitem{Gaiotto:2009tk}
D.~Gaiotto and D.~L. Jafferis, ``{Notes on adding D6 branes wrapping RP$^3$ in
  AdS(4) $\times$ CP$^3$},''
\href{http://arxiv.org/abs/0903.2175}{{\ttfamily arXiv:0903.2175 [hep-th]}}.

\bibitem{Aharony:1997gp}
O.~Aharony, ``{IR duality in d = 3 N = 2 supersymmetric USp(2N(c)) and U(N(c))
  gauge theories},''
  \href{http://dx.doi.org/10.1016/S0370-2693(97)00530-3}{{\em Phys. Lett.}
  {\bfseries B404} (1997) 71--76},
\href{http://arxiv.org/abs/hep-th/9703215}{{\ttfamily arXiv:hep-th/9703215}}.

\bibitem{Martelli:2008rt}
D.~Martelli and J.~Sparks, ``{Notes on toric Sasaki-Einstein seven-manifolds
  and AdS(4) / CFT(3)},''
  \href{http://dx.doi.org/10.1088/1126-6708/2008/11/016}{{\em JHEP} {\bfseries
  0811} (2008) 016}, \href{http://arxiv.org/abs/0808.0904}{{\ttfamily
  arXiv:0808.0904 [hep-th]}}.

\bibitem{Bashkirov:2011vy}
D.~Bashkirov, ``{Aharony duality and monopole operators in three dimensions},''
  \href{http://arxiv.org/abs/1106.4110}{{\ttfamily arXiv:1106.4110 [hep-th]}}.
  * Temporary entry *.

\bibitem{Marolf:2000cb}
D.~Marolf, ``{Chern-Simons terms and the three notions of charge},''
\href{http://arxiv.org/abs/hep-th/0006117}{{\ttfamily arXiv:hep-th/0006117}}.

\bibitem{Benini:2007gx}
F.~Benini, F.~Canoura, S.~Cremonesi, C.~Nunez, and A.~V. Ramallo,
  ``{Backreacting Flavors in the Klebanov-Strassler Background},''
  \href{http://dx.doi.org/10.1088/1126-6708/2007/09/109}{{\em JHEP} {\bfseries
  09} (2007) 109},
\href{http://arxiv.org/abs/0706.1238}{{\ttfamily arXiv:0706.1238 [hep-th]}}.

\bibitem{Freed:1999vc}
D.~S. Freed and E.~Witten, ``{Anomalies in string theory with D-branes},''
  \href{http://arxiv.org/abs/hep-th/9907189}{{\ttfamily arXiv:hep-th/9907189
  [hep-th]}}.

\bibitem{Minasian:1997mm}
R.~Minasian and G.~W. Moore, ``{K theory and Ramond-Ramond charge},'' {\em
  JHEP} {\bfseries 9711} (1997) 002,
  \href{http://arxiv.org/abs/hep-th/9710230}{{\ttfamily arXiv:hep-th/9710230
  [hep-th]}}.

\bibitem{Sharpe:1999qz}
E.~R. Sharpe, ``{D-branes, derived categories, and Grothendieck groups},''
  \href{http://dx.doi.org/10.1016/S0550-3213(99)00535-0}{{\em Nucl.Phys.}
  {\bfseries B561} (1999) 433--450},
  \href{http://arxiv.org/abs/hep-th/9902116}{{\ttfamily arXiv:hep-th/9902116
  [hep-th]}}.

\bibitem{Gaiotto:2009mv}
D.~Gaiotto and A.~Tomasiello, ``{The gauge dual of Romans mass},''
  \href{http://dx.doi.org/10.1007/JHEP01(2010)015}{{\em JHEP} {\bfseries 1001}
  (2010) 015}, \href{http://arxiv.org/abs/0901.0969}{{\ttfamily arXiv:0901.0969
  [hep-th]}}.

\bibitem{Fujita:2009kw}
M.~Fujita, W.~Li, S.~Ryu, and T.~Takayanagi, ``{Fractional Quantum Hall Effect
  via Holography: Chern- Simons, Edge States, and Hierarchy},''
  \href{http://dx.doi.org/10.1088/1126-6708/2009/06/066}{{\em JHEP} {\bfseries
  06} (2009) 066},
\href{http://arxiv.org/abs/0901.0924}{{\ttfamily arXiv:0901.0924 [hep-th]}}.

\bibitem{Petrini:2009ur}
M.~Petrini and A.~Zaffaroni, ``{N=2 solutions of massive type IIA and their
  Chern-Simons duals},''
  \href{http://dx.doi.org/10.1088/1126-6708/2009/09/107}{{\em JHEP} {\bfseries
  09} (2009) 107},
\href{http://arxiv.org/abs/0904.4915}{{\ttfamily arXiv:0904.4915 [hep-th]}}.

\bibitem{Berenstein:2011dr}
D.~Berenstein and M.~Romo, ``{Monopole operators, moduli spaces and
  dualities},'' \href{http://arxiv.org/abs/1108.4013}{{\ttfamily
  arXiv:1108.4013 [hep-th]}}.

\bibitem{Rickard1989}
J.~Rickard, ``{Morita Theory for Derived Categories},'' {\em {J. London Math.
  Soc. 39}} .

\bibitem{Mukhopadhyay:2003ky}
S.~Mukhopadhyay and K.~Ray, ``{Seiberg duality as derived equivalence for some
  quiver gauge theories},''
  \href{http://dx.doi.org/10.1088/1126-6708/2004/02/070}{{\em JHEP} {\bfseries
  0402} (2004) 070}, \href{http://arxiv.org/abs/hep-th/0309191}{{\ttfamily
  arXiv:hep-th/0309191 [hep-th]}}.

\bibitem{Sharpe:2003dr}
E.~Sharpe, ``{Lectures on D-branes and sheaves},''
  \href{http://arxiv.org/abs/hep-th/0307245}{{\ttfamily arXiv:hep-th/0307245
  [hep-th]}}.

\bibitem{Sen:1998sm}
A.~Sen, ``{Tachyon condensation on the brane antibrane system},'' {\em JHEP}
  {\bfseries 08} (1998) 012,
\href{http://arxiv.org/abs/hep-th/9805170}{{\ttfamily arXiv:hep-th/9805170}}.

\bibitem{Douglas:2000gi}
M.~R. Douglas, ``{D-branes, categories and N = 1 supersymmetry},''
  \href{http://dx.doi.org/10.1063/1.1374448}{{\em J. Math. Phys.} {\bfseries
  42} (2001) 2818--2843},
\href{http://arxiv.org/abs/hep-th/0011017}{{\ttfamily arXiv:hep-th/0011017}}.

\bibitem{1994alg.geom.11018K}
M.~{Kontsevich}, ``{Homological Algebra of Mirror Symmetry},'' in {\em eprint
  arXiv:alg-geom/9411018}, p.~11018.
\newblock Nov., 1994.

\bibitem{Harvey:1996gc}
J.~A. Harvey and G.~W. Moore, ``{On the algebras of BPS states},''
  \href{http://dx.doi.org/10.1007/s002200050461}{{\em Commun. Math. Phys.}
  {\bfseries 197} (1998) 489--519},
\href{http://arxiv.org/abs/hep-th/9609017}{{\ttfamily arXiv:hep-th/9609017}}.

\bibitem{Aspinwall:2008jk}
P.~S. Aspinwall, ``{D-Branes on Toric Calabi-Yau Varieties},''
\href{http://arxiv.org/abs/0806.2612}{{\ttfamily arXiv:0806.2612 [hep-th]}}.

\bibitem{Cheung:1997az}
Y.-K.~E. Cheung and Z.~Yin, ``{Anomalies, branes, and currents},''
  \href{http://dx.doi.org/10.1016/S0550-3213(98)00115-1}{{\em Nucl.Phys.}
  {\bfseries B517} (1998) 69--91},
  \href{http://arxiv.org/abs/hep-th/9710206}{{\ttfamily arXiv:hep-th/9710206
  [hep-th]}}.

\bibitem{Henningson:1998gx}
M.~Henningson and K.~Skenderis, ``{The Holographic Weyl anomaly},'' {\em JHEP}
  {\bfseries 9807} (1998) 023,
  \href{http://arxiv.org/abs/hep-th/9806087}{{\ttfamily arXiv:hep-th/9806087
  [hep-th]}}.

\bibitem{Franco:2005rj}
S.~Franco, A.~Hanany, K.~D. Kennaway, D.~Vegh, and B.~Wecht, ``{Brane dimers
  and quiver gauge theories},''
  \href{http://dx.doi.org/10.1088/1126-6708/2006/01/096}{{\em JHEP} {\bfseries
  0601} (2006) 096}, \href{http://arxiv.org/abs/hep-th/0504110}{{\ttfamily
  arXiv:hep-th/0504110 [hep-th]}}.

\end{thebibliography}\endgroup
\bibliographystyle{utphys}

\end{document}